\documentclass[preprintnumbers,secnumarabic,amsmath,amssymb,nofootinbib,floatfix]{revtex4}
\usepackage{graphics}
\usepackage{graphicx}
\usepackage{dcolumn}
\usepackage{bm}
\usepackage{mathrsfs}
\usepackage{slashed}
\usepackage{amsmath}
\usepackage{epsfig}
\usepackage{amsfonts}
\usepackage{amssymb}
\usepackage{tikz}
\usepackage{tikz-cd}
\begin{document}

\title{ Color-kinematics duality, double copy and the unitarity method \\
for higher-derivative QCD and quadratic gravity }
\author{Gabriel Menezes}
\email{gabrielmenezes@ufrrj.br}
\affiliation{~\\ Departamento de F\'{i}sica, Universidade Federal Rural do Rio de Janeiro, 23897-000, Serop\'{e}dica, RJ, Brazil \\ }
%

\begin{abstract}
Here we discuss color-kinematics duality for higher-derivative QCD-like amplitudes. We explicitly show that the duality still holds in this case and it can be instrumental in constructing the associated quadratic-gravity amplitudes by using the double-copy prescription. This allows one to drastically simplify calculations. We also evaluate some tree-level Compton scattering amplitudes in higher-derivative Yang-Mills and quadratic gravity coupled with matter. Furthermore, we illustrate the application of generalized unitarity method for both cases by studying a specific one-loop amplitude.
\end{abstract}

\maketitle

\section{Introduction}

Gauge theories and gravitational interactions form the basis of our current understanding of the universe. The Standard Model of elementary particle physics, which encapsulates our prevailing knowledge of fundamental physics, is a gauge theory, whereas gravity, in accordance with the tenets of general relativity, is the manifestation of the dynamics of spacetime. To realize them in a unified framework seems to be a formidable task, since they seem to be strikingly different. Nevertheless, gauge and gravity theories share many features. For instance, when studying their perturbative expansions one finds that the underlying dynamics of both theories provide identical kinematical building blocks. 

The foundations unveiling such features were embodied by the discovery of a duality between color and kinematics and the subsequent double-copy perspective. Scattering amplitudes associated with several quantum field theories expose a double-copy pattern. In particular, this is a key property of many supergravity theories whose amplitudes have been studied in detail. The double copy prescription identifies a sort of ``confederation" of gauge and gravity phenomena by demanding an identical set of basic units for both theories, which vastly streamlines calculations which otherwise would be impossibly burdensome. Indeed, the double-copy construction provides an alternative way to understand gravitation by displaying manifest evidence of powerful relationships between gravity and other fundamental forces via color-kinematics duality~\cite{CK,Bern:08,Bern:100,Bern:10,Bern:2015ooa,Bern:2017yxu}.

The double copy predictions establish a computational framework which enables one to study scattering amplitudes for a wide variety of quantum field theories, improving our technical ability to perform difficult calculations to higher loop orders. The first solid indication of the existence of a web of theories satisfying double-copy relations was uncovered when Kawaii, Lewellen and Tye realized that closed string tree-level amplitudes can be obtained from sums over permutations of Chan-Paton-stripped open-string amplitudes with the help of appropriate relations~\cite{KLT}. In the low energy limit, such results implied that tree-level graviton amplitudes can be written in terms of sums of products of color-ordered tree-level Yang-Mills amplitudes multiplied by kinematic factors. In the approach by Bern, Carrasco and Johansson, the idea is to suitably rearrange the kinematic building blocks of scattering amplitudes of theories with a given Lie-algebra symmetry, so that they satisfy identical algebraic relations as the associated color factors. With this reshuffle one is able to convert gauge-theory scattering amplitudes to corresponding gravity ones via the replacement of color factors by such kinematic elements. The duality also allows us to constrain the kinematic dependence of each Feynman diagram in perturbation theory. 

An important progress in the direction of gathering further evidence in favor of the existence of double copy was the realization that massive scalar QCD is compatible with color-kinematics duality~\cite{Johansson:15,Plefka:2019wyg,Carrasco:2020ywq} -- see also Ref.~\cite{Mastrolia:2015maa} for an off-shell investigation. Even though there exist only formal proofs for tree-level scattering amplitudes~\cite{Bjerrum-Bohr:2010pnr,Mafra:2011kj,Bjerrum-Bohr:2016axv,Du:2017kpo,delaCruz:2017zqr,Bridges:2019siz}, explicit calculations also furnish important corroboration for a wide class of theories at loop level~\cite{Bern:2010tq,Carrasco:2011mn,Oxburgh:2012zr,Bern:2012uf,Du:2012mt,Yuan:2012rg,Boels:2012ew,Boels:2013bi,Bjerrum-Bohr:2013iza,Bern:2013yya,Ochirov:2013xba,Mafra:2015mja,Mogull:2015adi,He:2015wgf,Yang:2016ear,Boels:2017skl,He:2017spx,Johansson:2017bfl,Jurado:2017xut,Primo:2016omk,Boels:2017ftb,Faller:2018vdz}. Recent interesting discussion concerning loop-level double copy can be found in Refs.~\cite{Borsten:2020zgj,Borsten:2021rmh}. Quite surprisingly, in the presence of adjoint fermions the duality implies supersymmetry~\cite{Chiodaroli:2013upa}.

Given all the aforementioned achievements, one may be tempted to jump to the conclusion that the double-copy structure naturally carries over to classical solutions. Here care must be exercised; even though scattering amplitudes are gauge invariant and independent of field reparametrizations, classical solutions in general depend on such choices, casting a shadow on the potential feasibility of relating gauge and gravity classical solutions. Notwithstanding such issues, there has been a substantial amount of recent progress in seeking explicit evidence of classical solutions related by double-copy features~\cite{CDC1,CDC2,CDC3,CDC4,CDC5,CDC6,CDC7,CDC8,CDC9,CDC10,CDC11,CDC12,CDC13,CDC14,CDC15,CDC16,CDC17,CDC18,Lescano:2021ooe,CDC19,CDC20,CDC21,CDC22,CDC23,CDC24,CDC25,CDC26,CDC27,CDC28}. Applications of the double-copy relations to gravitational-wave physics and related phenomena have been a promising avenue of research~\cite{Cheung:2018wkq,Kosower:2018adc,Bern:2019nnu,Antonelli:2019ytb,Bern:2019crd,Bern:2021dqo,Herrmann:2021tct,Herrmann:2021lqe,Brandhuber:2021eyq}.

In this paper we wish to report progress concerning color-kinematics duality for gauge theories and also gauge/gravity double-copy relations. Our aim here is the investigation of such dualities for a higher-derivative version of Yang-Mills~\cite{Grinstein:08}. The validity of color-kinematics duality for such a class of theories allows us to obtain, in a systematic way, associated amplitudes for quadratic gravity. Indeed, this is not entirely new; Refs.~\cite{Johansson:17,Johansson:18} has given an interesting account of double-copy relations associated with conformal gravity amplitudes -- see also Ref.~\cite{Azevedo:2019zbn} for an intriguing related study in the context of bosonic chiral strings. What is new here is that, for the first time, we calculate amplitudes of higher-derivative theories under the new understanding of the unstable ghostlike resonance which appears in such theories as discussed in Refs.~\cite{DM:19,Donoghue:2019ecz} (see also~\cite{Donoghue:2018izj}). This newly discovered perspective permits us to calculate scattering amplitudes involving the ghost resonances which were dubbed {\it Merlin modes} in previous works. This investigation provides a novel revenue for the study of Lee-Wick non-Abelian gauge and quadratic gravity theories. Here we will use units such that $\hbar=c=1$. We take the Minkowski metric as $\eta_{\mu\nu} = \textrm{diag}(1,-1,-1,-1)$ and the Riemann curvature tensor given by $R^{\lambda}_{\ \mu\nu\kappa} = \partial_{\kappa}\Gamma^{\lambda}_{\mu\nu} + \Gamma^{\eta}_{\mu\nu}\Gamma^{\lambda}_{\kappa\eta} - (\nu \leftrightarrow \kappa)$.

\section{Review of color-kinematics duality and the BCJ double copy}

As discussed, gravitational scattering amplitudes has a double-copy structure expressed in a suitable factorization in terms of gauge-theory amplitudes. Its first incarnation was derived by Kawai, Lewellen and Tye in a closed-string setting~\cite{KLT}. Nowadays it has become more of a systematic framework valid at tree and loop level with the inception of the Bern-Carrasco-Johansson (BCJ) double copy. The BCJ double copy relies upon a fundamental gauge-theory property -- the duality between color and kinematics~\cite{CK}. Such a duality can be found in several physically relevant gauge theories. It enables one to extend gauge symmetry to diffeomorphism invariance. There is a clear expectation that the double copy might be a general hallmark of gravitational theories. 

As well known, tree amplitudes of Yang-Mills theory can be written as a sum over distinct trivalent diagrams
\begin{equation}
A_n = \sum_{k} \frac{c_k n_k}{s_k}
\end{equation}
where $s_k$ are inverse propagators, $c_k$ is the group-theoretic color factor and $n_k$ is a kinematic part which is a polynomial of Lorentz-invariant contractions of polarization vectors and momenta. The amplitude is gauge invariant if the color factors obey the so-called Jacobi identity. Now, the color-kinematics duality asserts that gauge-theory amplitudes can be written in a representation such that all given numerators enjoy identical algebraic properties as the corresponding color factors.

The fact that the numerators are related to each other via a Jacobi-like relation, together with their known gauge-dependent property, implies that we can obtain a gauge-invariant homogeneous relation between color-ordered partial amplitudes~\cite{Bern:08}. Such BCJ amplitude relations are a direct consequence of the duality between color and kinematics. As stated above, in general the numerators are gauge dependent whereas the partial amplitudes are gauge invariant, and these features prevent the so-called propagator matrix to be inverted. 

In possession of the color-kinematics duality, one can prove that, when replacing color factors for kinematic factors in the above Yang-Mills amplitude, one obtains an amplitude for a gravity theory whose spectrum is given by the square of the Yang-Mills spectrum~\cite{Bern:100,Bern:10}:
\begin{equation}
M_n = \sum_{k} \frac{n_k n_k}{s_k}.
\end{equation}
This relation is called the BCJ double-copy relation. The squaring relation can be generalized to
\begin{equation}
M_n = \sum_{k} \frac{\widetilde{n}_k n_k}{s_k}
\end{equation}
where the gravity numerators are given as the product of two distinct Yang-Mills numerators, with only one set of such numerators constrained to satisfy color-kinematics duality.

One can also use the color-kinematics relation and the BCJ relations in order to rewrite gravity amplitudes as products of partial Yang-Mills amplitudes. In this way, it is possible to obtain a set of relations, called
Kawai-Lewellen-Tye (KLT) relations, between gravity and gauge-theory amplitudes as the upshot of color-kinematics duality and gauge-invariance constraints. The KLT relations furnishes a framework for obtaining gravity tree amplitudes in terms of gauge-theory partial amplitudes. Through four points these relations are~\footnote{We use the square brackets for color-ordered gauge-theory amplitudes to distinguish them from non-color-ordered amplitudes.}
\begin{eqnarray}
M^{\textrm{tree}}_{3}(1,2,3) &=& i A^{\textrm{tree}}_{3}[1,2,3] A^{\textrm{tree}}_{3}[1,2,3]
\nonumber\\
M^{\textrm{tree}}_{4}(1,2,3,4) &=& - i s_{12} A^{\textrm{tree}}_{4}[1,2,3,4] A^{\textrm{tree}}_{4}[1,2,4,3] 
\end{eqnarray}
where $M^{\textrm{tree}}_n$ are tree level gravity amplitudes. We have suppressed a factor of the coupling 
$(\kappa/2)^{n-2}$ at $n$ points, where $\kappa^2 = 32 \pi G$. As we can see, the basic input is given by gauge-theory tree-level scattering amplitudes. These can be converted into gravity tree amplitudes through the KLT and BCJ forms of the double copy. The KLT form is well suited when using four-dimensional helicity states. The BCJ double copy could be the preferred choice when disposing the computation in terms of diagrams. Some BCJ relations are given by
\begin{eqnarray}
s_{14} A^{\textrm{tree}}_{4}[1,2,3,4] &=& s_{13} A^{\textrm{tree}}_{4}[1,3,4,2]
\nonumber\\
s_{12} A^{\textrm{tree}}_{4}[1,2,3,4] &=& s_{13} A^{\textrm{tree}}_{4}[1,4,2,3]
\nonumber\\
s_{14} A^{\textrm{tree}}_{4}[1,4,2,3] &=& s_{12} A^{\textrm{tree}}_{4}[1,3,4,2] .
\label{BCJ}
\end{eqnarray}
One can also use the Kleiss-Kuijf relations for the left-hand side of the third equation in~(\ref{BCJ}) and making the switch $2 \leftrightarrow 3$ to convert this relation into
\begin{equation}
s_{14} A^{\textrm{tree}}_{4}[1,2,3,4] =  s_{13} A^{\textrm{tree}}_{4}[1,2,4,3] .
\end{equation}
A similar reasoning for the second equation leads us to the alternative form
\begin{equation}
s_{12} A^{\textrm{tree}}_{4}[1,2,3,4] = s_{13} A^{\textrm{tree}}_{4}[1,3,2,4] .
\end{equation}
These considerations lead us to rewrite the KLT relation at four points in alternative forms:
\begin{equation}
M^{\textrm{tree}}_{4}(1,2,3,4) = - i s_{23} A^{\textrm{tree}}_{4}[1,2,3,4] A^{\textrm{tree}}_{4}[1,3,2,4] 
\end{equation}
or
\begin{equation}
M^{\textrm{tree}}_{4}(1234) = - i \frac{s_{12} s_{14}}{s_{13}} A^{\textrm{tree}}_{4}[1234] 
A^{\textrm{tree}}_{4}[1234] .
\end{equation}
The KLT relations are usually formulated in terms of massless amplitudes, but a double-copy prescription for Weyl gravity has been constructed in Refs.~\cite{Johansson:17,Johansson:18}. Quite remarkable, BCJ relations and/or KLT relations have been relevant also for effective field theories, as highlighted by recent studies~\cite{Chen:2013fya,Kampf:2021jvf,Cachazo:2014xea,Elvang:2018dco,Du:2016tbc,Low:2019wuv,Low:2020ubn}. In the more general context of modern on-shell methods for effective field theories, see also Refs.~\cite{Aoude:19,Shadmi:19,Durieux:20}.

\section{Review of the generalized unitarity method for unstable particles}

We also wish to discuss one-loop amplitudes of higher-derivative Yang-Mills theories and quadratic gravity. We will employ the technique of the generalized unitarity method, which allows us to use tree-level amplitudes to reconstruct loop-level amplitudes~\cite{Bern:11,Ellis:12,Frellesvig:Thesis,Brandhuber:05}. In this formalism, there is an approach called the method of maximal cuts in which we start with generalized cuts that have the maximum number of cut propagators. We will use maximal cuts of only three-point tree amplitudes:
\begin{equation}
\sum_{\textrm{states}} A^{\textrm{tree}}_{(1)} A^{\textrm{tree}}_{(2)} A^{\textrm{tree}}_{(3)} 
\cdots A^{\textrm{tree}}_{(m)} .
\end{equation}
In $D$ dimensions all one-loop amplitudes can be written as a sum of one-loop scalar integrals $I_{m}$ for $m=1,2,3,\ldots,D$~\cite{Elvang:15,Henn:2014yza,vanNeerven:1983vr,Bern:2019crd92em,Bern:2019crd93kr,Brown:1952eu,thooft:79,Passarino:1978jh}:
\begin{equation}
A_{n}^{1-\textrm{loop}} = \sum_{i} C_{D}^{(i)} I_{D;n}^{(i)}
+ \sum_{j} C_{D-1}^{(j)} I_{D-1;n}^{(j)} + \cdots
+ \sum_{k} C_{2}^{(k)} I_{2;n}^{(k)} + \sum_{l} C_{1}^{(l)} I_{1;n}^{(l)} 
+ \mathcal{R}
\label{oneloop}
\end{equation}
where $\mathcal{R}$ denotes rational terms, $C_{D}^{(i)}$ are coefficients associated with tree-level amplitudes and $I_{m}^{(i)}$ are $m$-gon scalar integrals. In $D=4$, one-loop integrals reduce to a combination of box, triangle, bubble and tadpole scalar integrals. 

As shown in Refs.~\cite{Veltman:63,DM:19}, unitarity in theories with unstable particles is only warranted when unitarity cuts are not taken through the unstable particles. This raises a challenge for the unitarity method. Nevertheless, the method can still be applied, as extensively discussed in a recent paper~\cite{DM:21}. In summary, external momentum configurations of an amplitude may or may not permit the unstable propagator to become resonant. When it does, cut unstable propagator will have the correct cut structure. On the other hand, when one is off resonance, a possible way to deal with the problem of unstable particles is to eliminate them through the emergence of a non-local description, one which contains only stable modes -- one must just bear in mind that the only acceptable poles in amplitudes are the ones that come from propagators.

\section{Higher-derivative QCD-like amplitudes}

In this section we will discuss scattering amplitudes for higher-derivative QCD-like theories. We consider that only the gauge sector displays a higher-derivative contribution. The Lagrangian reads~\cite{Grinstein:08}
\begin{equation}
M^2 {\cal L} = - \frac{M^2}{4} F^{a}_{\mu\nu} F^{a \mu\nu} 
+ \frac{1}{2} D_{\mu} F^{a \mu\nu} D_{\lambda} F^{a \lambda}_{\ \ \nu}
\end{equation}
with the covariant derivative (in the adjoint representation)
\begin{equation}
D^{\mu} F_{\alpha\beta}^a = \partial^{\mu} F_{\alpha\beta}^a + g  f^{abc} A^{b \mu} F_{\alpha\beta}^c .
\end{equation}
Higher-derivative terms give rise to the so-called Merlin modes in the theory, which are unstable. Even though the asymptotic spectrum of higher-derivative Yang-Mills is the same as that of standard Yang-Mills, we will be interested in studying a somewhat intermediate regime in which the scattering of Merlin particles should be important.

We are interested in the cubic and quartic interaction terms. The cubic term reads (using, e.g., the Feynman gauge)
\begin{eqnarray}
M^2 {\cal L}_{\textrm{cubic}} &=& g f^{abc}
\Bigl( - M^2 \partial_{\mu} A_{\nu}^{a}  A^{b \mu} A^{c \nu}
+  \Box A^{a \nu}  \partial_{\lambda}(A^{b \lambda} A_{\nu}^{c})
+ \Box A^{a \nu}  A^{b}_{\lambda}\partial^{\lambda} A_{\nu}^{c}
-  \Box A^{a \nu}  A^{b}_{\lambda} \partial_{\nu} A^{c \lambda} 
\nonumber\\
&-& \partial_{\mu}\partial^{\nu} A^{a \mu}  \partial_{\lambda}(A^{b \lambda} A_{\nu}^{c})
- \partial_{\mu}\partial^{\nu} A^{a \mu}  A^{b}_{\lambda} \partial^{\lambda} A_{\nu}^{c}
+ \partial_{\mu}\partial^{\nu} A^{a \mu} A^{b}_{\lambda} \partial_{\nu} A^{c \lambda} \Bigr).
\end{eqnarray}
The quartic term is given by
\begin{eqnarray}
M^2 {\cal L}_{\textrm{quartic}} &=& g^2 f^{abe} f^{cde}  
\biggl[- \frac{M^2}{4}  A^{a}_{\mu}  A^{b}_{\nu} A^{c \mu} A^{d \nu}
+ \frac{1}{2} \Bigl( 
2 \Box A_{\nu}^{a} A^{b \lambda} A_{\lambda}^{c} A^{\nu d}
- 2 \partial^{\mu}\partial_{\nu} A_{\mu}^{a} A^{b \lambda} A_{\lambda}^{c} A^{\nu d}
\nonumber\\
&+&  2  A^{c \lambda} ( \partial_{\lambda} A^{\nu d} - \partial^{\nu} A_{\lambda}^{d} )
 \partial^{\mu} (A^{a}_{\mu} A^{b}_{\nu} )
 + \partial^{\mu} (A^{a}_{\mu} A^{b}_{\nu} )  \partial^{\lambda} ( A_{\lambda}^{c} A^{\nu d} )
 \nonumber\\
&+&  A^{a \mu} ( \partial_{\mu} A_{\nu}^{b} - \partial_{\nu} A_{\mu}^{b} )
 A^{c \lambda} ( \partial_{\lambda} A^{\nu d} - \partial^{\nu} A_{\lambda}^{d} )
 \Bigr)
\biggr] .
\end{eqnarray}
Alternatively, one can proceed as in Ref.~\cite{Grinstein:08} and define auxiliary gauge bosons with a very large mass M. The Lagrangian can then be written as ${\cal L} = {\cal L}_{A} + {\cal L}_{\textrm{Merlin}}$, where the first contribution is the usual Yang-Mills Lagrangian density and ${\cal L}_{\textrm{Merlin}}$ contains the contributions from the massive Merlin modes (free and interacting parts), which are at most quadratic in derivatives. Kinetic Merlin terms have an unusual overall minus sign. In this way we have switched from a higher-derivative description to a representation containing only second-order derivatives, albeit one containing unusual kinetic terms. Hence we have accomplished, at the Lagrangian level, the factorization of the gauge propagator which will be revealed in due course. Despite the presence of Merlin modes, higher-derivative Yang-Mills theories are unitary, as discussed with some detail in Appendix B -- see also Ref.~\cite{DM:19}.

\subsection{Tree-level amplitudes}

The triple-gluon vertex can be calculated as usual from
\begin{equation}
{\cal T}^{abc}_{\mu\nu\rho} = i \int d^4 x_1 \int d^4 x_2 \int d^4 x_3\,
e^{i(k \cdot x_{1} + p \cdot x_{2} + q \cdot x_3)}
\frac{\delta}{\delta A^{a \mu}(x_1)} \frac{\delta}{\delta A^{b \nu}(x_2)} 
\frac{\delta}{\delta A^{c \rho}(x_3)}	
\Bigl( S_{\textrm{cubic}}[A] \Bigr).
\end{equation}
A straightforward calculation leads us to
\begin{eqnarray}
\hspace{-8mm}
{\cal T}^{abc}_{\mu\nu\rho}(k,p,q) &=& \frac{g f^{abc}}{M^2} (2\pi)^4 \delta(k + p + q) \Bigl\{
(M^2 - k^2 - p^2 - q^2) \Bigl[ \eta_{\mu\nu} ( k_{\rho}  - p_{\rho} ) + \eta_{\nu\rho} ( p_{\mu} - q_{\mu}   )
+ \eta_{\rho\mu} (q_{\nu} - k_{\nu} ) \Bigr]
\nonumber\\
&+& k_{\mu}   \Bigl[ ( q_{\nu} - k_{\nu} ) k_{\rho} + k \cdot (p-q) \eta_{\nu\rho} 
 + ( k_{\rho} - p_{\rho} ) k_{\nu}   \Bigr] 
+ p_{\nu} \Bigl[ ( p_{\mu} - q_{\mu} ) p_{\rho}  + p \cdot (q-k) \eta_{\rho\mu} 
+ ( k_{\rho} - p_{\rho} ) p_{\mu}  \Bigr] 
\nonumber\\
&+& q_{\rho} \Bigl[ ( q_{\nu} - k_{\nu} ) q_{\mu}  + q \cdot (k-p) \eta_{\mu\nu}  
+ ( p_{\mu} - q_{\mu} ) q_{\nu}  \Bigr] 
\Bigr\} .
\end{eqnarray}
As we can easily see, $3$-particle amplitudes involving only physical gluons will not display contributions coming from higher-order derivative terms. So we can write
\begin{equation}
A^{(4)}_{3}[1^{h_1},2^{h_2},3^{h_3}] = {\cal T}_{\mu\nu\rho}[1,2,3] \epsilon^{\mu}_{h_1}(1)
\epsilon^{\nu}_{h_2}(2) \epsilon^{\rho}_{h_3}(3) = M^2 A^{(2)}_{3}[1^{h_1},2^{h_2},3^{h_3}]
\end{equation}
where the $h_{i}$s represent the different helicities of the gluons and the superscripts in the amplitudes are used to differentiate the four- and two-derivative theories. In addition, we used color decomposition in which $g /M^2$ is usually factored out, in compliance with the standard definition of color-ordered amplitudes. So it is clear that, in the absence of the usual Yang-Mills term, amplitudes involving only physical gluons are zero (for a fixed $g/M^2$), in accordance with earlier results~\cite{Johansson:18}. Using the BCFW recursion relations~\cite{BCFW1,BCFW2}, it is easy to show that this result carries out to an arbitrary number of gluons. In addition, using momentum conservation it is easy to see that the $3$-particle on-shell amplitude involving a single Merlin particle vanishes:
\begin{equation}
A^{(4)}_{3}(1^{h_1},2^{h_2}, 3^{IJ}) = {\cal T}_{\mu\nu\rho}[1,2,{\bf 3}] \epsilon^{\mu}_{h_1}(1)
\epsilon^{\nu}_{h_2}(2) \epsilon^{\rho IJ}(3) = 0 .
\end{equation}
This is a kind of an alternative form of Yang's theorem. Again using BCFW recursion relations one can easily construct an inductive proof for the result:
\begin{equation}
A^{(4)}_{n+1}(1^{h_1},2^{h_2},\ldots,n^{h_n},k^{IJ}) = 0 
\label{20}
\end{equation} 
since the right-hand side of the recursion relation always have fewer than $n$ gluons. To the best of this author's knowledge, Eq.~(\ref{20}) was first obtained in Ref.~\cite{Johansson:18}. Moreover, notice that massive gauge bosons carry explicit $SU(2)$ little-group indices. A quick review on the framework employed for massive particles is presented in the Appendix A.

Observe that we are employing the massive polarization vector for the Merlin modes. This means that we are using different propagators for the gluon and the Merlin particle. In the narrow-width approximation, the total propagator in momentum space has the form 
\begin{equation}
D^{ab}_{\mu\nu}(p) = - \frac{\delta^{ab}}{p^2} \left( \eta_{\mu\nu} - (1-\xi) \frac{p_{\mu} p_{\nu}}{p^2} \right)
+ \frac{\delta^{ab}}{p^2 - M^2 - i M \Gamma} \left( \eta_{\mu\nu} - \frac{p_{\mu} p_{\nu}}{M^2} \right) .
\end{equation}
where $\Gamma$ is the decay width of the Merlin particle. Unitarity and the Ward identity allows to rewrite the propagator in this form~\footnote{For unstable particles, one must consider a resummed form for the associated propagators since ordinary perturbation theory breaks down in the resonance region. One can construct a modified Lehmann representation for the propagators~\cite{DM:19,DM:19a} or else assume the validity of the narrow-width approximation. In what follows, unless otherwise stated, Merlin propagators will have the form expected by the narrow-width approximation, that is $1/(p^2 - M^2 - i M \Gamma)$, so at times the contribution of the width $\Gamma$ will be left implicit.}
.

Now let us consider the case with two Merlin particles. For the $3$-particle amplitude Feynman rules show that
\begin{eqnarray}
A_{3}[1^{+}, {\bf 2}, {\bf 3}] &=& \sqrt{2}  
\frac{\langle r| {\bf 3} | 1\bigr]}{\langle r1 \rangle} 
\langle {\bf 3} {\bf 2} \rangle^2
\nonumber\\
A_{3}[1^{-}, {\bf 2}, {\bf 3}] &=& \sqrt{2}  
\frac{\bigl[ r| {\bf 3} | 1\rangle}{\bigl[ 1r \bigr]} 
\bigl[ {\bf 3} {\bf 2} \bigr]^2 .
\end{eqnarray}
As explained in Appendix A, we can employ a bold notation to denote symmetric combinations of the $SU(2)$ little-group indices of massive spinors. The case with three Merlin particles can be obtained from the results worked out in Ref.~\cite{Durieux:20}:
\begin{eqnarray}
A_{3}[{\bf 1}, {\bf 2}, {\bf 3}] &=& - 2 M^2
\Bigl[ \eta_{\mu\nu} ( p_{1}  - p_2 )_{\rho} + \eta_{\nu\rho} ( 2 p_{2} + p_{1} )_{\mu}
- \eta_{\rho\mu} ( 2 p_{1} + p_2 )_{\nu} \Bigr]
\epsilon^{\mu}({\bf 1}) \epsilon^{\nu}({\bf 2}) \epsilon^{\rho}({\bf 3})
\nonumber\\
&=& 2 \sqrt{2} \Bigl( \bigl[ {\bf 1} {\bf 2} \bigr] \bigl[ {\bf 1} {\bf 3} \bigr]
\langle {\bf 2} {\bf 3} \rangle
+  \bigl[ {\bf 1} {\bf 2} \bigr] \langle {\bf 1} {\bf 3} \rangle \bigl[ {\bf 2} {\bf 3} \bigr] 
+ \langle {\bf 1} {\bf 2} \rangle \bigl[ {\bf 1} {\bf 3} \bigr]  \bigl[ {\bf 2} {\bf 3} \bigr] 
\Bigr) .
\end{eqnarray}
Amazingly, this amplitude involving only Merlin particles does not display contributions coming from higher-order derivative terms. Incidentally, given the above $3$-particle amplitudes, taken together with the results of Ref.~\cite{Durieux:20}, we see that the longitudinal scattering must obey a sort of Goldstone boson equivalence theorem, even though we are not dealing with a case of spontaneous symmetry breaking. For instance, the $2 \to 2$ scattering of Merlin particles with the exchange of a Merlin particle will display effectively the same behavior as the one for the longitudinal $W$-$Z$ scattering (in the limit of equal masses) since the associated $3$-point vertices are virtually the same (except for coupling constants and possible overall minus signs).

On the other hand, the four-gluon vertex can be calculated from
\begin{equation}
{\cal T}^{abcd}_{\mu\nu\rho\sigma} = i \int d^4 x_1 \int d^4 x_2 \int d^4 x_3 \int d^4 x_4\,
e^{i(k \cdot x_{1} + p \cdot x_{2} + q \cdot x_3 + r \cdot x_4)}
\frac{\delta}{\delta A^{a \mu}(x_1)} \frac{\delta}{\delta A^{b \nu}(x_2)} 
\frac{\delta}{\delta A^{c \rho}(x_3)} \frac{\delta}{\delta A^{d \sigma}(x_4)}	
\Bigl( S_{\textrm{quartic}}[A] \Bigr).
\end{equation}
After inserting the quartic term in the above equation, one obtains a rather awkward expression. We can benefit from the results of Ref.~\cite{Grinstein:08} which separates, at the level of the action, the contributions of the Merlin modes from the normal gauge field, as well as the mixings between them. The $4$-point contact amplitude involving only gluons can be easily obtained from the associated four-gluon interaction vertex:
\begin{eqnarray}
A_{4}(1^{h_1},2^{h_2},3^{h_3},4^{h_4})
&=& - M^2 \Bigl[ f^{abe} f^{cde} (\eta_{\mu\rho} \eta_{\nu\sigma} - \eta_{\mu\sigma} \eta_{\nu\rho})
+ f^{ace} f^{bde} (\eta_{\mu\nu} \eta_{\rho\sigma} - \eta_{\mu\sigma} \eta_{\nu\rho})
+ f^{ade} f^{bce} (\eta_{\mu\nu} \eta_{\rho\sigma} - \eta_{\mu\rho} \eta_{\nu\sigma})
\Bigr]
\nonumber\\
&\times& 
\epsilon^{\mu}_{h_1}(1) \epsilon^{\nu}_{h_2}(2) \epsilon^{\rho}_{h_3}(3) \epsilon^{\sigma}_{h_4}(4)
\end{eqnarray}
where we factored out $g^2/M^2$. So again we see that, for a fixed $g^2/M^2$ this amplitude will vanish in the absence of the usual Yang-Mills term. But it also vanishes with an adequate choice of reference momenta for the polarizations, which is the standard result: Diagrams involving the four-gluon vertex never contribute to MHV amplitudes. The $4$-point contact amplitude involving only Merlins can be obtained in a similar fashion:
\begin{eqnarray}
A_{4}({\bf 1},{\bf 2},{\bf 3},{\bf 4})
&=& 3 M^2 \Bigl[ f^{abe} f^{cde} (\eta_{\mu\rho} \eta_{\nu\sigma} - \eta_{\mu\sigma} \eta_{\nu\rho})
+ f^{ace} f^{bde} (\eta_{\mu\nu} \eta_{\rho\sigma} - \eta_{\mu\sigma} \eta_{\nu\rho})
+ f^{ade} f^{bce} (\eta_{\mu\nu} \eta_{\rho\sigma} - \eta_{\mu\rho} \eta_{\nu\sigma})
\Bigr]
\nonumber\\
&\times& 
\epsilon^{\mu}({\bf 1}) \epsilon^{\nu}({\bf 2}) \epsilon^{\rho}({\bf 3})
\epsilon^{\sigma}({\bf 4}) .
\end{eqnarray}
In general, this amplitude does not vanish and should be included in the scattering involving only Merlin particles. For the amplitude involving two gluons and two Merlins, the result will depend on the position of the particles due to the presence of color factors. Hence the total amplitude is given by the sum of all the possible permutations, divided by the number of total permutations. We find that
\begin{eqnarray}
A_{4}(1^{h_1},{\bf 2},{\bf 3},4^{h_4})
&=& M^2 \Bigl[ f^{abe} f^{cde} (\eta_{\mu\rho} \eta_{\nu\sigma} - \eta_{\mu\sigma} \eta_{\nu\rho})
+ f^{ace} f^{bde} (\eta_{\mu\nu} \eta_{\rho\sigma} - \eta_{\mu\sigma} \eta_{\nu\rho})
+ f^{ade} f^{bce} (\eta_{\mu\nu} \eta_{\rho\sigma} - \eta_{\mu\rho} \eta_{\nu\sigma})
\Bigr]
\nonumber\\
&\times& \epsilon^{\mu}_{h_1}(1)  \epsilon^{\nu}({\bf 2}) \epsilon^{\rho}({\bf 3})
\epsilon^{\sigma}_{h_4}(4)
\end{eqnarray}
where again we factored out $g^2/M^2$. Finally, the case of three Merlins and one gluon reads
\begin{eqnarray}
A_{4}({\bf 1},{\bf 2},{\bf 3},4^{h_4})
&=& 2 M^2 \Bigl[ f^{abe} f^{cde} (\eta_{\mu\rho} \eta_{\nu\sigma} - \eta_{\mu\sigma} \eta_{\nu\rho})
+ f^{ace} f^{bde} (\eta_{\mu\nu} \eta_{\rho\sigma} - \eta_{\mu\sigma} \eta_{\nu\rho})
+ f^{ade} f^{bce} (\eta_{\mu\nu} \eta_{\rho\sigma} - \eta_{\mu\rho} \eta_{\nu\sigma})
\Bigr]
\nonumber\\
&\times& \epsilon^{\mu}({\bf 1})  \epsilon^{\nu}({\bf 2}) \epsilon^{\rho}({\bf 3})
\epsilon^{\sigma}_{h_4}(4)
\end{eqnarray}
In order to calculate $n$-point tree-level amplitudes involving two Merlin particles we can use these $3$-particle amplitudes as the building blocks of the amplitudes and resort to the BCFW recursion relation with massive particles~\cite{Badger:05,Johansson:18,Ballav:21}. In this case, the recursion formula has an additional contribution given by~\footnote{This is a good point to observe that the terminology ``tree-level" to those amplitudes calculated henceforth that involve a Merlin propagator is a rather abuse of language as we are considering a resummed form for their propagator, as explained in the previous footnote. In this way, amplitudes involving a Merlin propagator have only a tree-level-like structure. In any case, many of the results to be presented here are independent of the explicit inclusion of the width.}
\begin{equation}
A^{\textrm{tree}}_{n}(1,\cdots,n) = \sum_{s_{I}, \textrm{diagrams} \, I} 
\widehat{A}^{\textrm{tree}}_{L}(1,\cdots, \widehat{i}, \cdots, a-1, \widehat{P}_{I}^{s_I},b+1,\cdots,n)
\frac{1}{P^2_{I} - m_{I}^2}
\widehat{A}^{\textrm{tree}}_{R}(-  \widehat{P}_{I}^{s_I}, a, \cdots, \widehat{j}, \cdots, b)
\end{equation}
where the sum is over all possible factorization channels $I$ and $\widehat{P}_{I} = \widehat{P}(z_{I})$ (a boundary term is absent by resorting to a valid shift). The quantity $z_{I}$ is determined by the condition $\widehat{P}_{I}^2 = m_{I}^2$ and is different for each diagram. In the above expression, $m_{I}$ is the mass of the intermediate state. Here we shall consider that the subamplitudes in the formula are connected to each other with Merlin propagators, so that $m_{I} = M, \forall I$. Furthermore, there should be an overall minus sign as well concerning the Merlin contribution~\cite{Johansson:18}. Moreover, we only consider shifts for the momenta associated with the gluons. Using the recursion relation $n - 3$ times gives a representation of the $n$-point amplitude utterly in terms of the $3$-point vertices~\cite{Badger:05}.

In principle, one could wonder why we should bother at all to calculate amplitudes that have Merlin modes as external states since, as asserted above, these do not appear in the asymptotic spectrum. The main reason behind these calculations is the use of the unitarity method to calculate loop amplitudes. In this method, we recycle the information coming from lower-order amplitudes to evaluate higher-order ones. Hence a tree-level amplitude is potentially useful in the calculation of a certain one-loop process. Therefore, it is important to understand in detail some features of tree-level amplitudes involving Merlins since in principle these can display undesirable behavior in some situations -- and if a pathology arises already at tree level, we must definitely worry about the associated loop amplitudes.

\subsubsection{Compton scattering: The gluon case}

In what follows we will be interested in the color-ordered amplitude $A_{4}[2^{KL}, 1^{+},4^{+}, 3^{IJ}]$ that appears in the analysis of Compton scattering. Using a $\bigl[ 1,4 \rangle$-shift, defined as
\begin{eqnarray}
| \widehat{1} \bigr] &=& | 1 \bigr] + z | 4 \bigr], \,\,\,
| \widehat{1} \rangle = | 1 \rangle
\nonumber\\
| \widehat{4} \rangle &=& | 4 \rangle - z | 1 \rangle, \,\,\,
| \widehat{4} \bigr] = | 4\bigr]
\end{eqnarray}
where $z$ is some complex number, one finds that
\begin{eqnarray}
A_{4}[{\bf 2},1^{+},4^{+},  {\bf 3}] &=& A_{4}[{\bf 2},\widehat{1}^{+},  - \widehat{P}^{MN}]
\frac{-1}{P^2 - M^2} A_{4}[\widehat{P}_{MN},\widehat{4}^{+}, {\bf 3}]
\nonumber\\
&=& 2 M^4 \frac{\bigl[ 14 \bigr]}{\langle 14 \rangle}
\frac{\langle {\bf 3} {\bf 2} \rangle^2}{s_{12} - M^2}
\end{eqnarray}
where $P = p_1 + p_2 = -p_3 - p_4$ and we chose $r_1 = \widehat{p}_4$ and $r_4 = \widehat{p}_1$. As usual, $s_{ij} = (p_i + p_j)^2$. The on-shell condition $\widehat{P}^2 = M^2$ determines $z$:
\begin{eqnarray}
\widehat{P} &=& \widehat{p}_1 + p_2
\nonumber\\
&\therefore& \langle \widehat{1} {\bf 2}^{I} \rangle \bigl[ {\bf 2}_{I} \widehat{1} \bigr] 
= 0
\end{eqnarray}
and also
\begin{eqnarray}
\widehat{P} &=& - \widehat{p}_4 - p_3
\nonumber\\
&\therefore& \langle \widehat{4} {\bf 3}^{I} \rangle \bigl[ {\bf 3}_{I} \widehat{4} \bigr]  = 0.
\end{eqnarray}
We also employed the following convention for analytic continuation
\begin{eqnarray}
|-p\rangle &=& - |p\rangle, \,\,\, \langle-p| = -\langle p|
\nonumber\\
|-p\bigr] &=& |p\bigr], \,\,\, \bigl[-p| = \bigl[p| .
\end{eqnarray}
The case with all-minus gluons can be easily obtained from the above result:
\begin{equation}
A_{4}[{\bf 2}, 1^{-},4^{-}, {\bf 3}] = 2 M^4 \frac{\langle 14 \rangle}{\bigl[ 14 \bigr]}
\frac{\bigl[ {\bf 3} {\bf 2} \bigr]^2}{s_{12} - M^2} .
\end{equation}
Such results can also be obtained by employing the same considerations as in Ref.~\cite{Arkani-Hamed:17}. To see that both amplitudes have well-behaved high-energy limit, first recall that the coupling constant goes as $g/M^2$, so restoring explicitly the $M^2$ factors in the denominators cancel the $M^4$ factors in the numerators. Furthermore, using the unbold prescription described in Ref.~\cite{Arkani-Hamed:17}, we obtain, for transverse polarizations a result that agrees with the usual Parke-Taylor formula, as expected. 

The other amplitude $A_{4}[2^{KL}, 1^{-},4^{+}, 3^{IJ}]$ reads
\begin{eqnarray}
A_{4}[{\bf 2}, 1^{-},4^{+}, {\bf 3}] &=& A_{4}[{\bf 2},\widehat{1}^{-},  -\widehat{P}^{MN}]
\frac{-1}{P^2 - M^2} A_{4}[\widehat{P}_{MN},\widehat{4}^{+}, {\bf 3}]
\nonumber\\
&=& 2 {M^4} \frac{1}{s_{14}(s_{12} - M^2)}
\Bigl(  \bigl[ 4 {\bf 3} \bigr] \langle 1 {\bf 2} \rangle  + \langle 1 {\bf 3}  \rangle \bigl[ 4 {\bf 2} \bigr]  \Bigr)^2
\end{eqnarray}
where we have made the same choices as above for $r_1, r_4$ and made use of the Schouten identities. Again we find that our results are in accordance with the ones derived in Ref.~\cite{Arkani-Hamed:17}. Moreover, due to the fact that amplitudes with a single Merlin particle vanishes, one should interpret the pole from $s_{14}$ as coming from the massive Merlin particle in the other channel. In this case, we should write $- s_{14} = s_{13} - M^2$ when $s_{12} = M^2$. In this way, the amplitude has the nice property of being manifestly symmetric under ${\bf 2} \leftrightarrow {\bf 3}$. In turn, the high-energy limit is also well behaved. For instance, with $K=L=2$ and $I=J=1$, we again obtain agreement with the Parke-Taylor formula. 

To see more clearly that longitudinal modes will not generate a pathological high-energy behavior for the above amplitudes, use expressions for the above $3$-point amplitudes in terms of the polarization vectors and analyze the residue of the massive pole with longitudinal Merlins. One finds that
\begin{equation}
\textrm{Res}(s) \approx  M^2 s
\Bigl( \epsilon(4) \cdot p_3 \epsilon(1) \cdot p_2 + \epsilon(4) \cdot \epsilon(1) s \Bigr) 
\end{equation}
and, with a similar calculation
\begin{equation}
\textrm{Res}(t) \approx \frac{M^2}{2} t
\Bigl( \epsilon(4) \cdot \epsilon(1) ( u  - s ) 
+ 2 (p_3-p_2) \cdot \epsilon(4) p_{4} \cdot \epsilon(1)
- 2 (p_3-p_2) \cdot \epsilon(1) p_{1} \cdot \epsilon(4) \Bigr)
\end{equation}
at leading order, where $p = p_1 + p_2 = - p_3 - p_4$. Here $s = (p_1 + p_2)^2$, $t = (p_2 + p_3)^2$ and $u = (p_1 + p_3)^2$ are usual Mandelstam variables satisfying $s+t+u= 2 M^2$. At high energy $E \approx |{\bf p}|$, so that $\epsilon({\bf p}) \approx p/M + {\cal O}(M)$. We now see clearly that we need to consider the contact term to have a nice high-energy behavior. We find that
\begin{eqnarray}
A_{\textrm{contact}}(1,{\bf 2}^L,{\bf 3}^L,4)
&\approx& 
f^{abe} f^{cde} (\epsilon(1) \cdot p_3 \epsilon(4) \cdot p_2  - \epsilon(1) \cdot \epsilon(4) p_3 \cdot p_2 )
+ f^{ace} f^{bde} (\epsilon(1) \cdot p_2 \epsilon(4) \cdot p_3  - \epsilon(1) \cdot \epsilon(4) p_3 \cdot p_2)
\nonumber\\
&+& 
f^{ade} f^{bce} (\epsilon(1) \cdot p_2 \epsilon(4) \cdot p_3 - \epsilon(1) \cdot p_3 \epsilon(4) \cdot p_2 ) .
\end{eqnarray}
So, restoring the propagators for the $s$ and $t$ channels as well as the $1/M^2$ factors, we obtain that
\begin{eqnarray}
A_{4}[{\bf 2}^L,1,4,{\bf 3}^L] &\approx& - \frac{1}{M^2}
\Bigl( \epsilon(4) \cdot p_3 \epsilon(1) \cdot p_2 
+ \epsilon(4) \cdot \epsilon(1) s \Bigr) 
\nonumber\\
&-& \frac{1}{2 M^2}\Bigl( \epsilon(4) \cdot \epsilon(1) ( u  - s ) 
+ 2 (p_3-p_2) \cdot \epsilon(4) p_{4} \cdot \epsilon(1)
- 2 (p_3-p_2) \cdot \epsilon(1) p_{1} \cdot \epsilon(4) \Bigr)
\nonumber\\
&-& \frac{1}{M^2}
\Bigl( - \epsilon(1) \cdot p_3 \epsilon(4) \cdot p_2  + \frac{1}{2} \epsilon(1) \cdot \epsilon(4) t
+  \epsilon(1) \cdot p_2 \epsilon(4) \cdot p_3 - \epsilon(1) \cdot p_3 \epsilon(4) \cdot p_2 \Bigr)
\nonumber\\
&=& - \frac{1}{2 M^2} \epsilon(1) \cdot \epsilon(4) (u+s+t)
\end{eqnarray}
at leading order. Notice the overall change of sign -- the Merlin propagator has an overall minus sign. So we see that our result for longitudinal polarizations does not grow with energy. 

The color-ordered amplitudes discussed above obey the standard BCJ relations. For instance, it is easy to see that~\cite{Johansson:19}:
\begin{equation}
( s_{12} - M^2 ) A_{4}[{\bf 2},1^{+},4^{+}, {\bf 3}]
= ( s_{13} - M^2 ) A_{4}[{\bf 2},4^{+},1^{+}, {\bf 3}] .
\end{equation}
Similar relations hold for the other helicity configurations. As a result, the associated numerators will satisfy color-kinematics duality.

\subsubsection{Compton scattering: The scalar case}

We start with the calculation of the $3$-particle amplitude involving two massive complex scalars and one Abelian Merlin particle. It can be obtained as follows:
\begin{equation}
A_{3}^{\textrm{tree}}[\ell_A^{+},{\bf p},\ell_B^{-}]  
= A_{3}^{\textrm{tree}}[\ell_A^{-},{\bf p},\ell_B^{+}] 
= ( \ell^{\mu}_{A} - \ell^{\mu}_{B} ) 
\epsilon_{\mu}({\bf p})
= \sqrt{2} \frac{ \langle {\bf p} | \ell_A | {\bf p} \bigr]  }{M} .
\end{equation}
The $+$ and $-$ indices in the scalar momenta are labels for a scalar and an anti-scalar. It is important for the scalar to be complex; the vector coupling to two identical scalars vanishes because of Bose symmetry. 

In order to build up the Compton scattering amplitude involving two Merlin particles and two massive scalars, we will first consider the contributions due to consistent factorization in all possible channels given the three-particle amplitudes. We follow closely the technique presented in Ref.~\cite{Arkani-Hamed:17}. The residue in the $s$-channel is given by
\begin{equation}
\textrm{Res}_{s} = 2 \frac{ \langle {\bf 1} | \ell_A | {\bf 1} \bigr]  }{M}
\frac{ \langle {\bf 2} | \ell_B | {\bf 2} \bigr]  }{M} .
\end{equation}
The residue in the $u$-channel is obtained by swapping ${\bf 1}$ and ${\bf 2}$. For the residue in the $t$-channel, we have two possibilities, which must produce equal numerators. By interpreting the pole as coming from a massless particle, the factorization demands that the residue is given by the product of a two-scalar-one-gluon amplitude and two-Merlin-one gluon amplitude:
\begin{eqnarray}
\textrm{Res,massless}_{t} &=& - 2  \left( \frac{\langle r | \ell_A | p \bigr]}{ \langle r p \rangle}
  \frac{\langle p | {\bf 1} | q \bigr]}{\bigl[ pq \bigr]} \bigl[ {\bf 1} {\bf 2} \bigr]^2
+  \frac{\langle r| {\bf 1} | p\bigr]}{\langle rp \rangle} \frac{\langle p | \ell_A | q \bigr]}{\bigl[ pq \bigr]}  
\langle {\bf 1} {\bf 2} \rangle^2  \right)
\nonumber\\
&=& - 4 ( \ell_{A} \cdot {\bf 1} )
\langle {\bf 1} {\bf 2} \rangle \bigl[ {\bf 2} {\bf 1} \bigr]
- 2 \Bigl( \langle {\bf 2} | \ell_{A} | {\bf 2} \bigr] \langle {\bf 1} | {\bf 2} | {\bf 1} \bigr]
- \langle {\bf 1} | \ell_{A} | {\bf 1} \bigr] \langle {\bf 2} | {\bf 1} | {\bf 2} \bigr] \Bigr)  
\end{eqnarray}
where $p = {\bf 1} + {\bf 2} = - \ell_A - \ell_B$ and we set $r = q$. Several results collected in Ref.~\cite{Durieux:20} were used here. Notice that the first term does not have the symmetry under ${\bf 1} \leftrightarrow {\bf 2}$ as the second one has. Indeed, from minimal coupling one would expect to get 
${\bf 1} \cdot \ell_A - {\bf 2} \cdot \ell_A$~\cite{Arkani-Hamed:17}. The difference is given by (restoring the propagator)
$$
\frac{2 ( \ell_{A} \cdot {\bf 1} ) - ( {\bf 1} \cdot \ell_A - {\bf 2} \cdot \ell_A )}{s_{12}} = - \frac{s_{12}}{2 s_{12}}.
$$
It has no factorization poles. Taking this into account, one gets
\begin{equation}
\textrm{Res,massless}_{t} = 
- 2 \langle {\bf 1} {\bf 2} \rangle \bigl[ {\bf 2} {\bf 1} \bigr] ( {\bf 1} - {\bf 2} ) \cdot \ell_A 
- 2 \Bigl( \langle {\bf 2} | \ell_{A} | {\bf 2} \bigr] \langle {\bf 1} | {\bf 2} | {\bf 1} \bigr]
- \langle {\bf 1} | \ell_{A} | {\bf 1} \bigr] \langle {\bf 2} | {\bf 1} | {\bf 2} \bigr] \Bigr)  .
\end{equation}
This, in fact, has the right symmetry under ${\bf 1} \leftrightarrow {\bf 2}$ exchange. However, the residue in the t-channel is not complete; one still have the possibility that the pole may come from the massive Merlin particle. In this case, we still have to consider the contribution given by the product of a two-scalar-one-Merlin amplitude and a $3$-Merlin amplitude:
\begin{eqnarray}
\textrm{Res,massive}_{t} &=&
- 2 M^2
\Bigl[ \eta_{\mu\nu} ( p_{1}  - p_2 )_{\rho} + \eta_{\nu\rho} ( 2 p_{2} + p_{1} )_{\mu}
- \eta_{\rho\mu} ( 2 p_{1} + p_2 )_{\nu} \Bigr]
\epsilon^{\mu}({\bf 1}) \epsilon^{\nu}({\bf 2}) \epsilon^{\rho MN}(p)
2 \ell_{A {\alpha}}  [\epsilon^{ \alpha}_{MN}(p)]^{*}
\nonumber\\
&=& 4 M^2
\Bigl[   ( \epsilon({\bf 1}) \cdot \epsilon({\bf 2}) )  ( {\bf 1}  - {\bf 2} ) \cdot \ell_{A}
- 2 \Bigl( \ell_{A} \cdot \epsilon({\bf 1}) {\bf 1} \cdot \epsilon({\bf 2})
- \ell_{A} \cdot \epsilon({\bf 2}) {\bf 2} \cdot \epsilon({\bf 1})  \Bigr) \Bigr]
\end{eqnarray}
where we used that $2 \ell_A \cdot {\bf p} = M^2$. As expected, we essentially obtain the same residue found for the massless $t$-channel. Hence by putting back the $s$-, $t$- and $u-$channel propagators, we obtain the final answer for the color-ordered amplitudes associated with Compton scattering involving Merlin and scalar particles:
\begin{eqnarray}
A_{4}[\ell_A, {\bf 1}, {\bf 2}, \ell_B] &=& 2 \frac{ \langle {\bf 1} | \ell_A | {\bf 1} \bigr]  
\langle {\bf 2} | \ell_B | {\bf 2} \bigr] }{M^2} \frac{1}{s - m^2}
\nonumber\\
&-& \frac{2}{M^2} \left[ \langle {\bf 1} {\bf 2} \rangle \bigl[ {\bf 2} {\bf 1} \bigr] ( {\bf 1} - {\bf 2} ) \cdot \ell_A 
-  \Bigl( \langle {\bf 1} | \ell_{A} | {\bf 1} \bigr] \langle {\bf 2} | {\bf 1} | {\bf 2} \bigr] 
- \langle {\bf 2} | \ell_{A} | {\bf 2} \bigr] \langle {\bf 1} | {\bf 2} | {\bf 1} \bigr] \Bigr)
\right]
\left( \frac{ 1 }{t} - \frac{ 2 }{t - M^2} \right)
\nonumber\\
A_{4}[\ell_B, {\bf 1}, {\bf 2}, \ell_A] &=& 2 \frac{ \langle {\bf 1} | \ell_B | {\bf 1} \bigr] 
\langle {\bf 2} | \ell_A | {\bf 2} \bigr] }{M^2} \frac{1}{u - m^2}
\nonumber\\
&+& \frac{2}{M^2} \left[ \langle {\bf 1} {\bf 2} \rangle \bigl[ {\bf 2} {\bf 1} \bigr] ( {\bf 1} - {\bf 2} ) \cdot \ell_A 
-  \Bigl( \langle {\bf 1} | \ell_{A} | {\bf 1} \bigr] \langle {\bf 2} | {\bf 1} | {\bf 2} \bigr] 
- \langle {\bf 2} | \ell_{A} | {\bf 2} \bigr] \langle {\bf 1} | {\bf 2} | {\bf 1} \bigr] \Bigr)
\right]
\left( \frac{ 1 }{t} - \frac{ 2 }{t - M^2} \right)
\end{eqnarray}
where $s = (\ell_A + p_1)^2$, $t = (p_1 + p_2)^2$ and $u = (\ell_A + p_2)^2$ are Mandelstam variables that satisfy $s+t+u = 2 m^2 + 2 M^2$. We also have restored the $1/M^2$ factors in the $t$-channel. So we see that our result manifestly matches the poles in all the channels. Since the three-point amplitudes were local, the resulting four-point amplitude can be written in a manifest local way with two separate channels~\cite{Arkani-Hamed:17}. Notice the overall minus sign in front of the last term due to the usual ghost feature of the Merlin mode. Furthermore, when scalars transform in the fundamental representation of the gauge group, the ``string" of fields in the amplitudes must always start and end with the scalar~\cite{Badger:05}. 

Let us analyze the high-energy limit. In this limit all particles participating in the scattering process are essentially massless, so effectively $s+t+u=0$ at leading order. It is easy to see that one recovers the standard two-gluon two-scalar tree amplitudes for transverse polarizations. On the other hand, once more the longitudinal terms produce contributions proportional to powers of $M/E$. For instance, take $\epsilon({\bf 1})$ transverse and $\epsilon({\bf 2})$ longitudinal. The leading term produces
\begin{eqnarray}
\textrm{t-channel} &\approx& - \frac{1}{M}  (2 \ell_{A} + p_2) \cdot \epsilon(1) 
\nonumber\\
\textrm{s-channel} &\approx& 2  \ell_{A} \cdot \epsilon(1) \frac{1}{M} .
\end{eqnarray}
Hence we find that 
\begin{equation}
A_{4}[\ell_A, 1^{T}, 2^{L}, \ell_B] \xrightarrow[\text{HE}]{} - \frac{1 }{M} p_2 \cdot \epsilon(1) .
\end{equation}
where the superscript $T$ ($L$) means transverse (longitudinal). The same result also holds for $A_{4}[\ell_B, 1^{T}, 2^{L}, \ell_A]$. So it seems that we have a potential problem with such color-ordered amplitudes. The solution is to consider the contribution of the contact term:
\begin{equation}
\textrm{Contact-Term} = \epsilon({\bf 1}) \cdot \epsilon({\bf 2}) = 
\frac{\langle {\bf 1} {\bf 2} \rangle \bigl[ {\bf 2} {\bf 1} \bigr] }{M^2} .
\end{equation}
In the present situation we find that
\begin{equation}
\textrm{Contact-Term} \xrightarrow[\text{HE}]{} \frac{1 }{M} p_2 \cdot \epsilon(1)
\end{equation}
which cancels with the aforementioned term. So such amplitudes involving one longitudinal polarization behaves at least as ${\cal O}(M/E)$ at high energy. On the other hand, if we consider longitudinal polarizations, we find that
\begin{eqnarray}
\textrm{t-channel} &\approx& 
- 2 ( \epsilon({\bf 1}) \cdot \epsilon({\bf 2}) )  ( {\bf 1}  - {\bf 2} ) \cdot \ell_{A} 
\left( \frac{1}{t} + \frac{2 M^2}{t^2} + \frac{2 M^4}{t^3}  \right)
\approx - \frac{( s - u ) }{2 M^2} 
\nonumber\\
\textrm{s-channel} &\approx& \frac{s}{M^2}
\nonumber\\
\textrm{Contact-Term}  &\approx& \frac{t}{2 M^2}
\end{eqnarray}
at leading order. Summing up the contributions, we find that
\begin{equation}
A_{4}[\ell_A, 1^{L}, 2^{L}, \ell_B] \xrightarrow[\text{HE}]{} \frac{s+t+u}{2 M^2} 
\end{equation}
at leading order. The same holds for $A_{4}[\ell_B, 1^{L}, 2^{L}, \ell_A]$. So our amplitudes with longitudinal modes do not seem to grow with energy. 

Let us discuss color-kinematics duality. It is easy to see that the color factors must obey the following Jacobi identity
$$
c_s - c_u = c_t .
$$
On the other hand, as discussed we must consider the contact term in order to have a nice behavior at high energies. Such contact terms can be blown-up into $s$- and $u$-channel $3$-vertex diagrams by trivial multiplications by $1 = (s-m^2)/(s-m^2) = (u-m^2)/(u-m^2)$. Hence we can construct the following numerators for the physical channels:
\begin{eqnarray}
n_s &=& \frac{ 2 }{M^2} \langle {\bf 1} | \ell_A | {\bf 1} \bigr]  
\langle {\bf 2} | \ell_B | {\bf 2} \bigr]
+ \frac{\langle {\bf 1} {\bf 2} \rangle \bigl[ {\bf 2} {\bf 1} \bigr] }{M^2} [2 ({\bf 1} \cdot \ell_A) + M^2]
\nonumber\\
n_t &=& \frac{2}{M^2} 
\Bigl( \langle {\bf 1} {\bf 2} \rangle \bigl[ {\bf 2} {\bf 1} \bigr] ( {\bf 1} - {\bf 2} ) \cdot \ell_A 
+  \langle {\bf 1} | \ell_{A} | {\bf 1} \bigr] \langle {\bf 2} | \ell_B | {\bf 2} \bigr] 
- \langle {\bf 2} | \ell_{A} | {\bf 2} \bigr] \langle {\bf 1} | \ell_B | {\bf 1} \bigr] 
\Bigr)
\nonumber\\
n_u &=& \frac{2}{M^2}  \langle {\bf 1} | \ell_B | {\bf 1} \bigr] 
\langle {\bf 2} | \ell_A | {\bf 2} \bigr]
+ \frac{\langle {\bf 1} {\bf 2} \rangle \bigl[ {\bf 2} {\bf 1} \bigr] }{M^2} [2 ({\bf 2} \cdot \ell_A) + M^2] .
\end{eqnarray}
Observe that $c_t$ is antisymmetric under an interchange of two legs, and the corresponding numerator $n_t$ also obeys this property. More importantly:
$$
n_s - n_u = n_t .
$$
It is easy to see that such a representation is unique. For instance, concerning the two color-ordered amplitudes given above, we have that
\begin{equation}
\begin{pmatrix}
A_{4}[\ell_A, 1^{IJ}, 2^{KL}, \ell_B] \\
A_{4}[\ell_B, 1^{IJ}, 2^{KL}, \ell_A]
\end{pmatrix}
=
\begin{pmatrix}
\frac{1}{s- m^2} && - \frac{ 1 }{t} + \frac{ 2 }{t - M^2}  \\
\frac{1}{u - m^2} && - \frac{1}{u - m^2} 
+ \left( \frac{ 1 }{t} - \frac{ 2 }{t - M^2}\right)
\end{pmatrix}
\begin{pmatrix}
n_1 \\
n_2
\end{pmatrix}
\end{equation}
where $n_1 = n_s$ and $n_2 = n_t$. The fact that the propagator matrix
$$
\Theta = \begin{pmatrix}
\frac{1}{s- m^2} && - \frac{ 1 }{t} + \frac{ 2 }{t - M^2}  \\
\frac{1}{u - m^2} && - \frac{1}{u - m^2} 
+ \left( \frac{ 1 }{t} - \frac{ 2 }{t - M^2}\right)
\end{pmatrix}
$$
has full rank (non-vanishing determinant) implies that it is invertible and hence we obtain a linearly independent system of two equations with a unique solution. This means that we have obtained a representation that trivially satisfies color-kinematics duality. Indeed, in the present case each kinematic numerator is gauge invariant by itself and thus the amplitude representation is necessarily unique. In order to have homogeneous linear relations among the color-ordered amplitudes (the BCJ relations), we need to have gauge redundancy. We can thus at most build one gauge-invariant quantity out of these, and hence all partial amplitudes must be related. In the present case, we still have a relation between the color-ordered amplitudes, but a ``non-homogeneous" one -- a generalized BCJ relation, that is:
$$
( u - m^2 ) A_{4}[\ell_B, 1^{IJ}, 2^{KL}, \ell_A] - ( s - m^2 ) A_{4}[\ell_A, 1^{IJ}, 2^{KL}, \ell_B]
= \left[ -1 + \left(2 M^2-t\right) \left(\frac{1}{t}-\frac{2}{t-M^2}\right) \right] n_2
$$
where we used that $s+t+u=2 m^2+2 M^2$. The fact that $n_2$ does not drop out of this equation implies that we cannot take $n_2$ to be ``anything" without affecting the physical amplitude -- the numerators are gauge invariant, in contradiction with the well-known redundancy of gauge numerators of pure Yang-Mills theory. So, for the standard BCJ relations to hold one would have to consider additional fields with new interactions that generate additional contributions to the above amplitudes that could systematically cancel the non-homogeneous term in the above equation. Indeed, we can envisage the amplitudes of the higher-derivative Yang-Mills theory as an appropriate limit of a more ``complete" theory studied in detail in Refs.~\cite{Johansson:18,Johansson:17}. In turn, another possibility is to impose that $n_2 \equiv 0$, which implies demanding the absence of a physical $t$-channel for this Compton amplitude, i.e., no exchange of gluons or Merlins between massive scalars and Merlin particles. From the analysis depicted above, we see this is a powerful constraint, as it might amount to choosing special kinematical points at which the amplitude has also a nice high-energy behavior

In any case, we know that the BCJ relations are valid for $n$-point amplitudes with massive particles with at most a pair of massive fundamental fields of arbitrary spin and $n-2$ external massless adjoint particles~\cite{Naculich:14}, so the fact that here color-kinematics duality does not imply homogeneous BCJ relations cannot come as a surprise. Furthermore, it is possible to construct BCJ numerators for the pure Yang-Mills case that obey color-kinematics and are themselves gauge invariant~\cite{Brandhuber:21,Brandhuber:2021bsf}. Nevertheless, in the high-energy limit where $M^2 \to 0$, the above result demonstrates that the standard BCJ relations for the amplitudes involving scalars and gluons are recovered.

\subsubsection{Compton scattering: The fermion case}

Now let us discuss amplitudes involving Merlin particles and massless fermions. In turn, due to the form of the vertex, one must consider a fermion incoming and the other outgoing (or an anti-fermion incoming). Moreover, for $3$-particle amplitudes, both need to have opposite helicity to give a non-vanishing result. One finds
\begin{eqnarray}
i A_{3}^{\textrm{tree}}[f^{-},f^{+},{\bf p}] &=& \bar{u}_{-}(-1) ig \gamma^{\mu} u_{+}(2) \epsilon_{\mu}
\nonumber\\
&=& - \sqrt{2} i g \frac{\langle 1 {\bf p} \rangle \bigl[ {\bf p} 2 \bigr]}{M}
\end{eqnarray}
where we used that $\eta^{\mu\nu} \sigma_{\mu}^{\alpha \dot{\alpha}} \sigma_{\nu}^{\beta \dot{\beta}}
= 2 \varepsilon^{\alpha\beta} \varepsilon^{\dot{\alpha}\dot{\beta}}$ and $g$ is the associated coupling constant. Since $\langle p| \gamma^{\mu} |q \bigr] = \bigr[q| \gamma^{\mu} |p\rangle$, one also obtains
\begin{equation}
A_{3}^{\textrm{tree}}[f^{+},f^{-},{\bf p}] = \sqrt{2} g \frac{\langle 2 {\bf p} \rangle \bigl[ {\bf p} 1 \bigr]}{M} .
\end{equation}
Now let us consider the Compton scattering amplitude involving two Merlin particles and two massless fermions. We follow the same idea as above. Consider fermion $4$ as outgoing and fermion $3$ as incoming. It is easy to see that once more the fermions must have opposite helicity for a non-vanishing amplitude. We consider that $h_3 = - 1/2$ and $h_4 = + 1/2$. The residue in the $s$-channel reads
\begin{equation}
\textrm{Res}_{s} = 2 \frac{\langle 3 {\bf 1} \rangle \bigl[ {\bf 1} p \bigr]}{M}
\frac{\langle p {\bf 2} \rangle \bigl[ {\bf 2} 4 \bigr]}{M}
= 2 \frac{\langle 3 {\bf 1} \rangle \bigl[ {\bf 1}| ({\bf 1} + 3) |{\bf 2} \rangle \bigl[ {\bf 2} 4 \bigr]}{M^2}
\end{equation}
where we suppressed the coupling constant $g$ and used momentum conservation. As above, the residue in the $u$-channel is obtained with the replacement ${\bf 1} \leftrightarrow {\bf 2}$. We can build up an expression for the residue of the $t$-channel by noting the similarity of the present case with the corresponding scalar expressions. That is, by formally defining the following auxiliary momentum $\ell =  |3 \rangle \bigl[ 4|$, one can use the same steps as those previously employed to construct the analogous residue in the scalar case. Hence by putting back the $s$-, $t$- and $u-$channel propagators, we obtain the final answer for the color-ordered amplitudes associated with Compton scattering involving Merlin and massless fermion particles:
\begin{eqnarray}
A_{4}[3^{-1/2}, {\bf 1}, {\bf 2}, 4^{+1/2}] &=& 
2 \frac{\langle 3 {\bf 1} \rangle \bigl[ {\bf 1} | ({\bf 1} + 3) |{\bf 2} \rangle \bigl[ {\bf 2} 4 \bigr]}{M^2} \frac{1}{s}
\nonumber\\
&+&  \frac{1}{M^2} \left[  \langle {\bf 1} {\bf 2} \rangle \bigl[ {\bf 2} {\bf 1} \bigr] 
\langle 3 | ( {\bf 1} - {\bf 2} ) | 4 \bigr]
-  2 \Bigl( \langle {\bf 1} 3 \rangle \bigl[ 4 {\bf 1} \bigr] \langle {\bf 2} | {\bf 1} | {\bf 2} \bigr] 
- \langle {\bf 2} 3 \rangle \bigl[ 4 {\bf 2} \bigr] \langle {\bf 1} | {\bf 2} | {\bf 1} \bigr] \Bigr)
\right]
\left( \frac{ 1 }{t} - \frac{ 2 }{t - M^2} \right)
\nonumber\\
A_{4}[3^{-1/2}, {\bf 2}, {\bf 1}, 4^{+1/2}] &=& 
2\frac{\langle 3 {\bf 2} \rangle \bigl[ {\bf 2} | ({\bf 2} + 3) |{\bf 1} \rangle \bigl[ {\bf 1} 4 \bigr]}{M^2} 
\frac{1}{u}
\nonumber\\
&-&  \frac{1}{M^2} \left[  \langle {\bf 1} {\bf 2} \rangle \bigl[ {\bf 2} {\bf 1} \bigr] 
\langle 3 | ( {\bf 1} - {\bf 2} ) | 4 \bigr]
-  2 \Bigl( \langle {\bf 1} 3 \rangle \bigl[ 4 {\bf 1} \bigr] \langle {\bf 2} | {\bf 1} | {\bf 2} \bigr] 
- \langle {\bf 2} 3 \rangle \bigl[ 4 {\bf 2} \bigr] \langle {\bf 1} | {\bf 2} | {\bf 1} \bigr] \Bigr)
\right]
\left( \frac{ 1 }{t} - \frac{ 2 }{t - M^2} \right) 
\nonumber\\
\end{eqnarray}
where $s = (p_1 + p_3)^2$, $t = (p_1 + p_2)^2$ and $u = (p_1 + p_4)^2$. We have restored $1/M^2$ factors in the $t$-channel. In the high-energy limit, the above terms reproduce the standard Compton scattering of massless fermions and gluons for transerse polarizations. As before, the longitudinal terms produce contributions proportional to powers of $M/E$. To see this, take $\epsilon({\bf 1})$ transverse and $\epsilon({\bf 2})$ longitudinal. The leading term produces
\begin{eqnarray}
\textrm{t-channel} &\approx&  \frac{1}{M}  2 \ell \cdot \epsilon(1) 
\nonumber\\
\textrm{s-channel} &\approx& - \frac{1}{M} 2  \ell \cdot \epsilon(1) 
\end{eqnarray}
where, as stated above, $\ell =  |3 \rangle \bigl[ 4|$. So we see that the amplitude 
$A_{4}[3^{-1/2},1^{T}, 2^{L}, 4^{+1/2}]$ is ${\cal O}(M/E)$. On the other hand, if we consider longitudinal polarizations, we find that
\begin{eqnarray}
\textrm{t-channel} &\approx& 
\frac{1}{M^2}  ( p_1  - p_2 ) \cdot \ell = \frac{2 p_1 \cdot \ell}{M^2}  
\nonumber\\
\textrm{s-channel} &\approx& - \frac{2 \ell \cdot p_1}{M^2}
\end{eqnarray}
at leading order. So again we see that the amplitude does not grow with energy. Concerning color-kinematics duality, we identify the following numerators:
\begin{eqnarray}
n_s &=& \frac{2}{M^2}
\langle 3 {\bf 1} \rangle \bigl[ {\bf 1} | ({\bf 1} + 3) |{\bf 2} \rangle \bigl[ {\bf 2} 4 \bigr]
\nonumber\\
n_t &=& - \frac{1}{M^2}
 \left[  \langle {\bf 1} {\bf 2} \rangle \bigl[ {\bf 2} {\bf 1} \bigr] 
\langle 3 | ( {\bf 1} - {\bf 2} ) | 4 \bigr]
+  2 \Bigl( \langle 3 {\bf 1} \rangle \bigl[ {\bf 1} 4 \bigr] \langle {\bf 2} | (3-4) | {\bf 2} \bigr] 
- \langle 3 {\bf 2} \rangle \bigl[ {\bf 2} 4 \bigr] \langle {\bf 1} | (3-4) | {\bf 1} \bigr] \Bigr)
\right]
\nonumber\\
n_u &=& \frac{2}{M^2}
\langle 3 {\bf 2} \rangle \bigl[ {\bf 2} | ({\bf 2} + 3) |{\bf 1} \rangle \bigl[ {\bf 1} 4 \bigr] .
\end{eqnarray}
It is easy to see that $n_s - n_u = n_t$, confirming the Jacobi identity for the numerators. As in the scalar case, $c_t \to - c_t$ under an interchange of two legs, implying that $n_t \to - n_t$. Finally, for the same reason encountered in the scalar case, such a representation is also unique. Again the usual BCJ relations can be obtained by demanding that $n_t \equiv 0$. As above, in the limit where there are no Merlin modes, we recover the standard BCJ relations for the amplitudes involving fermions and gluons.

\subsubsection{Scattering of Merlin particles}

Let us evaluate the tree-level $2 \to 2$ scattering amplitude involving solely Merlin particles. This is important since there could be a potential issue involving the contribution of extra longitudinal modes which can grow fast with energy. The associated expression for the residue from the $s$-channel reads
\begin{eqnarray}
\textrm{Res}(s) &=& M^4 \Pi^{\rho \kappa}(p)
\Bigl[ \eta_{\mu\nu} ( p_{1}  - p_2 )_{\rho} + \eta_{\nu\rho} ( 2 p_{2} + p_{1} )_{\mu}
- \eta_{\rho\mu} ( 2 p_{1} + p_2 )_{\nu} \Bigr]
\epsilon^{\mu }({\bf 1}) \epsilon^{\nu }({\bf 2}) 
\nonumber\\
&\times&
\Bigl[ \eta_{\alpha\beta} ( p_{3}  - p_4 )_{\kappa} + \eta_{\beta\kappa} ( 2 p_{4} + p_{3} )_{\alpha}
- \eta_{\kappa\alpha} ( 2 p_{3} + p_4 )_{\beta} \Bigr]
\epsilon^{\alpha AB}({\bf 3}) \epsilon^{\beta CD}({\bf 4}) 
\end{eqnarray}
where
\begin{eqnarray}
\Pi^{\rho \kappa}(p) \Bigg|_{\textrm{massless}} & \to& - \eta^{\rho\kappa}
\nonumber\\
\Pi^{\rho \kappa}(p) \Bigg|_{\textrm{massive}} &=& - 4 \left( \eta^{\rho\kappa} 
- \frac{p^{\rho} p^{\kappa}}{M^2} \right) .
\end{eqnarray}
But since $p = - (p_1 + p_2) = p_3 + p_4$, clearly the second term of the massive polarization sum produces a vanishing result when contracted with the remaining contribution. So both residues produce essentially the same result, as expected:
\begin{eqnarray}
\textrm{Res}(s) &=& - 
(4) \Bigl[ \langle {\bf 1} {\bf 2} \rangle \bigl[ {\bf 2} {\bf 1} \bigr] ( p_{1}  - p_2 )^{\kappa} 
+ \sqrt{2} M \langle {\bf 1} | {\bf 2} | {\bf 1} \bigr] \epsilon^{\kappa}({\bf 2})
- \sqrt{2} M   \langle {\bf 2} | {\bf 1} | {\bf 2} \bigr]  \epsilon^{\kappa}({\bf 1}) \Bigr]
\nonumber\\
&\times&
\Bigl[ \langle {\bf 3} {\bf 4} \rangle \bigl[ {\bf 4} {\bf 3} \bigr] ( p_{3}  - p_4 )_{\kappa} 
+ \sqrt{2} M \langle {\bf 3} | {\bf 4} | {\bf 3} \bigr] \epsilon_{\kappa}({\bf 4})
- \sqrt{2}  M \langle {\bf 4} | {\bf 3} | {\bf 4} \bigr]  \epsilon_{\kappa}({\bf 3})
 \Bigr]
\end{eqnarray}
where the factor of $4$ is only present for the massive case. The residue associated with the $t$-channel is obtained through the replacement $2 \leftrightarrow 4$ whereas the $u$-channel is given by 
$2 \leftrightarrow 3$ in the above expression (and associated replacements in the color factors). Since such residues are local, it is easy to display the final form of the amplitude:
\begin{eqnarray}
A_{4}( {\bf 1}, {\bf 2}, {\bf 3}, {\bf 4}) &=& A_{4,s}( {\bf 1}, {\bf 2}, {\bf 3}, {\bf 4})
+ A_{4,t}( {\bf 1}, {\bf 2}, {\bf 3}, {\bf 4}) + A_{4,u}( {\bf 1}, {\bf 2}, {\bf 3}, {\bf 4})
+ A_{4,\textrm{ct}}( {\bf 1}, {\bf 2}, {\bf 3}, {\bf 4})
\nonumber\\
A_{4,s}( {\bf 1}, {\bf 2}, {\bf 3}, {\bf 4}) &=& - f^{abe} f^{cde}
\Bigl[ \langle {\bf 1} {\bf 2} \rangle \bigl[ {\bf 2} {\bf 1} \bigr] ( p_{1}  - p_2 )^{\kappa} 
+ \sqrt{2} M \langle {\bf 1} | {\bf 2} | {\bf 1} \bigr] \epsilon^{\kappa}({\bf 2})
- \sqrt{2} M  \langle {\bf 2} | {\bf 1} | {\bf 2} \bigr]  \epsilon^{\kappa}({\bf 1}) \Bigr]
\nonumber\\
&\times&
\Bigl[ \langle {\bf 3} {\bf 4} \rangle \bigl[ {\bf 4} {\bf 3} \bigr] ( p_{3}  - p_4 )_{\kappa} 
+ \sqrt{2} M  \langle {\bf 3} | {\bf 4} | {\bf 3} \bigr] \epsilon_{\kappa}({\bf 4})
- \sqrt{2}  M \langle {\bf 4} | {\bf 3} | {\bf 4} \bigr]  \epsilon_{\kappa}({\bf 3})
 \Bigr]
\left( \frac{1}{s + i \epsilon} - \frac{4}{s - M^2 - i M \Gamma} \right)
\nonumber\\
A_{4,t}( {\bf 1}, {\bf 2}, {\bf 3}, {\bf 4}) &=& f^{ade} f^{bce}
\Bigl[ \langle {\bf 1} {\bf 4} \rangle \bigl[ {\bf 4} {\bf 1} \bigr] ( p_{1}  - p_4 )^{\kappa} 
+ \sqrt{2} M \langle {\bf 1} | {\bf 4} | {\bf 1} \bigr] \epsilon^{\kappa}({\bf 4})
- \sqrt{2} M  \langle {\bf 4} | {\bf 1} | {\bf 4} \bigr]  \epsilon^{\kappa}({\bf 1}) \Bigr]
\nonumber\\
&\times&
\Bigl[ \langle {\bf 3} {\bf 2} \rangle \bigl[ {\bf 2} {\bf 3} \bigr] ( p_{3}  - p_2 )_{\kappa} 
+ \sqrt{2} M  \langle {\bf 3} | {\bf 2} | {\bf 3} \bigr] \epsilon_{\kappa}({\bf 2})
- \sqrt{2}  M \langle {\bf 2} | {\bf 3} | {\bf 2} \bigr]  \epsilon_{\kappa}({\bf 3})
 \Bigr]
\left( \frac{1}{t + i \epsilon} - \frac{4}{t - M^2 - i M \Gamma} \right)
\nonumber\\
A_{4,u}( {\bf 1}, {\bf 2}, {\bf 3}, {\bf 4}) &=& - f^{ace} f^{bde}
\Bigl[ \langle {\bf 1} {\bf 3} \rangle \bigl[ {\bf 3} {\bf 1} \bigr] ( p_{1}  - p_3 )^{\kappa} 
+ \sqrt{2} M \langle {\bf 1} | {\bf 3} | {\bf 1} \bigr] \epsilon^{\kappa}({\bf 3})
- \sqrt{2} M  \langle {\bf 3} | {\bf 1} | {\bf 3} \bigr]  \epsilon^{\kappa}({\bf 1}) \Bigr]
\nonumber\\
&\times&
\Bigl[ \langle {\bf 2} {\bf 4} \rangle \bigl[ {\bf 4} {\bf 2} \bigr] ( p_{2}  - p_4 )_{\kappa} 
+ \sqrt{2} M  \langle {\bf 2} | {\bf 4} | {\bf 2} \bigr] \epsilon_{\kappa}({\bf 4})
- \sqrt{2}  M \langle {\bf 4} | {\bf 2} | {\bf 4} \bigr]  \epsilon_{\kappa}({\bf 2})
 \Bigr]
\left( \frac{1}{u + i \epsilon} - \frac{4}{u - M^2 - i M \Gamma} \right)
\nonumber\\
A_{4,\textrm{ct}}({\bf 1},{\bf 2},{\bf 3},{\bf 4})
&=& \frac{3}{M^4} 
\Bigl[ f^{abe} f^{cde} \Bigl( \langle {\bf 1} {\bf 3} \rangle \bigl[ {\bf 3} {\bf 1} \bigr] 
\langle {\bf 2} {\bf 4} \rangle \bigl[ {\bf 4} {\bf 2} \bigr]
-  \langle {\bf 1} {\bf 4} \rangle \bigl[ {\bf 4} {\bf 1} \bigr] 
\langle {\bf 2} {\bf 3} \rangle \bigl[ {\bf 3} {\bf 2} \bigr] \Bigr)
\nonumber\\
&+& f^{ace} f^{bde} \Bigl( \langle {\bf 1} {\bf 2} \rangle \bigl[ {\bf 2} {\bf 1} \bigr] 
\langle {\bf 3} {\bf 4} \rangle \bigl[ {\bf 4} {\bf 3} \bigr] 
- \langle {\bf 1} {\bf 4} \rangle \bigl[ {\bf 4} {\bf 1} \bigr] 
\langle {\bf 2} {\bf 3} \rangle \bigl[ {\bf 3} {\bf 2} \bigr] \Bigr)
\nonumber\\
&+& f^{ade} f^{bce} \Bigl( \langle {\bf 1} {\bf 2} \rangle \bigl[ {\bf 2} {\bf 1} \bigr] 
\langle {\bf 3} {\bf 4} \rangle \bigl[ {\bf 4} {\bf 3} \bigr] 
-  \langle {\bf 1} {\bf 3} \rangle \bigl[ {\bf 3} {\bf 1} \bigr] 
\langle {\bf 2} {\bf 4} \rangle \bigl[ {\bf 4} {\bf 2} \bigr] \Bigr)
\Bigr] .
\end{eqnarray}
where $s = (p_1 + p_2)^2$, $t = (p_2 + p_3)^2$ and $u = (p_1 + p_3)^2$. With the above amplitudes we can construct the following color-ordered amplitude (without the contribution of contact terms)
\begin{eqnarray}
A_{4}[ {\bf 1}, {\bf 2}, {\bf 3}, {\bf 4}] &=& - 
\Bigl[ \langle {\bf 1} {\bf 2} \rangle \bigl[ {\bf 2} {\bf 1} \bigr] ( p_{1}  - p_2 )^{\kappa} 
+ \sqrt{2} M \langle {\bf 1} | {\bf 2} | {\bf 1} \bigr] \epsilon^{\kappa}({\bf 2})
- \sqrt{2} M  \langle {\bf 2} | {\bf 1} | {\bf 2} \bigr]  \epsilon^{\kappa}({\bf 1}) \Bigr]
\nonumber\\
&\times&
\Bigl[ \langle {\bf 3} {\bf 4} \rangle \bigl[ {\bf 4} {\bf 3} \bigr] ( p_{3}  - p_4 )_{\kappa} 
+ \sqrt{2} M  \langle {\bf 3} | {\bf 4} | {\bf 3} \bigr] \epsilon_{\kappa}({\bf 4})
- \sqrt{2}  M \langle {\bf 4} | {\bf 3} | {\bf 4} \bigr]  \epsilon_{\kappa}({\bf 3})
 \Bigr]
\left( \frac{1}{s + i \epsilon} - \frac{4}{s - M^2 - i M \Gamma} \right)
\nonumber\\
&-&
\Bigl[ \langle {\bf 1} {\bf 4} \rangle \bigl[ {\bf 4} {\bf 1} \bigr] ( p_{1}  - p_4 )^{\kappa} 
+ \sqrt{2} M \langle {\bf 1} | {\bf 4} | {\bf 1} \bigr] \epsilon^{\kappa}({\bf 4})
- \sqrt{2} M  \langle {\bf 4} | {\bf 1} | {\bf 4} \bigr]  \epsilon^{\kappa}({\bf 1}) \Bigr]
\nonumber\\
&\times&
\Bigl[ \langle {\bf 3} {\bf 2} \rangle \bigl[ {\bf 2} {\bf 3} \bigr] ( p_{3}  - p_2 )_{\kappa} 
+ \sqrt{2} M  \langle {\bf 3} | {\bf 2} | {\bf 3} \bigr] \epsilon_{\kappa}({\bf 2})
- \sqrt{2}  M \langle {\bf 2} | {\bf 3} | {\bf 2} \bigr]  \epsilon_{\kappa}({\bf 3})
 \Bigr]
\left( \frac{1}{t + i \epsilon} - \frac{4}{t - M^2 - i M \Gamma} \right)
\label{67}
\end{eqnarray}
where we considered a resummed form for the Merlin propagator and the narrow-width approximation. 

Let us now discuss the high-energy limit of this amplitude. Recall that, in this limit, $s+t+u=0$ at leading order. For the transverse polarizations, it is easy to see that we obtain the standard color-ordered amplitude involving only gluons (if we restore the $g/M^2$ factors). As for the longitudinal terms, things could get more complicated. In order to obtain a nice behavior at high energies, one must also take into account the contribution of $4$-point contact amplitude. We find that
\begin{eqnarray}
\textrm{s-channel} &\approx& \frac{3 s}{4 M^4} (u-t) 
\nonumber\\
\textrm{t-channel} &\approx& \frac{3 t}{4 M^4} (u-s)
\end{eqnarray}
at leading order for the color-stripped amplitudes, where we have reestablished the $1/M^4$ factors. As for the contact term, we find that
\begin{eqnarray}
A_{\textrm{contact}}[{\bf 1}^L,{\bf 2}^L,{\bf 3}^L,{\bf 4}^L]
&\approx& \frac{3}{4 M^4} \Bigl[ f^{abe} f^{cde} ( u^2 - t^2 )
+ f^{ace} f^{bde} ( s^2 - t^2 )
+ f^{ade} f^{bce} ( s^2 - u^2 )
\Bigr]
\end{eqnarray}
again at leading order. So we find that
\begin{equation}
A_{4}[ {\bf 1}^L, {\bf 2}^L, {\bf 3}^L, {\bf 4}^L] \approx  
\frac{3}{4 M^4} ( 2 u - s - t ) (s+t+u) 
\end{equation}
at leading order. So the possible high-energy growth is canceled just by summing these amplitudes. This is a consequence of gauge invariance, which relates the four-boson and three-boson gauge couplings. For three longitudinal polarizations and one transversal, one gets
\begin{eqnarray}
\textrm{s-channel} &\approx& - \frac{3 s}{2 M^3} (p_1-p_2) \cdot \epsilon(4)
\nonumber\\
\textrm{t-channel} &\approx& - \frac{3 t}{2 M^3} (p_3-p_2) \cdot \epsilon(4)
\end{eqnarray}
at leading order. The contact term produces
\begin{eqnarray}
A_{4}[{\bf 1}^{L},{\bf 2}^{L},{\bf 3}^{L},{\bf 4}^{T}]
&\approx& \frac{3}{2 M^3} \Bigl[ f^{abe} f^{cde} ( u p_2 \cdot \epsilon(4) - t p_1 \cdot \epsilon(4) )
+ f^{ace} f^{bde} ( s p_3 \cdot \epsilon(4) - t p_1 \cdot \epsilon(4) )
\nonumber\\
&+& f^{ade} f^{bce} ( s p_3 \cdot \epsilon(4) - u p_2 \cdot \epsilon(4) )
\Bigr]
\end{eqnarray}
at leading order. So we find that
\begin{equation}
A_{4}[ {\bf 1}^L, {\bf 2}^L, {\bf 3}^L, {\bf 4}^T] \approx  
\frac{3}{M^3} p_2 \cdot \epsilon(4) (s+t+u)  
\end{equation}
at leading order. Finally, for two longitudinal polarizations and two transversals, one gets
\begin{eqnarray}
\textrm{s-channel} &\approx& - \frac{3}{2 M^2} 
\Bigl( \epsilon(3) \cdot \epsilon(4) ( u  - t ) 
+ 2 (p_1-p_2) \cdot \epsilon(4) p_{4} \cdot \epsilon(3)
- 2 (p_1-p_2) \cdot \epsilon(3) p_{3} \cdot \epsilon(4) \Bigr)
\nonumber\\
\textrm{t-channel} &\approx& - \frac{3}{M^2}    
\Bigl( p_{2} \cdot \epsilon(3) p_1 \cdot \epsilon(4) + \epsilon(3) \cdot \epsilon(4) t \Bigr)
\end{eqnarray}
at leading order, whereas the contact term yields
\begin{eqnarray}
A_{4}[{\bf 1}^{L},{\bf 2}^{L},{\bf 3}^{T},{\bf 4}^{T}]
&\approx& \frac{3}{M^2} \Bigl[ f^{abe} f^{cde} 
( p_1 \cdot \epsilon(3)p_2 \cdot \epsilon(4) - p_1 \cdot \epsilon(4)p_2 \cdot \epsilon(3) )
+ f^{ace} f^{bde} 
\Bigl( \frac{s}{2} \epsilon(3) \cdot \epsilon(4) -  p_1 \cdot \epsilon(4)p_2 \cdot \epsilon(3) \Bigr)
\nonumber\\
&+& f^{ade} f^{bce} 
\Bigl( \frac{s}{2} \epsilon(3) \cdot \epsilon(4)  - p_1 \cdot \epsilon(3)p_2 \cdot \epsilon(4)  \Bigr)
\Bigr] .
\end{eqnarray}
Hence
\begin{equation}
A_{4}[ {\bf 1}^L, {\bf 2}^L, {\bf 3}^T, {\bf 4}^T] \approx  
- \frac{3}{2M^2} \epsilon(3) \cdot \epsilon(4) (s+t+u)  
\end{equation}
at leading order. These results explicitly demonstrate the assertions put forward in Ref.~\cite{Wise:2009mi}.

Now let us discuss the color-kinematics duality. Here the discussion is more subtle due to the presence of the Merlin propagator, which displays a different numerator in comparison with the gluon propagator. We find the following numerators for the gluon $A^{a}_{\mu}$
\begin{eqnarray}
n^{A}_s &=&  - \frac{1}{M^4}
\Bigl[ \langle {\bf 1} {\bf 2} \rangle \bigl[ {\bf 2} {\bf 1} \bigr]
 \langle {\bf 3} {\bf 4} \rangle \bigl[ {\bf 4} {\bf 3} \bigr] (u-t)
+ 2 \langle {\bf 3} | {\bf 4} | {\bf 3} \bigr] \langle {\bf 4} | {\bf 1} | {\bf 4} \bigr]
\langle {\bf 1} {\bf 2} \rangle \bigl[ {\bf 2} {\bf 1} \bigr]
-  2 \langle {\bf 3} | {\bf 1}   | {\bf 3} \bigr] \langle {\bf 4} | {\bf 3} | {\bf 4} \bigr] 
\langle {\bf 1} {\bf 2} \rangle \bigl[ {\bf 2} {\bf 1} \bigr]
\nonumber\\
&+& 2 \langle {\bf 1} | {\bf 2} | {\bf 1} \bigr] \langle {\bf 2} | {\bf 3} | {\bf 2} \bigr]
\langle {\bf 3} {\bf 4} \rangle \bigl[ {\bf 4} {\bf 3} \bigr]
+ 2 \langle {\bf 1} | {\bf 2} | {\bf 1} \bigr] \langle {\bf 3} | {\bf 4} | {\bf 3} \bigr]
 \langle {\bf 2} {\bf 4} \rangle \bigl[ {\bf 4} {\bf 2} \bigr] 
- 2 \langle {\bf 1} | {\bf 2} | {\bf 1} \bigr] \langle {\bf 4} | {\bf 3} | {\bf 4} \bigr]   
\langle {\bf 3} {\bf 2} \rangle \bigl[ {\bf 2} {\bf 3} \bigr]
\nonumber\\
&-&  2 \langle {\bf 1} | {\bf 3} | {\bf 1} \bigr] \langle {\bf 2} | {\bf 1} | {\bf 2} \bigr]
\langle {\bf 3} {\bf 4} \rangle \bigl[ {\bf 4} {\bf 3} \bigr]
- 2 \langle {\bf 2} | {\bf 1} | {\bf 2} \bigr]  \langle {\bf 3} | {\bf 4} | {\bf 3} \bigr]
 \langle {\bf 1} {\bf 4} \rangle \bigl[ {\bf 4} {\bf 1} \bigr]
+ 2 \langle {\bf 2} | {\bf 1} | {\bf 2} \bigr] \langle {\bf 4} | {\bf 3} | {\bf 4} \bigr]  
\langle {\bf 1} {\bf 3} \rangle \bigl[ {\bf 3} {\bf 1} \bigr]
\nonumber\\
&+& s \Bigl( \langle {\bf 1} {\bf 3} \rangle \bigl[ {\bf 3} {\bf 1} \bigr] 
\langle {\bf 2} {\bf 4} \rangle \bigl[ {\bf 4} {\bf 2} \bigr]
-  \langle {\bf 1} {\bf 4} \rangle \bigl[ {\bf 4} {\bf 1} \bigr] 
\langle {\bf 2} {\bf 3} \rangle \bigl[ {\bf 3} {\bf 2} \bigr] \Bigr)
 \Bigr]
\nonumber\\
n^{A}_t &=&  \frac{1}{M^4}
\Bigl[ \langle {\bf 1} {\bf 4} \rangle \bigl[ {\bf 4} {\bf 1} \bigr]
 \langle {\bf 3} {\bf 2} \rangle \bigl[ {\bf 2} {\bf 3} \bigr] (u-s)
+ 2 \langle {\bf 2} | {\bf 1} | {\bf 2} \bigr] \langle {\bf 3} | {\bf 2} | {\bf 3} \bigr] 
\langle {\bf 1} {\bf 4} \rangle \bigl[ {\bf 4} {\bf 1} \bigr]
-  2 \langle {\bf 2} | {\bf 3} | {\bf 2} \bigr] \langle {\bf 3} | {\bf 1}  | {\bf 3} \bigr]
\langle {\bf 1} {\bf 4} \rangle \bigl[ {\bf 4} {\bf 1} \bigr]
\nonumber\\
&+& 2 \langle {\bf 1} | {\bf 4} | {\bf 1} \bigr] \langle {\bf 4} | {\bf 3} | {\bf 4} \bigr]
\langle {\bf 3} {\bf 2} \rangle \bigl[ {\bf 2} {\bf 3} \bigr]
+ 2 \langle {\bf 1} | {\bf 4} | {\bf 1} \bigr] \langle {\bf 3} | {\bf 2} | {\bf 3} \bigr]
 \langle {\bf 2} {\bf 4} \rangle \bigl[ {\bf 4} {\bf 2} \bigr] 
- 2 \langle {\bf 1} | {\bf 4} | {\bf 1} \bigr] \langle {\bf 2} | {\bf 3} | {\bf 2} \bigr]   
\langle {\bf 3} {\bf 4} \rangle \bigl[ {\bf 4} {\bf 3} \bigr]
\nonumber\\
&-&  2  \langle {\bf 1} | {\bf 3} | {\bf 1} \bigr] \langle {\bf 4} | {\bf 1} | {\bf 4} \bigr] 
\langle {\bf 3} {\bf 2} \rangle \bigl[ {\bf 2} {\bf 3} \bigr]
- 2 \langle {\bf 3} | {\bf 2} | {\bf 3} \bigr] \langle {\bf 4} | {\bf 1} | {\bf 4} \bigr]  
 \langle {\bf 1} {\bf 2} \rangle \bigl[ {\bf 2} {\bf 1} \bigr]
+ 2  \langle {\bf 2} | {\bf 3} | {\bf 2} \bigr]  \langle {\bf 4} | {\bf 1} | {\bf 4} \bigr]
\langle {\bf 1} {\bf 3} \rangle \bigl[ {\bf 3} {\bf 1} \bigr]
 \nonumber\\
&-& t \Bigl( \langle {\bf 1} {\bf 2} \rangle \bigl[ {\bf 2} {\bf 1} \bigr] 
\langle {\bf 3} {\bf 4} \rangle \bigl[ {\bf 4} {\bf 3} \bigr] 
-  \langle {\bf 1} {\bf 3} \rangle \bigl[ {\bf 3} {\bf 1} \bigr] 
\langle {\bf 2} {\bf 4} \rangle \bigl[ {\bf 4} {\bf 2} \bigr] \Bigr)
 \Bigr]
\nonumber\\
n^{A}_u &=&  - \frac{1}{M^4}
\Bigl[ \langle {\bf 1} {\bf 3} \rangle \bigl[ {\bf 3} {\bf 1} \bigr]
 \langle {\bf 2} {\bf 4} \rangle \bigl[ {\bf 4} {\bf 2} \bigr] (s-t)
+ 2 \langle {\bf 2} | {\bf 4} | {\bf 2} \bigr] \langle {\bf 4} | {\bf 1} | {\bf 4} \bigr]
\langle {\bf 1} {\bf 3} \rangle \bigl[ {\bf 3} {\bf 1} \bigr]
-  2  \langle {\bf 2} | {\bf 1}  | {\bf 2} \bigr] \langle {\bf 4} | {\bf 2} | {\bf 4} \bigr]
\langle {\bf 1} {\bf 3} \rangle \bigl[ {\bf 3} {\bf 1} \bigr]
\nonumber\\
&+& 2  \langle {\bf 1} | {\bf 3} | {\bf 1} \bigr] \langle {\bf 3} | {\bf 2}  | {\bf 3} \bigr]
\langle {\bf 2} {\bf 4} \rangle \bigl[ {\bf 4} {\bf 2} \bigr]
+ 2 \langle {\bf 1} | {\bf 3} | {\bf 1} \bigr] \langle {\bf 2} | {\bf 4} | {\bf 2} \bigr]
 \langle {\bf 3} {\bf 4} \rangle \bigl[ {\bf 4} {\bf 3} \bigr] 
- 2 \langle {\bf 1} | {\bf 3} | {\bf 1} \bigr] \langle {\bf 4} | {\bf 2} | {\bf 4} \bigr]   
\langle {\bf 3} {\bf 2} \rangle \bigl[ {\bf 2} {\bf 3} \bigr]
\nonumber\\
&-&  2  \langle {\bf 1} | {\bf 2} | {\bf 1} \bigr] \langle {\bf 3} | {\bf 1} | {\bf 3} \bigr] 
\langle {\bf 2} {\bf 4} \rangle \bigl[ {\bf 4} {\bf 2} \bigr]
- 2   \langle {\bf 2} | {\bf 4} | {\bf 2} \bigr] \langle {\bf 3} | {\bf 1} | {\bf 3} \bigr]
 \langle {\bf 1} {\bf 4} \rangle \bigl[ {\bf 4} {\bf 1} \bigr]
+ 2 \langle {\bf 3} | {\bf 1} | {\bf 3} \bigr] \langle {\bf 4} | {\bf 2} | {\bf 4} \bigr]  
\langle {\bf 1} {\bf 2} \rangle \bigl[ {\bf 2} {\bf 1} \bigr]
\nonumber\\
&+& u \Bigl( \langle {\bf 1} {\bf 2} \rangle \bigl[ {\bf 2} {\bf 1} \bigr] 
\langle {\bf 3} {\bf 4} \rangle \bigl[ {\bf 4} {\bf 3} \bigr] 
- \langle {\bf 1} {\bf 4} \rangle \bigl[ {\bf 4} {\bf 1} \bigr] 
\langle {\bf 2} {\bf 3} \rangle \bigl[ {\bf 3} {\bf 2} \bigr] \Bigr)
 \Bigr]
\label{numm1}
\end{eqnarray}
and for the Merlin $\widetilde{A}^{a}_{\mu}$
\begin{eqnarray}
n^{\widetilde{A}}_s &=&  - \frac{1}{M^4}
\Bigl[ \langle {\bf 1} {\bf 2} \rangle \bigl[ {\bf 2} {\bf 1} \bigr]
 \langle {\bf 3} {\bf 4} \rangle \bigl[ {\bf 4} {\bf 3} \bigr] (u-t)
+ 2 \langle {\bf 3} | {\bf 4} | {\bf 3} \bigr] \langle {\bf 4} | {\bf 1} | {\bf 4} \bigr]
\langle {\bf 1} {\bf 2} \rangle \bigl[ {\bf 2} {\bf 1} \bigr]
-  2 \langle {\bf 3} | {\bf 1}   | {\bf 3} \bigr] \langle {\bf 4} | {\bf 3} | {\bf 4} \bigr] 
\langle {\bf 1} {\bf 2} \rangle \bigl[ {\bf 2} {\bf 1} \bigr]
\nonumber\\
&+& 2 \langle {\bf 1} | {\bf 2} | {\bf 1} \bigr] \langle {\bf 2} | {\bf 3} | {\bf 2} \bigr]
\langle {\bf 3} {\bf 4} \rangle \bigl[ {\bf 4} {\bf 3} \bigr]
+ 2 \langle {\bf 1} | {\bf 2} | {\bf 1} \bigr] \langle {\bf 3} | {\bf 4} | {\bf 3} \bigr]
 \langle {\bf 2} {\bf 4} \rangle \bigl[ {\bf 4} {\bf 2} \bigr] 
- 2 \langle {\bf 1} | {\bf 2} | {\bf 1} \bigr] \langle {\bf 4} | {\bf 3} | {\bf 4} \bigr]   
\langle {\bf 3} {\bf 2} \rangle \bigl[ {\bf 2} {\bf 3} \bigr]
\nonumber\\
&-&  2 \langle {\bf 1} | {\bf 3} | {\bf 1} \bigr] \langle {\bf 2} | {\bf 1} | {\bf 2} \bigr]
\langle {\bf 3} {\bf 4} \rangle \bigl[ {\bf 4} {\bf 3} \bigr]
- 2 \langle {\bf 2} | {\bf 1} | {\bf 2} \bigr]  \langle {\bf 3} | {\bf 4} | {\bf 3} \bigr]
 \langle {\bf 1} {\bf 4} \rangle \bigl[ {\bf 4} {\bf 1} \bigr]
+ 2 \langle {\bf 2} | {\bf 1} | {\bf 2} \bigr] \langle {\bf 4} | {\bf 3} | {\bf 4} \bigr]  
\langle {\bf 1} {\bf 3} \rangle \bigl[ {\bf 3} {\bf 1} \bigr]
\nonumber\\
&+& (s-M^2) \Bigl( \langle {\bf 1} {\bf 3} \rangle \bigl[ {\bf 3} {\bf 1} \bigr] 
\langle {\bf 2} {\bf 4} \rangle \bigl[ {\bf 4} {\bf 2} \bigr]
-  \langle {\bf 1} {\bf 4} \rangle \bigl[ {\bf 4} {\bf 1} \bigr] 
\langle {\bf 2} {\bf 3} \rangle \bigl[ {\bf 3} {\bf 2} \bigr] \Bigr)
 \Bigr]
\nonumber\\
n^{\widetilde{A}}_t &=&  \frac{1}{M^4}
\Bigl[ \langle {\bf 1} {\bf 4} \rangle \bigl[ {\bf 4} {\bf 1} \bigr]
 \langle {\bf 3} {\bf 2} \rangle \bigl[ {\bf 2} {\bf 3} \bigr] (u-s)
+ 2 \langle {\bf 2} | {\bf 1} | {\bf 2} \bigr] \langle {\bf 3} | {\bf 2} | {\bf 3} \bigr] 
\langle {\bf 1} {\bf 4} \rangle \bigl[ {\bf 4} {\bf 1} \bigr]
-  2 \langle {\bf 2} | {\bf 3} | {\bf 2} \bigr] \langle {\bf 3} | {\bf 1}  | {\bf 3} \bigr]
\langle {\bf 1} {\bf 4} \rangle \bigl[ {\bf 4} {\bf 1} \bigr]
\nonumber\\
&+& 2 \langle {\bf 1} | {\bf 4} | {\bf 1} \bigr] \langle {\bf 4} | {\bf 3} | {\bf 4} \bigr]
\langle {\bf 3} {\bf 2} \rangle \bigl[ {\bf 2} {\bf 3} \bigr]
+ 2 \langle {\bf 1} | {\bf 4} | {\bf 1} \bigr] \langle {\bf 3} | {\bf 2} | {\bf 3} \bigr]
 \langle {\bf 2} {\bf 4} \rangle \bigl[ {\bf 4} {\bf 2} \bigr] 
- 2 \langle {\bf 1} | {\bf 4} | {\bf 1} \bigr] \langle {\bf 2} | {\bf 3} | {\bf 2} \bigr]   
\langle {\bf 3} {\bf 4} \rangle \bigl[ {\bf 4} {\bf 3} \bigr]
\nonumber\\
&-&  2  \langle {\bf 1} | {\bf 3} | {\bf 1} \bigr] \langle {\bf 4} | {\bf 1} | {\bf 4} \bigr] 
\langle {\bf 3} {\bf 2} \rangle \bigl[ {\bf 2} {\bf 3} \bigr]
- 2 \langle {\bf 3} | {\bf 2} | {\bf 3} \bigr] \langle {\bf 4} | {\bf 1} | {\bf 4} \bigr]  
 \langle {\bf 1} {\bf 2} \rangle \bigl[ {\bf 2} {\bf 1} \bigr]
+ 2  \langle {\bf 2} | {\bf 3} | {\bf 2} \bigr]  \langle {\bf 4} | {\bf 1} | {\bf 4} \bigr]
\langle {\bf 1} {\bf 3} \rangle \bigl[ {\bf 3} {\bf 1} \bigr]
 \nonumber\\
&-& (t-M^2) \Bigl( \langle {\bf 1} {\bf 2} \rangle \bigl[ {\bf 2} {\bf 1} \bigr] 
\langle {\bf 3} {\bf 4} \rangle \bigl[ {\bf 4} {\bf 3} \bigr] 
-  \langle {\bf 1} {\bf 3} \rangle \bigl[ {\bf 3} {\bf 1} \bigr] 
\langle {\bf 2} {\bf 4} \rangle \bigl[ {\bf 4} {\bf 2} \bigr] \Bigr)
 \Bigr]
\nonumber\\
n^{\widetilde{A}}_u &=& - \frac{1}{M^4}
\Bigl[ \langle {\bf 1} {\bf 3} \rangle \bigl[ {\bf 3} {\bf 1} \bigr]
 \langle {\bf 2} {\bf 4} \rangle \bigl[ {\bf 4} {\bf 2} \bigr] (s-t)
+ 2 \langle {\bf 2} | {\bf 4} | {\bf 2} \bigr] \langle {\bf 4} | {\bf 1} | {\bf 4} \bigr]
\langle {\bf 1} {\bf 3} \rangle \bigl[ {\bf 3} {\bf 1} \bigr]
-  2  \langle {\bf 2} | {\bf 1}  | {\bf 2} \bigr] \langle {\bf 4} | {\bf 2} | {\bf 4} \bigr]
\langle {\bf 1} {\bf 3} \rangle \bigl[ {\bf 3} {\bf 1} \bigr]
\nonumber\\
&+& 2  \langle {\bf 1} | {\bf 3} | {\bf 1} \bigr] \langle {\bf 3} | {\bf 2}  | {\bf 3} \bigr]
\langle {\bf 2} {\bf 4} \rangle \bigl[ {\bf 4} {\bf 2} \bigr]
+ 2 \langle {\bf 1} | {\bf 3} | {\bf 1} \bigr] \langle {\bf 2} | {\bf 4} | {\bf 2} \bigr]
 \langle {\bf 3} {\bf 4} \rangle \bigl[ {\bf 4} {\bf 3} \bigr] 
- 2 \langle {\bf 1} | {\bf 3} | {\bf 1} \bigr] \langle {\bf 4} | {\bf 2} | {\bf 4} \bigr]   
\langle {\bf 3} {\bf 2} \rangle \bigl[ {\bf 2} {\bf 3} \bigr]
\nonumber\\
&-&  2  \langle {\bf 1} | {\bf 2} | {\bf 1} \bigr] \langle {\bf 3} | {\bf 1} | {\bf 3} \bigr] 
\langle {\bf 2} {\bf 4} \rangle \bigl[ {\bf 4} {\bf 2} \bigr]
- 2   \langle {\bf 2} | {\bf 4} | {\bf 2} \bigr] \langle {\bf 3} | {\bf 1} | {\bf 3} \bigr]
 \langle {\bf 1} {\bf 4} \rangle \bigl[ {\bf 4} {\bf 1} \bigr]
+ 2 \langle {\bf 3} | {\bf 1} | {\bf 3} \bigr] \langle {\bf 4} | {\bf 2} | {\bf 4} \bigr]  
\langle {\bf 1} {\bf 2} \rangle \bigl[ {\bf 2} {\bf 1} \bigr]
\nonumber\\
&+& (u-M^2) \Bigl( \langle {\bf 1} {\bf 2} \rangle \bigl[ {\bf 2} {\bf 1} \bigr] 
\langle {\bf 3} {\bf 4} \rangle \bigl[ {\bf 4} {\bf 3} \bigr] 
- \langle {\bf 1} {\bf 4} \rangle \bigl[ {\bf 4} {\bf 1} \bigr] 
\langle {\bf 2} {\bf 3} \rangle \bigl[ {\bf 3} {\bf 2} \bigr] \Bigr)
 \Bigr] .
 \label{numm2}
\end{eqnarray}
Therefore, we can write that
\begin{equation}
A_{4,s}( {\bf 1}, {\bf 2}, {\bf 3}, {\bf 4}) =  f^{abe} f^{cde}
\left( \frac{n^{A}_s}{s_{12} + i \epsilon} - \frac{4 n^{\widetilde{A}}_s}{s_{12} - M^2 - i \Gamma} \right)
\end{equation}
and similarly for the other physical channels. Each set of numerators will obey separately color-kinematics duality, i.e.
$$
n^{k}_s + n^{k}_t - n^{k}_u = 0
$$
with $k = A, \widetilde{A}$. Furthermore, here all color factors are antisymmetric under an interchange of two legs, and the corresponding numerators all satisfy the same antisymmetry relations
$$
c_i \to - c_i \Longrightarrow n_i \to - n_i.
$$
Hence, concerning the color-ordered amplitude~(\ref{67}), we can write that
\begin{equation}
\begin{pmatrix}
A_{4}[ {\bf 1}, {\bf 2}, {\bf 3}, {\bf 4}]  \\
A_{4}[ {\bf 1}, {\bf 3},{\bf 2}, {\bf 4}] 
\end{pmatrix}
=
\begin{pmatrix}
\frac{1}{s} && - \frac{1}{t}  \\
\frac{1}{u} && \frac{1}{u} 
+ \frac{1}{t}
\end{pmatrix}
\begin{pmatrix}
n_1 \\
n_2
\end{pmatrix}
+
\begin{pmatrix}
-\frac{4}{s-M^2} &&  \frac{4}{t-M^2}  \\
- \frac{4}{u-M^2} && - \frac{4}{u-M^2} 
- \frac{4}{t-M^2}
\end{pmatrix}
\begin{pmatrix}
\widetilde{n}_1 \\
\widetilde{n}_2
\end{pmatrix}
\label{80}
\end{equation}
where $n_1 = n^{A}_s$, $\widetilde{n}_1 = n^{\widetilde{A}}_s$, $n_2 = n^{A}_t$ and 
$\widetilde{n}_2 = n^{\widetilde{A}}_t$. We can put this equation in a more convenient form by noting that $n^{\widetilde{A}}_s = n^{A}_s + S$, $n^{\widetilde{A}}_t = n^{A}_t + T$ and 
$n^{\widetilde{A}}_u = n^{A}_u + U$, where $S,U,T$ can be easily identified from the above equations that define the numerators. We get:
\begin{equation}
\begin{pmatrix}
A_{4}[ {\bf 1}, {\bf 2}, {\bf 3}, {\bf 4}]  \\
A_{4}[ {\bf 1}, {\bf 3},{\bf 2}, {\bf 4}] 
\end{pmatrix}
=
\begin{pmatrix}
\frac{1}{s} - \frac{4}{s-M^2} && - \frac{1}{t} + \frac{4}{t-M^2} \\
\frac{1}{u} - \frac{4}{u-M^2} && \frac{1}{u} - \frac{4}{u-M^2} + \frac{1}{t} - \frac{4}{t-M^2}
\end{pmatrix}
\begin{pmatrix}
n_1 \\
n_2
\end{pmatrix}
+
\begin{pmatrix}
-\frac{4}{s-M^2} &&  \frac{4}{t-M^2}  \\
- \frac{4}{u-M^2} && - \frac{4}{u-M^2} 
- \frac{4}{t-M^2}
\end{pmatrix}
\begin{pmatrix}
S \\
T
\end{pmatrix} .
\end{equation}
The associated propagator matrix
$$
\Theta = \begin{pmatrix}
\frac{1}{s} - \frac{4}{s-M^2} && - \frac{1}{t} + \frac{4}{t-M^2} \\
\frac{1}{u} - \frac{4}{u-M^2} && \frac{1}{u} - \frac{4}{u-M^2} + \frac{1}{t} - \frac{4}{t-M^2}
\end{pmatrix}
$$
has full rank and hence can be invertible. In other words, the system has a unique solution and this implies that we have found a unique representation for the numerators. On the other hand, one possibility to preserve traditional BCJ relations is to impose that the scattering of Merlin particles should proceed only via the exchange of Merlin modes, which would lead us to
$$
(u-M^2) A_{4}[ {\bf 1}, {\bf 3},{\bf 2}, {\bf 4}]  =  (s-M^2) A_{4}[ {\bf 1}, {\bf 2}, {\bf 3}, {\bf 4}]
- \frac{4 M^2}{t-M^2} \widetilde{n}_2.
$$
As above, we would also need to impose that $\widetilde{n}_2 \equiv 0$. Notice that in the high-energy limit we recover the standard BCJ relations for the color-ordered amplitudes.

\subsection{A one-loop example}

Now we proceed to calculate the color-ordered one-loop amplitude associated with the process $g^{+}g^{+} \to g^{+}g^{+}$. We will use the generalized unitarity method. We reconstruct the one-loop amplitude by using the information provided by the set of generalized unitarity cuts. One can choose different directions to extract the pertinent information for engineering the amplitudes in the most efficient way. Here we intend to use the method of maximal cuts.

The case of only physical gluons running in the loop has already been consider in many places in the literature, see for instance Refs.~\cite{Bern:97,Frellesvig:Thesis,Brandhuber:05}. On the other hand, we know that all internal propagators have the generic form (suppressing Lorentz indices)
\begin{equation}
iD(q) = \frac{i}{q^2 +i \epsilon - \frac{q^4}{M^2} + \Sigma (q) }  \ \  .
\end{equation}
The pole at $q^2=0$ is the gluon of the theory. The function $\Sigma (q)$ is the vacuum polarization function. The Merlin resonance appears for timelike values of $q^2$. Expanding near that resonance, we get the form
\begin{equation}
iD(q) \bigg|_{q^2\sim M^2} \sim \frac{-i}{q^2-M^2 -i \gamma}  \ \   .
\end{equation}
This is the form of the internal propagators near the Merlin resonance.

The one-loop amplitude for the standard Yang-Mills case is both ultraviolet and infrared finite~\cite{Bern:97}. There is only a contribution coming from a box-integral., i.e., the coefficients associated with triangles and bubbles vanish. We expect the same to happen here. Let us begin our calculation with the quadruple cut, which will produce the coefficient associated with the box integral. Since the amplitudes involving one single external Merlin particle vanish, the only contribution will come from diagrams with only Merlin particles running in the loop. Notice that it does not necessarily mean that this will be true for {\it any} four-derivative theory; indeed, in a scalar four-derivative theory one may find a non-vanishing result by calculating a two-particle cut of an one-loop self-energy function comprised of two internal lines, one of which is associated with a Merlin propagator. 

So we must consider only cuts from the unstable Merlin modes. As discussed above, in general cuts from unstable states should not be considered and the one-loop amplitude will be proportional to the amplitude of the standard Yang-Mills, as in the tree-level case. Hence for simplicity we shall consider the narrow-width approximation in what follows, so that we can freely cut the unstable internal lines associated with Merlin propagators.

\begin{figure}[htb]
\begin{center}
\includegraphics[height=80mm,width=85mm]{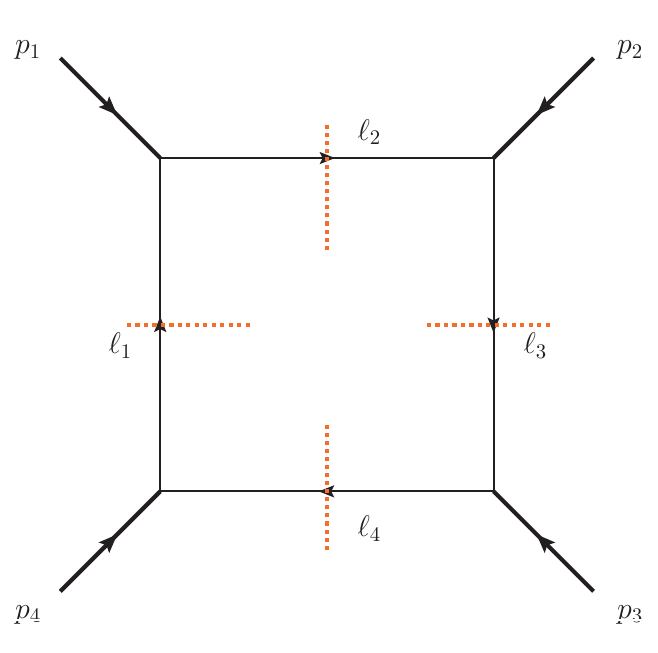}
\caption{A quadruple-cut diagram for the all-plus amplitude discussed in the text. Bold external lines represent physical gluons or gravitons, depending on the context.}
\label{quad}
\end{center}
\end{figure}

The quadruple-cut reads (see Fig.~\ref{quad})
\begin{eqnarray}
A^{1-\textrm{loop integrand}}_{4}(\ell) \bigg|_{\textrm{quadruple cut}} &=& 
A^{\textrm{tree}}_{3}[p^{+}_1, - \ell_{2KL}, \ell^{IJ}_1] 
A^{\textrm{tree}}_{3}[p^{+}_2, -\ell_{3MN}, \ell^{KL}_2] 
\nonumber\\
&\times&
A^{\textrm{tree}}_{3}[p^{+}_3, -\ell_{4PQ}, \ell^{MN}_3] 
A^{\textrm{tree}}_{3}[p^{+}_4, -\ell_{1IJ}, \ell^{PQ}_4] .
\end{eqnarray}
The cut conditions are given by
\begin{eqnarray}
\ell_1^2 &=& \ell^2 = M^2 
\nonumber\\
\ell_2^2 &=& (\ell + p_{1})^2 = M^2
\nonumber\\ 
\ell_3^2 &=& (\ell + p_1 + p_2)^2 = M^2 
\nonumber\\
\ell_4^2 &=& (\ell_3 + p_3)^2 = (\ell - p_4)^2 = M^2 . 
\end{eqnarray}
A straightforward calculation leads us to
\begin{equation}
A^{1-\textrm{loop integrand}}_{4}(\ell) \bigg|_{\textrm{quadruple cut}} = 
- 16 M^{12} \frac{ s_{12} s_{23} }
{\langle 1 2 \rangle \langle 23 \rangle \langle 34 \rangle \langle 41 \rangle} .
\end{equation}
\begin{figure}
\centering
\begin{tabular}{cc}
\includegraphics[height=90mm,width=90mm]{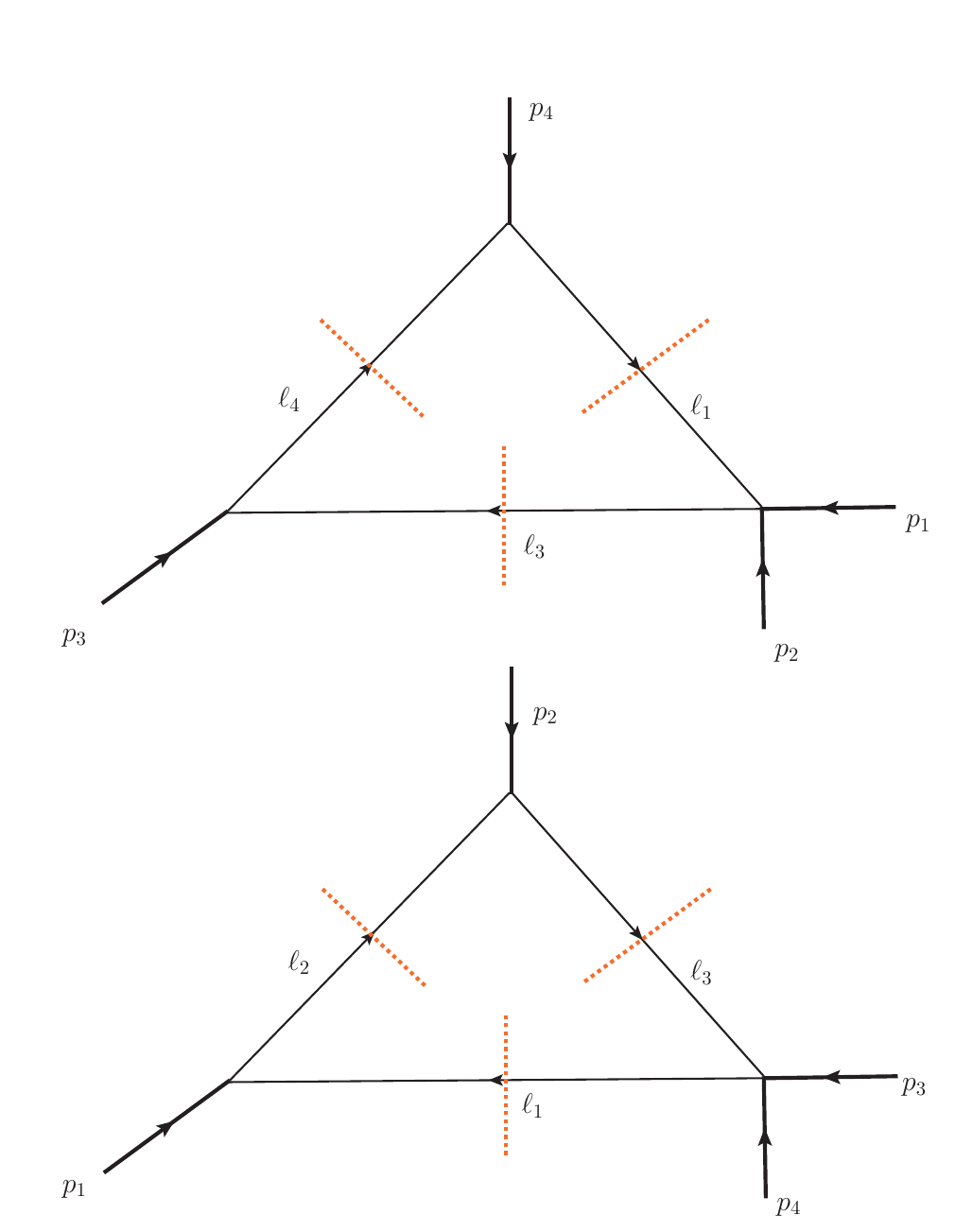}&
\includegraphics[height=90mm,width=90mm]{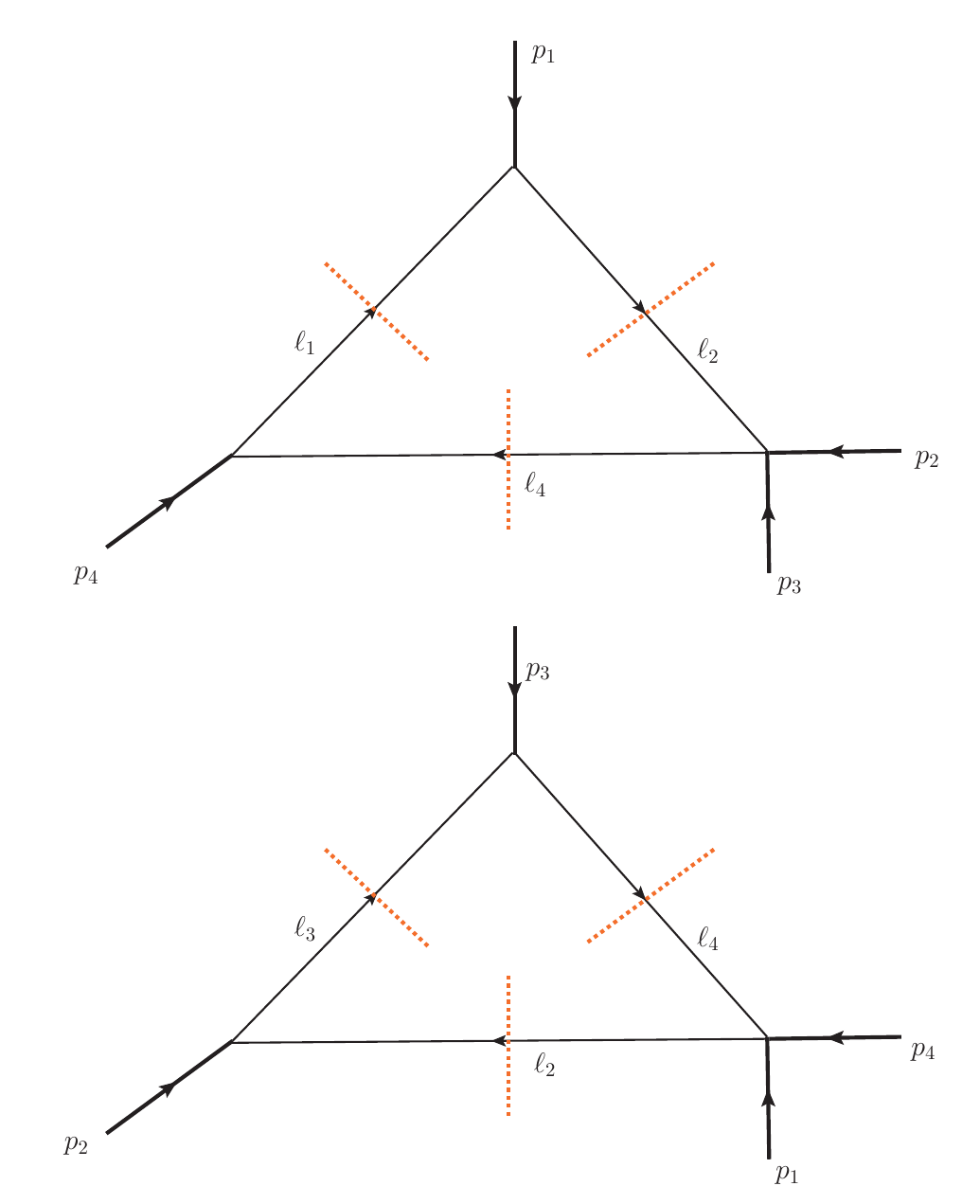}
\end{tabular}
\caption{Three-particle cut diagrams for the all-plus amplitude discussed in the text. Bold external lines represent physical gluons or gravitons, depending on the context.}
\label{triple}
\end{figure}

In turn, the triple-cut reads (see Fig.~\ref{triple})
\begin{equation}
A^{1-\textrm{loop integrand}}_{4}(\ell) \bigg|_{\textrm{triple cut}} = 
A^{\textrm{tree}}_{3}[p^{+}_1, - \ell_{2KL}, \ell^{IJ}_1] 
A^{\textrm{tree}}_{3}[p^{+}_2, -\ell_{3MN}, \ell^{KL}_2] 
A^{\textrm{tree}}_{4}[p^{+}_4, -\ell_{1IJ}, \ell^{MN}_3, p^{+}_3].
\end{equation}
The other possible three-particle cut diagrams are obtained from this one by cyclic relabeling of the external particles. The cut conditions are given by
\begin{eqnarray}
\ell_1^2 &=& \ell^2 = M^2 
\nonumber\\
\ell_2^2 &=& (\ell + p_{1})^2 = M^2
\nonumber\\ 
\ell_3^2 &=& (\ell + p_1 + p_2)^2 = M^2 . 
\end{eqnarray}
Notice that these imply that $\ell \cdot p_1 = 0$. One finds that
\begin{equation}
A^{1-\textrm{loop integrand}}_{4}(\ell) \bigg|_{\textrm{triple cut}} = - 16 M^{12} \frac{ s_{12} s_{23} }
{\langle 1 2 \rangle \langle 23 \rangle \langle 34 \rangle \langle 41 \rangle}
\frac{1}{(\ell - p_4)^2 - M^2}
\end{equation}
where we have chosen $r_1 = 2$ and $r_2 = 1$ and we used the cut conditions. 

\begin{figure}[htb]
\begin{center}
\includegraphics[height=80mm,width=85mm]{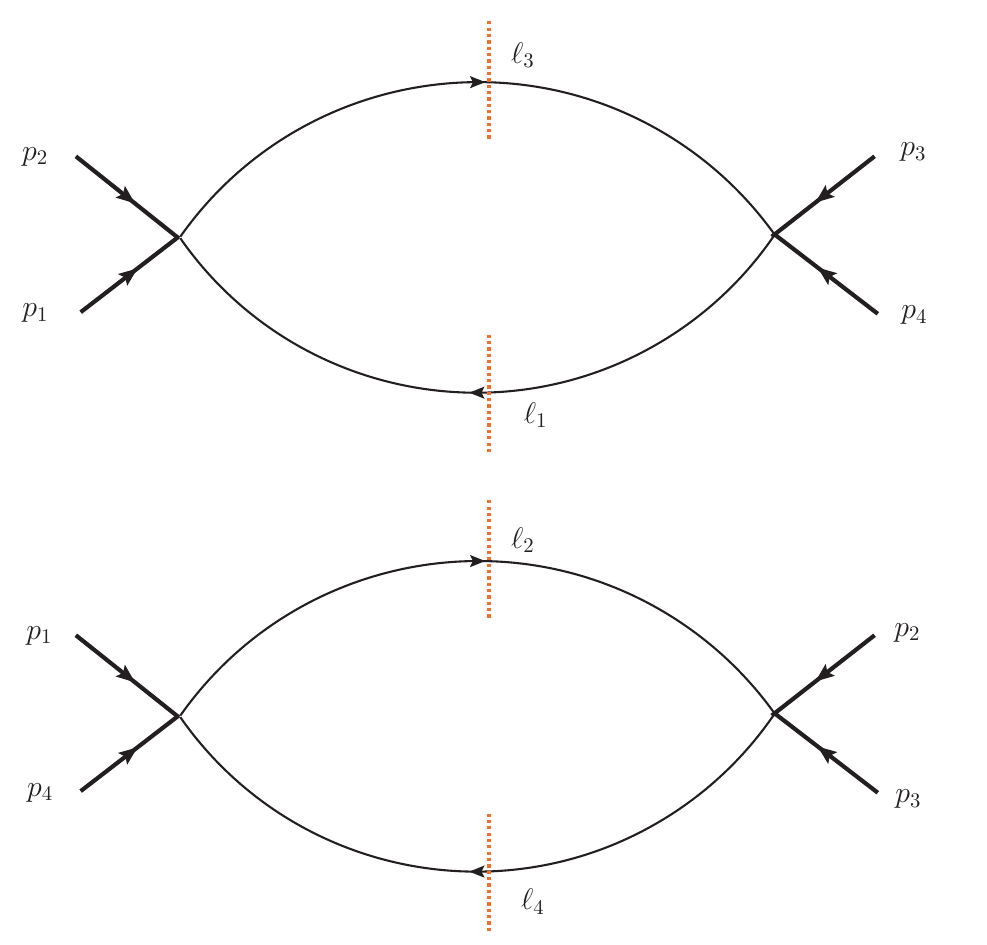}
\caption{Two-particle cut diagrams for the all-plus amplitude discussed in the text. Bold external lines represent physical gluons or gravitons, depending on the context. There are a total of six different bubbles.}
\label{double}
\end{center}
\end{figure}

Now let us evaluate the two-particle cut. We have that (see Fig.~\ref{double})
\begin{equation}
A^{1-\textrm{loop integrand}}_{4}(\ell) \bigg|_{\textrm{double cut}} = 
A^{\textrm{tree}}_{3}[p^{+}_2, -\ell_{3MN}, \ell^{IJ}_1,p^{+}_1] 
A^{\textrm{tree}}_{4}[p^{+}_4, -\ell_{1IJ}, \ell^{MN}_3, p^{+}_3].
\end{equation}
As above, the other possible two-particle cut diagrams are obtained from this one by cyclic relabeling of the external particles. The cut conditions are given by
\begin{eqnarray}
\ell_1^2 &=& \ell^2 = M^2 
\nonumber\\
\ell_3^2 &=& (\ell + p_{1} + p_{2})^2 = M^2 . 
\end{eqnarray}
One finds that
\begin{equation}
A^{1-\textrm{loop integrand}}_{4}(\ell) \bigg|_{\textrm{double cut}} = - 16 M^{12}
\frac{ s_{12} s_{23} }{\langle 1 2 \rangle \langle 23 \rangle \langle 34 \rangle \langle 41 \rangle}
\frac{1}{(\ell+p_1)^2 - M^2}
\frac{1}{(\ell - p_4)^2 - M^2} 
\end{equation}
where we used momentum conservation and the cut conditions.
 
Let us take a look at our results more carefully. For instance, for the triple cut there are two integrals that can contribute: The box integral and the triangle integral. However, in the expression for the triple cut one finds one uncut propagator. This would exclude triangle integrals from the expansion of the loop integral. 
This is not enough for a final conclusion; one must analyze the $2$-particle cut. So let us turn to this cut.
Box and triangle integrals generally contribute, as do bubble integrals. However, the double cut has two uncut propagators. This endorses the exclusion of triangle integrals, and also indicates that bubble integrals might be absent. By taking a single cut one finds three uncut propagators in the result. Therefore the final answer is that only the box integral should be present.

As well known, rational terms cannot be detected by unitarity cuts in four dimensions. One may consider dimensionally regularized forms for the tree-level amplitudes. In other words, we consider $d$-dimensional cuts, with $d = 4-2\epsilon$. Gluons live in four dimensions, but the loop momentum is $d$-dimensional. For an alternative and interesting formulation, see Ref.~\cite{Fazio:2014xea}.

Now a key component to be considered is the sum over physical states. The sum over the physical states of a gluon in general dimension is less straightforward than summing over positive and negative helicities in four dimensions~\cite{Bern:2019crd}. It is given by the so-called physical state projector:
\begin{equation}
P_{\mu\nu} = \sum_{\textrm{pols.}} \epsilon_{\mu}(-p) \epsilon_{\nu}(p) = - \eta_{\mu\nu} + \cdots
\end{equation}
where the ellipsis represent terms comprising of an arbitrary null reference momentum (for massless particles) or of the mass of the particle (for massive particles). We will focus on the maximal cut as only the box integral is present. Here only the $\eta_{\mu\nu}$-term in the physical state projector is important here. This significantly simplifies calculations because it allows us to use a simpler physical state projector. 

In the calculation of the maximal cut with the aforementioned physical state projector, we have to distinguish between the dimension of loop momenta and the space where physical states live -- we treat any factor of $D$ arising from contracting Lorentz indices ($\delta^{\mu}_{\mu} = D$) differently from loop momenta which we take to live in $d = 4-2\epsilon$ dimensions. In the limit that $D \to d$ one gets the same result as before, but one must consider the following shift in the mass: $M^2 \to M^2 + \mu^2$, where $\mu^{\alpha}$ is a vector representing the $(-2\epsilon)-$dimensional part of the loop momentum. The modified cut conditions are given by
\begin{eqnarray}
\ell_1^2 &=& \ell^2 = M^2 + \mu^2
\nonumber\\
\ell_2^2 &=& (\ell + p_{1})^2 = M^2 + \mu^2
\nonumber\\ 
\ell_3^2 &=& (\ell + p_1 + p_2)^2 = M^2 + \mu^2 
\nonumber\\
\ell_4^2 &=& (\ell_3 + p_3)^2 = (\ell - p_4)^2 = M^2 + \mu^2 . 
\end{eqnarray}
The final result, including the contribution coming from gluon internal lines, is given by
\begin{eqnarray}
A^{1-\textrm{loop}}_{4}[1^{+}, 2^{+}, 3^{+}, 4^{+}] &=&
- \frac{2i}{(4\pi)^{2-\epsilon}} \frac{ s_{12} s_{23} }{\langle 1 2 \rangle \langle 23 \rangle \langle 34 \rangle \langle 41 \rangle}
\Bigl[ I_{4}[\mu^4] + 8 {\cal I}_{4}[(M^2 + \mu^2)^2] \Bigr]
\end{eqnarray}
where we explicitly included the term $1/M^8 \to 1/(M^2 + \mu^2)^4$ in the second term, and~\cite{Brandhuber:05,Bern:96,Bern:97}
\begin{eqnarray}
I^{d}_{n}[\mu^{2r}] &=& i (-1)^{n+1} (4\pi)^{d/2} \int \frac{d^{-2\epsilon}\mu}{(2\pi)^{-2\epsilon}}
\int \frac{d^4 \ell}{(2\pi)^4} \frac{\mu^{2r}}{(\ell^2 - \mu^2)((\ell+p_1)^2 - \mu^2) \cdots 
((\ell+\sum_{i=1}^{n-1} p_i)^2 - \mu^2)}
\nonumber\\
&=& - \epsilon(1-\epsilon) \cdots (r-1-\epsilon) I^{d+2r}_{n}[1]
\nonumber\\
I^{d}_{n}[1] &=& i (-1)^{n+1} (4\pi)^{d/2} \int \frac{d^{-2\epsilon}\mu}{(2\pi)^{-2\epsilon}}
\int \frac{d^4 \ell}{(2\pi)^4} \frac{1}{(\ell^2 - \mu^2)((\ell+p_1)^2 - \mu^2) \cdots 
((\ell+\sum_{i=1}^{n-1} p_i)^2 - \mu^2)}
\nonumber\\
{\cal I}^{d}_{n}[\mu^{2r}] &=& i (-1)^{n+1} (4\pi)^{d/2} \int \frac{d^{-2\epsilon}\mu}{(2\pi)^{-2\epsilon}}
\int \frac{d^4 \ell}{(2\pi)^4} \frac{\mu^{2r}}{(\ell^2 - M^2 - \mu^2)((\ell+p_1)^2 - M^2 - \mu^2) \cdots 
((\ell+\sum_{i=1}^{n-1} p_i)^2 - M^2 - \mu^2)}
\nonumber\\
&=& - \epsilon(1-\epsilon) \cdots (r-1-\epsilon) {\cal I}^{d+2r}_{n}[1]
\nonumber\\
{\cal I}^{d}_{n}[1] &=& i (-1)^{n+1} (4\pi)^{d/2} \int \frac{d^{-2\epsilon}\mu}{(2\pi)^{-2\epsilon}}
\int \frac{d^4 \ell}{(2\pi)^4} \frac{1}{(\ell^2 - M^2 - \mu^2)((\ell+p_1)^2 - M^2 - \mu^2) \cdots 
((\ell+\sum_{i=1}^{n-1} p_i)^2 - M^2 - \mu^2)}
\nonumber\\
\end{eqnarray}
and therefore
\begin{eqnarray}
I_{4}[\mu^4] &=& - \epsilon(1-\epsilon) I^{d=8-2\epsilon}_{4}[1] = - \frac{1}{6} + {\cal O}(\epsilon)
\nonumber\\
{\cal I}_{4}[(M^2 + \mu^2)^2] &=& M^4 {\cal I}^{d=4-2\epsilon}_{4}[1] 
- 2 M^2 \epsilon {\cal I}^{d=6-2\epsilon}_{4}[1]
- \epsilon(1-\epsilon) {\cal I}^{d=8-2\epsilon}_{4}[1] 
\nonumber\\
- \epsilon {\cal I}^{d=6-2\epsilon}_{4}[1] &=& 0 + {\cal O}(\epsilon)
\nonumber\\
- \epsilon(1-\epsilon) {\cal I}^{d=8-2\epsilon}_{4}[1] &=& - \frac{1}{6} + {\cal O}(\epsilon).
\end{eqnarray}
The integrals with Merlin propagators are to be evaluated using the Lee-Wick prescription for integration in the complex $\ell^{0}$ plane. Scalar box integrals were explicitly calculated in Refs.~\cite{Bern:96,thooft:79,Denner:91}.

For the full elimination of ambiguities associated with rational terms, additional procedures should be taken into account~\cite{Bern:96}. Such strategies are, however, trivial here as the associated tree-level amplitude vanishes and there are no ultraviolet divergences.

\section{Quadratic gravity amplitudes}

We now discuss quadratic gravity amplitudes using modern on-shell amplitude methods for some interesting cases. Recent studies have discussed that ultra-Planckian scattering in the context of quantum quadratic gravity~\footnote{This terminology is used to reinforce the fact that, in the context of Refs.~\cite{Holdom:2021hlo,Holdom:2021oii}, one must still keep the Einstein-Hilbert term in the gravitational action at ultra-Planckian energies. We do not pursue this reasoning here; in this paper we are still going to name our theory as simply quadratic gravity.} might involve gravitational parton showers~\cite{Holdom:2021hlo}. Tree-level photon-photon scattering amplitude was also explored in the same gravitational context and was found that it is well-behaved at all energies~\cite{Holdom:2021oii}. In summary, the conclusion in such works is that, in an off-shell description, when all the exclusive cross sections are added for gravitational scattering processes, then one obtains a nice high-energy behavior.

The generalized unitarity method is particularly efficient in gravity theories concerning double-copy constructions~\cite{Bern:08,Bern:100,Bern:10}.  As well known, when the double copy is applied to gauge-theory tree-level amplitudes with external massless adjoint particles, the result is equivalent to the KLT formula; in this case, we can also consider KLT relations to the Weyl gravity.

The idea here is to employ the map
\begin{equation}
(\textrm{Higher-derivative YM}) \otimes \textrm{YM}
= \textrm{Weyl-Einstein}
\label{doublecopy}
\end{equation}
which is motivated by Refs.~\cite{Johansson:18,Johansson:17}. The Weyl-Einstein theory is described by the action
\begin{equation}
S = \int d^4x \sqrt{-g}
\left[\frac{2}{\kappa^2} R 
- \frac{1}{2\xi^2} C_{\mu\nu\alpha\beta}C^{\mu\nu\alpha\beta} \right]
\label{einsteinf}
\end{equation}
where $\kappa^2 = 32 \pi G$, $C_{\mu\nu\alpha\beta}$ is the Weyl tensor and we are disregarding the contributions coming from the cosmological constant, a surface term $\Box R$ and the Gauss-Bonnet term. This is the Einstein frame action, which can be obtained through a simple field redefinition from the more general Jordan-frame action~\cite{Salvio:18}
\begin{equation}
S = \int d^4x \sqrt{-g}
\left[\frac{2}{\kappa^2} R + \frac{1}{6 f_0^2} R^2 - \frac{1}{2\xi^2} C_{\mu\nu\alpha\beta}C^{\mu\nu\alpha\beta} \right].
\label{jordanf}
\end{equation}
In the Einstein frame there is an extra scalar field (with a specific potential) whose contribution can be added to the matter sector. This contribution is highly relevant in discussions of inflationary cosmology. Unitarity of quadratic gravity in a Minkowski background is discussed in Appendix B.

The discussion here has some dissimilarities in comparison with the one put forward in Refs.~\cite{Johansson:18,Johansson:17}. As in the previous section, we again introduce an auxiliary field with a very large mass $M$ (here $M$ is proportional to the Planck mass) and rewrite the action in terms of two quadratic derivative fields (and their interactions). As above, we find an unusual kinetic term that has an overall minus sign -- this is associated with the gravitational Merlin modes and the known backwards-in-time feature in the propagator. The other important difference has to do with the content of the map~(\ref{doublecopy}) itself; in Refs.~\cite{Johansson:18,Johansson:17}, a dimension-six gauge theory was constructed out of non-Abelian gauge fields and scalars with additional particular interactions; hence the field content was composed of a gauge field transforming in the adjoint representation of the gauge group and a scalar field transforming in a given real representation. The spectrum and associated interactions were established in such a way that could allow one to obtain non-trivial tree-level amplitudes that satisfied color-kinematics duality as well as standard BCJ relations. This in turn implies in a double copy prescription (including or not supersymmetry) that enables one to recast the BCJ double copy in terms of appropriate KLT relations. In fact, their map provides us a way to calculate tree-level amplitudes of a particular conformal (super)gravity theory, known as the (non-minimal) Berkovits-Witten theory. The present paper deals only with the mass-deformed minimal $(DF)^2$ theory, which is their terminology for the case discussed in the previous section -- the associated Lagrangian consists in the standard Yang-Mills term supplemented by a specific higher derivative contribution, with no extra scalars nor $F^3$ terms. As demonstrated above, at least up to four points this gauge theory has tree-level amplitudes consistent with color-kinematics duality, however in some cases standard BCJ relations are not satisfied. As will be discussed in due course, this means that the associated BCJ double copy cannot be rephrased in terms of traditional KLT relations in such situations. In addition, the dimension-six gauge theory considered in Refs.~\cite{Johansson:18,Johansson:17} is a four-derivative theory, as the one studied here, but without an explicit extra massive degree of freedom (the Merlin mode, in the present description); hence the YM part of the map is just a regular pure Yang-Mills theory. By contrast, in our case the YM part of our map cannot be a simple Yang-Mills theory for reasons that we will soon unveil.

Actually, the map given by Eq.~(\ref{doublecopy}) does not produce a pure Weyl-Einstein gravity. Typically the double copy of two vector fields contains more than just the graviton. For example, a gluon in four dimension has two helicities $\pm 1$, so the square has four states: the $\pm 2$ correspond to graviton, and the zero-helicity states which are identified as the dilaton and axion. So we write that $2 \times 2 = 2 + 1^{*} + 1$; $1^{*}$ here represents the massless two-form field generated by the double copy, which has one degree of freedom and as such should be dual to a scalar. This also happens for the Weyl-Einstein gravity: Here the spectrum also consists of additional five physical degrees of freedom associated with a gravitational Merlin, three states associated with a Merlin $2$-form field and a Merlin scalar -- that is $3 \times 3 = 5 + 3 + 1$ -- now the $3$ corresponds to the $2$-form field which, being massive, has three degrees of freedom and therefore is dual to a massive vector. The projection to pure gravity, which eliminates the dilaton and axion, can be performed in any case: In a unitarity cut, we apply the projection to graviton on each massless cut leg, which means simply correlating the helicities in the two gauge-theory copies. This works in a similar fashion for quadratic gravity: In order to work with only gravitational Merlins, we take the symmetric tensor product of gauge-theory Merlins. This is similar to the massive-gravity case, see Ref.~\cite{Momeni:20}. When using the four-dimensional helicity method, this means that we simply correlate the helicities of the two copies of gauge-theory amplitudes (including the Merlin particle). On the other hand, for $d$-dimensional constructions, as in the gauge-theory case we can use a suitable defined physical state projector~\cite{Bern:2019crd} (again including the Merlin particle).

As remarked right above, there is a subtlety associated with the map~(\ref{doublecopy}). When discussing amplitudes with gravitational Merlin particles on-shell, we should consider double copies using spontaneously broken gauge theories. This is important for obtaining a scattering amplitude that is well-behaved in the massless limit and which is compatible with spontaneous symmetry breaking~\cite{Johansson:19}.~\footnote{ Since QCD amplitudes double copy to gravity amplitudes containing massive abelian vectors, one may resort to a suitable covariantization of the Proca action. For the higher-derivative case the same phenomenon is expected to happen. For more details in the case of standard QCD, see Ref.~\cite{Johansson:19}.} Hence the pure YM theory of the double copy should contain massive gauge fields. Furthermore, one must be careful with the Merlin propagator, which has a different overall sign in comparison with the analogous massive propagator in the spontaneously broken gauge theory of the double copy. For a thorough discussion of the double-copy construction for spontaneously broken gauge theories, see Refs.~\cite{Chiodaroli:17,Chiodaroli:18,Chiodaroli:19}.

There is another subtlety involving the map~(\ref{doublecopy}). This map was shown to be correct (at least up to four points) in Refs.~\cite{Johansson:17,Johansson:18,Johansson:19} when at least one of the particles is a physical graviton (or gluon). Hence, we believe that a double-copy procedure for at least $3$-particle amplitudes involving Merlin particles and other particles (scalars, photons and fermions) also exists. There are a number of reasons to argue along these lines. First of all, the map~(\ref{doublecopy}) relates the spectrum of two theories; hence one could expect that the map holds for every particle in each spectrum in a suitable context. On the other hand, a double-copy construction of scattering amplitudes for massive gravity from massive Yang-Mills was explicitly realized in Refs.~\cite{Johnson:20,Momeni:20} up to four-point amplitudes.

Here we take the polarization tensors of massive Merlin higher-spin particles as tensor products of the spin-$1$ polarization vectors~\cite{Guevara:19a}:
\begin{equation}
\bigl[ \epsilon^{\alpha \dot{\alpha}} \bigr]^{(S)} = \bigl[ \epsilon^{\alpha \dot{\alpha}} \bigr]^{\otimes S}
= \left( \frac{ \sqrt{2} }{M} \right)^{S} \Bigl( | {\bf p} \rangle \bigl[ {\bf p} | \Bigr)^{S} 
\end{equation}
where symmetrization in the spinor products is implicitly understood (anti-symmetrization and the trace part corresponds to polarization of the $2$-form Merlin field and the Merlin scalar field, respectively). Henceforth we will consider this in the calculations that follow.

\subsection{Tree-level amplitudes}

Now let us calculate gravity amplitudes.~\footnote{See footnote 3.} For the pure-gravity case, we will use the KLT relations given above. As in the YM case, $3$-particle amplitudes involving only physical gravitons do not display contributions coming from higher-order derivative terms~\cite{Johansson:18,Dona:2015tra}, 
\begin{equation}
M^{(4)}_{3}[1^{h_1},2^{h_2},3^{h_3}] = M^2 M^{(2)}_{3}[1^{h_1},2^{h_2},3^{h_3}]
\end{equation}
and this result generalizes to an arbitrary number of gravitons by using BCFW recursion relations. Therefore, in the absence of the Einstein-Hilbert term, amplitudes involving only physical gravitons are zero. This is an interesting result; indeed, it is well known that tree scattering amplitudes involving gravitons from general relativity have a very bad high-energy behavior~\cite{Berends:1974gk}. But here high-energy behavior also means $M \to 0$ and hence it turns out that such amplitudes are indeed well behaved at high energies. Another way to obtain this result is to calculate explicitly the triple graviton coupling between an off-shell graviton and two on-shell gravitons as given by conformal gravity -- this direct computation shows us accurately the vanishing of such vertex and, as a result, the vanishing of the associated $3$-particle amplitude~\cite{Donoghue:2018izj,Dona:2015tra,Maldacena:2011mk,Adamo:2013tja,Adamo:2016ple}.~\footnote{Bob Holdom has done this calculation also, and has found the same result -- see Ref.~\cite{Holdomtalk}.}

The reader might feel uneasy with this statement since $M \to 0$ could amount to considering that the Einstein-Hilbert term is not present at high energies. But this is actually a key feature in induced gravity theories; for instance, in the recent proposed GAQ gravity, gravitational interactions are weakly coupled at all energy scales and gets assisted by a Yang-Mills theory which has a two-fold role -- it determines the Planck scale and induces the Einstein-Hilbert action at low energies~\cite{Donoghue:2018izj}. So the high-energy Lagrangian density consists simply of the Yang-Mills term plus the Weyl tensor term, that is $F^2 + C^2$ (suppressing couplings and Lorentz indices); in other words, it amounts to taking $M \to 0$ in the expressions for the amplitudes. On the other hand, one can still have the Einstein-Hilbert term and assume that the energy scale of the intrinsic $G^{-1}$ is much smaller than the Planck scale as determined by the helper gauge theory.

Amplitudes involving a single gravitational Merlin particle vanishes:  
\begin{equation}
M^{(4)}_{n+1}(1^{h_1},2^{h_2},\ldots,n^{h_n},{\bf k}) = 0
\end{equation} 
where $h_i$ represents a given graviton helicity. Once more we recover the results of Ref.~\cite{Johansson:18}. 

As in the gauge-theory case, we will be employing different propagators for the physical graviton and the gravitational Merlin, which implies the following form for the total gravitational propagator in momentum space:
\begin{equation}
D_{\mu\nu\rho\sigma}(p) = 
\frac{1}{2 p^2} \Bigl( \eta_{\mu\rho} \eta_{\nu\sigma} + \eta_{\mu\sigma} \eta_{\nu\rho} 
- \eta_{\mu\nu} \eta_{\rho\sigma} \Bigr) 
- \frac{1}{2 (p^2 - M^2 - i M \Gamma)} \Bigl( \eta_{\mu\rho} \eta_{\nu\sigma} 
+ \eta_{\mu\sigma} \eta_{\nu\rho} 
- \frac{2}{3} \eta_{\mu\nu} \eta_{\rho\sigma} \Bigr) .
\label{prop_grav}
\end{equation}
There would be, of course, numerator terms containing projectors with explicit $p_{\mu}$ dependence, but these are irrelevant since they cancel out in connected correlators~\cite{Johansson:18}.

The $3$-particle amplitude involving two gravitational Merlin particles may be obtained from the YM case by using the KLT relations~\cite{Johansson:19}:
\begin{eqnarray}
M_{3}(1^{++},{\bf 2},{\bf 3}) &=& i A^{\textrm{tree, HD}}_{3}[1^{+},{\bf 2},{\bf 3}] 
A^{\textrm{tree,YM}}_{3}[1^{+},{\bf 2},{\bf 3}] = - 2i
\frac{\langle r| {\bf 3} | 1\bigr]^2}{M^4 \langle r1 \rangle^2} 
\langle {\bf 3} {\bf 2} \rangle^4
\nonumber\\
M_{3}(1^{--},{\bf 2},{\bf 3}) &=& i A^{\textrm{tree, HD}}_{3}[1^{-},{\bf 2},{\bf 3}] 
A^{\textrm{tree,YM}}_{3}[1^{-},{\bf 2},{\bf 3}] = - 2i \frac{\bigl[ r| {\bf 3} | 1\rangle^2}{M^4 \bigl[ 1r \bigr]^2} 
\bigl[ {\bf 3} {\bf 2} \bigr]^4 .
\end{eqnarray}
We are suppressing factors of $(\kappa/2)^{m-2}$ at $m$ points. A similar relation also holds for the case involving only Merlin particles; one should just recall that, for a non-Abelian spontaneous broken gauge theory, the number of massive and massless bosons will depend on the representation of the group in which the order parameter transform. So, if in the higher-derivative theory one has $N$ gluons and $N$ Merlin particles, in the spontaneously broken gauge theory one must also have $N$ gluons and $N$ massive vector particles. All massive particles have the same mass $M$, including the gravitational Merlins. We find
\begin{eqnarray}
M_{3}({\bf 1},{\bf 2},{\bf 3}) &=& i A^{\textrm{tree, HD}}_{3}[{\bf 1},{\bf 2},{\bf 3}] 
A^{\textrm{tree,YM}}_{3}[{\bf 1},{\bf 2},{\bf 3}]
\nonumber\\
&=& - 2 i
\Bigl[ \eta_{\mu\nu} ( p_{1}  - p_2 )_{\rho} + \eta_{\nu\rho} ( 2 p_{2} + p_{1} )_{\mu}
- \eta_{\rho\mu} (2 p_{1} + p_2 )_{\nu} \Bigr]
\nonumber\\
&\times& \Bigl[ \eta_{\alpha\beta} ( p_{1}  - p_2 )_{\gamma} + \eta_{\beta\gamma} ( 2 p_{2} + p_{1} )_{\alpha}
- \eta_{\gamma\alpha} (2 p_{1} + p_2 )_{\beta} \Bigr]
 \epsilon^{\mu\alpha}({\bf 1})\epsilon^{\nu\beta}({\bf 2}) \epsilon^{\rho\gamma}({\bf 3}) 
\end{eqnarray}
Moreover, notice that this is consistent with the amplitudes presented in Refs.~\cite{Arkani-Hamed:17,Johansson:19}. In addition, recall that in the gauge theory we have defined the coupling constant as $g/M^2$ (in the amplitudes involving solely gluons and Merlin particles).

\subsubsection{Compton scattering involving gravitons and matter particles}

As demonstrated above, graviton-graviton scattering in quadratic gravity has the same form as calculated in general relativity (except for an overall factor of $M^2$)~\cite{Dona:2015tra}. Can we expect this to hold for the gravitational Compton scattering involving matter particles?

The answer is in the affirmative. In fact, this is true for all tree-level processes that do not carry external Merlins~\cite{Holdom:2021hlo}. In our perspective, this comes from a simple analysis of $3$-particle amplitudes. Indeed, when using recursive relations to build up the $t$-channel amplitude, one would use a $3$-particle amplitude involving three gravitons (for a process running through graviton exchange) and a $3$-particle amplitude involving a Merlin mode and two gravitons (for a process running through Merlin exchange). As aforementioned, the latter identically vanishes, whereas the former is just 
$M^2 M^{(2)}_{3}[1^{h_1},2^{h_2},3^{h_3}]$. Hence in the high-energy limit it vanishes. However, when building up the complete amplitude using KLT relations, one finds that it is expressed as a product of Yang-Mills amplitudes without the $t$-channel contribution. Moreover, since the couplings of gravitons with matter are the same in both theories, $s$-channel and $u$-channel contributions will actually be the same in quadratic gravity as the ones calculated within general relativity. Accordingly the total Compton amplitude will be the same also. Within the induced-gravity context, the Einstein-Hilbert term would be absent in the ultra-Planckian regime, and then we would expect that the tree amplitudes involving gravitons and conformally coupled scalars, fermions and/or ($4$-dimensional) photons vanish at high energies as the S-matrix in a conformal field theory ought to be trivial.

\subsubsection{Compton scattering: The Merlin-graviton case}

Amplitudes involving gravitational Merlin particles and gravitons can be constructed from the corresponding gauge-theory ones by using the double-copy procedure. Since in this case the color-ordered gauge-theory amplitudes obey the standard BCJ relations, standard KLT relations are available to construct the gravity amplitudes.  So let us display the amplitude $M_{4}[1^{++}, {\bf 2}, {\bf 3},4^{++}]$ which involves two gravitational Merlin particles. One finds
\begin{equation}
M^{\textrm{tree}}_{4}({\bf 2},1^{++},4^{++},{\bf 3}) = - i s_{23} 
A^{\textrm{tree, HD}}_{4}[{\bf 2},1^{+},4^{+},{\bf 3}] A^{\textrm{tree, YM}}_{4}[{\bf 2},4^{+},1^{+},{\bf 3}] 
= 4i   \frac{\bigl[ 14 \bigr]^4}{s_{23}}
\frac{\langle {\bf 3} {\bf 2} \rangle^4}{(s_{12} - M^2)(s_{13} - M^2)}
\end{equation}
which agrees with the results presented in~\cite{Johansson:19}. The other Compton scattering amplitude reads
\begin{eqnarray}
M^{\textrm{tree}}_{4}({\bf 2},1^{--},4^{++},{\bf 3}) &=& - i s_{23} 
A^{\textrm{tree, HD}}_{4}[{\bf 2},1^{-},4^{+},{\bf 3}] A^{\textrm{tree, YM}}_{4}[{\bf 2},4^{+},1^{-},{\bf 3}] 
\nonumber\\
&=& 4 i  \frac{1}{s_{23} (s_{12} - M^2) (s_{13} - M^2) }
\Bigl(  \bigl[ 4 {\bf 3} \bigr] \langle 1 {\bf 2} \rangle  + \langle 1 {\bf 3}  \rangle \bigl[ 4 {\bf 2} \bigr]  \Bigr)^4 .
\end{eqnarray}

\subsubsection{Compton scattering: The Merlin-scalar case}

The $3$-particle amplitude with two massive complex scalars and one Merlin particle can be obtained as follows:
\begin{eqnarray}
M_{3}^{\textrm{tree}}(\ell_A^{+},{\bf p},\ell_B^{-})  
&=& M_{3}^{\textrm{tree}}(\ell_A^{-},{\bf p},\ell_B^{+}) 
=  i A^{\textrm{tree, HD}}_{3}[\ell_A^{+},{\bf p},\ell_B^{-}] 
A^{\textrm{tree,YM}}_{3}[\ell_A^{+},{\bf p},\ell_B^{-}]
\nonumber\\
&=& 2 i \frac{ \langle {\bf p} | \ell_A | {\bf p} \bigr]^2  }{M^2} .
\end{eqnarray}
This can be explicitly verified by employing the vertices derived in Ref.~\cite{Johansson:19}. 

Let us evaluate the Compton scattering amplitude involving two Merlin particles and two massive scalars. We can now apply fully the BCJ double-copy prescription. This means that we can employ a similar relation as in~\cite{Johansson:19}, but now applied for the numerators:
\begin{equation}
M(1_s,2,3,4_s) = i \sum_{k} \frac{n^{(s_1)}_k \tilde{n}^{(s_2)}_k }{s_k}
\end{equation}
where $s=s_1+s_2$, $2,3$ are graviton or Merlin particles, $\tilde{n}_k$ are numerators belonging to the spontaneously broken gauge theory of the double copy described earlier and $s_k$ are inverse propagators (they could be massive). We emphasize that the general formula 
\begin{equation}
M^{\textrm{tree}}(1_{s},2,3,4_{s}) = i \left( \frac{\kappa}{2} \right)^2 s_{23} 
(-1)^{ \left \lfloor{s}\right \rfloor - \left \lfloor{s_1}\right \rfloor - \left \lfloor{s_2}\right \rfloor + 1}
A^{\textrm{tree}}[1_{s_1},2,3,4_{s_1}] 
A^{\textrm{tree}}[1_{s_2},3,2,4_{s_2}] , \,\,\,
s = s_1 + s_2 
\end{equation}
as put forward in Ref.~\cite{Johansson:19} implicitly relies on the existence of the BCJ relations. Indeed, the fact that the color-ordered amplitudes obey the BCJ relations imply that gravity amplitudes should satisfy a KLT-like formula, exactly as the one above, since BCJ relations are a manifestation of the color-kinematics duality and the duality implies the BCJ double copy. Hence the BCJ double copy is equivalent to the tree-level KLT formulas. But since the KLT relations are valid for external massless adjoint particles, or at most two non-adjoint ones~\cite{Johansson:15}, again we have a similar situation found when we discussed the absence of homogeneous BCJ relations between the amplitudes, so here we should expect to find a ``non-homogeneous" KLT relation. Schematically:
\begin{equation}
\begin{tikzcd}
  \textrm{Color-kinematics duality} \arrow[r] \arrow[d, red]
    & \textrm{BCJ double copy} \arrow[d, red] \\
  \textrm{(Non-Hom) BCJ relations} \arrow[r, red,black]
& |[black]| \textrm{(Non-Hom) KLT relations} 
\end{tikzcd}
\end{equation}
As in the discussion of color-kinematics duality, we might expect the emergence of KLT relations in a standard form when demanding the absence of physical $t$-channel contributions. As argued, this is a strong constraint in that it imposes a kind of special kinematics for the amplitudes.

By resorting to recursion relations involving massive particles~\cite{Badger:05,Franken:20,Ballav:21} and applying BCJ relations with massive particles~\cite{Naculich:14}, the tree-level proof of the squaring relations given in Ref.~\cite{Bern:100} can be generalized in a straightforward way -- one just have to remember that the gauge numerators now are gauge invariant, so the product of left and right numerators of subamplitudes in the BCFW approach must be equal to the numerator of the original representation of the amplitude, but evaluated at shifted momenta, that is (in the notation of Ref.~\cite{Bern:100})
$$
i \hat{n}^{\alpha}_{L,i} \hat{n}^{\alpha}_{R,i} = \hat{n}_i(z_{\alpha}).
$$
In other words, the product of left and right numerators of subamplitudes in the BCFW formalism for the associated gravity amplitudes should inherit this property. In addition, in the course of their proof, the authors in Ref.~\cite{Bern:100} used the fact that the gravity amplitude ${\cal M}$, constructed from recursion relations by assuming the validity of the double copy for the subamplitudes, and the gravity amplitude ${\cal M}^{\prime}$, obtained by using directly the double copy, differ by a function ${\cal P}$ of momenta and polarization vectors that is polynomial in $z$ under the associated complex shift. This polynomial turned out to be a gauge-invariant, local expression quadratic in momenta. However, in four dimensions a local, gauge-invariant expression must be expressible as a polynomial in the angle and square brackets. As noted by the authors, no expression quadratic in angle and square brackets can have the correct little-group scaling property, which implies that ${\cal P} \equiv 0$. One may be worried that in our case this is no longer valid since we are dealing with a higher-derivative gravity theory and hence ${\cal P}$ could contain contributions that are higher order in momenta; moreover it would be expressible as a polynomial in the angle and square brackets of massive spinors. Notwithstanding, the same conclusions can be obtained here by noting that: (1) A Lee-Wick-type theory can always be recast in terms of a theory containing only quadratic propagators, in which one of the fields describe a Merlin particle, and hence our gravity theory can be rewritten as a quadratic theory, which implies that ${\cal P}$ must be again quadratic in momenta, and (2) In the massless limit, the massive spinors reproduce the usual helicity-formalism amplitudes for transverse polarizations, and hence in order to have a correct high-energy limit, ${\cal P}$ again must be expressible as a polynomial in the angle and square brackets of the spinor-helicity formalism in the massless limit, which implies that, in order for the high-energy limit to reproduce the previous result of~\cite{Bern:100}, the associated expression cannot transform correctly under the action of the little group and therefore ${\cal P} \equiv 0$. 

Coming back to the scalar case, since in the gauge-theory case we had something like
\begin{equation}
A(1,2,3,4) = \sum_{k} \frac{c_k n^{(0)}_k}{s_k}
\end{equation}
where $c_k$ are color factors and $n^{(0)}_k$ are numerators associated with the scattering with scalars that share the same algebraic properties of the $c_k$, we can employ the double-copy prescription in order to calculate the associated gravity amplitudes:
\begin{equation}
M(1,2,3,4) = i \sum_{k} \frac{n^{(0)}_k \tilde{n}^{(0)}_k }{s_k} .
\end{equation}
Hence we find that
\begin{eqnarray}
M_{4}(\ell_A, {\bf 1}, {\bf 2}, \ell_B) &=& 
\left[ \frac{ 2 }{M^2} \langle {\bf 1} | \ell_A | {\bf 1} \bigr]  
\langle {\bf 2} | \ell_B | {\bf 2} \bigr]
+ \frac{\langle {\bf 1} {\bf 2} \rangle \bigl[ {\bf 2} {\bf 1} \bigr] }{M^2} (2 ({\bf 1} \cdot \ell_A) + M^2) \right]^2
 \frac{i}{s - m^2}
\nonumber\\
&+&
\left[ \frac{2}{M^2}  \langle {\bf 1} | \ell_B | {\bf 1} \bigr] 
\langle {\bf 2} | \ell_A | {\bf 2} \bigr]
+ \frac{\langle {\bf 1} {\bf 2} \rangle \bigl[ {\bf 2} {\bf 1} \bigr] }{M^2} (2 ({\bf 2} \cdot \ell_A) + M^2) \right]^2
\frac{i}{u - m^2}
\nonumber\\
&-& \frac{4 i}{M^4} 
\Bigl( \langle {\bf 1} {\bf 2} \rangle \bigl[ {\bf 2} {\bf 1} \bigr] ( {\bf 1} - {\bf 2} ) \cdot \ell_A 
+ \langle {\bf 1} | \ell_{A} | {\bf 1} \bigr] \langle {\bf 2} | \ell_B | {\bf 2} \bigr] 
- \langle {\bf 2} | \ell_{A} | {\bf 2} \bigr] \langle {\bf 1} | \ell_B | {\bf 1} \bigr] 
\Bigr)^2
\left( \frac{ 1 }{t} - \frac{ 2 }{t - M^2} \right) .
\label{Mampscalar}
\end{eqnarray}
However, as extensively discussed in the literature, the double-copy procedure produces states corresponding to a graviton, an antisymmetric tensor and dilaton -- such contributions are typically given by the crossed terms in the $t$-channel above. If we wish to allow for only gravitons and Merlins to flow through the cuts for the $t$-channel, we must use the following (gauge-dependent) four-dimensional physical state projectors
\begin{eqnarray}
\sum_{\lambda = \pm 2} \epsilon^{\mu\nu}_{\lambda}(p;r) \epsilon^{\rho\sigma *}_{\lambda}(p;r)&\to&
\frac{1}{2} \Bigl( \eta^{\mu\rho} \eta^{\nu\sigma} + \eta^{\mu\sigma} \eta^{\nu\rho} 
- \eta^{\mu\nu} \eta^{\rho\sigma} \Bigr) \,\,\,\ (\textrm{Graviton})
\nonumber\\
\sum_{\lambda = -2}^{2} \epsilon^{\mu\nu}_{\lambda} \epsilon^{\rho\sigma *}_{\lambda} &\to&
\frac{1}{2} \Bigl( \eta_{\mu\rho} \eta_{\nu\sigma}
+ \eta_{\mu\sigma} \eta_{\nu\rho} \Bigr)
- \frac{1}{3} \eta_{\mu\nu} \eta_{\rho\sigma} \,\,\,\ (\textrm{Merlin}) .
\label{projectors}
\end{eqnarray}
These are the forms consistent with the expression of the gravitational propagator given in Eq.~(\ref{prop_grav}). The state projector projects out non-gravitational states. This modifies the numerator of the $t$-channel to (up to an overall sign)
\begin{eqnarray}
\textrm{Res,massless}_{t}  &=& 
\frac{4}{M^4} \Bigl( \langle {\bf 1} {\bf 2} \rangle \bigl[ {\bf 2} {\bf 1} \bigr] ( {\bf 1} - {\bf 2} ) \cdot \ell_A
+ \langle {\bf 2} | \ell_{A} | {\bf 2} \bigr] \langle {\bf 1} | {\bf 2} | {\bf 1} \bigr]
- \langle {\bf 1} | \ell_{A} | {\bf 1} \bigr] \langle {\bf 2} | {\bf 1} | {\bf 2} \bigr]  \Bigr)^2
- \frac{8 m^2}{M^2} \bigl[ {\bf 1} {\bf 2} \bigr]^2 \langle {\bf 1} {\bf 2} \rangle^2
\nonumber\\
\textrm{Res,massive}_{t} &=& 8
\Bigl[  ( \epsilon({\bf 1}) \cdot \epsilon({\bf 2}) )  ( {\bf 1}  - {\bf 2} ) \cdot \ell_{A}
- 2 \Bigl( \ell_{A} \cdot \epsilon({\bf 1}) {\bf 1} \cdot \epsilon({\bf 2})
- \ell_{A} \cdot \epsilon({\bf 2}) {\bf 2} \cdot \epsilon({\bf 1})  \Bigr) \Bigr]^2
\nonumber\\
&-& 8 m^2 M^2 \Bigl( \epsilon({\bf 1}) \cdot \epsilon({\bf 2}) \Bigr)^2.
\end{eqnarray}
after demanding consistent factorization in the physical channels in terms of the three-particle amplitudes.

Let us study the high-energy behavior of the amplitude~(\ref{Mampscalar}). At leading order, $s+t+u=0$ as all particles involved can be taken to be approximately massless. First consider $\epsilon({\bf 1})$ transverse and $\epsilon({\bf 2})$ longitudinal. We find that
\begin{eqnarray}
\textrm{s-channel} &\approx& \Bigl( (2 \ell_{A} +  p_2) \cdot \epsilon(1) \Bigr)^2 
\frac{i s}{M^2}
\nonumber\\
\textrm{u-channel} &\approx& \Bigl( 2 \ell_{B} \cdot \epsilon(1) +  p_2 \cdot \epsilon(1) \Bigr)^2 
\frac{i u}{M^2}
= \Bigl( (2 \ell_{A} +  p_2) \cdot \epsilon(1) \Bigr)^2 \frac{i u}{M^2}
\nonumber\\
\textrm{t-channel} &\approx&   \Bigl( (2 \ell_{A} + p_2) \cdot \epsilon(1) \Bigr)^2 \frac{i t}{M^2}
\end{eqnarray}
and therefore
\begin{equation}
M_{4}(\ell_A,1^{T}, 2^{L},  \ell_B) \xrightarrow[\text{HE}]{} 
\Bigl( (2 \ell_{A} + p_2) \cdot \epsilon(1) \Bigr)^2  \frac{i (s+t+u)}{M^2}.
\end{equation}
On the other hand, if we consider longitudinal polarizations, we find that
\begin{eqnarray}
\textrm{s-channel} &\approx&  (2 s + t)^2 
\frac{i s}{4 M^4}
\nonumber\\
\textrm{u-channel} &\approx& (2 u + t)^2 
\frac{i u}{4 M^4}
\nonumber\\
\textrm{t-channel} &\approx&   (s - u)^2 \frac{i t}{4 M^4} .
\end{eqnarray}
Hence
\begin{equation}
M_{4}(\ell_A,1^{L}, 2^{L} , \ell_B) \xrightarrow[\text{HE}]{} 
i (s+t+u) \frac{(4 s^2+s (t-4 u)+u (t+4 u))}{4 M^4}.
\end{equation}
So our gravitational amplitudes with longitudinal modes do not seem to grow with energy. As for the transverse polarization, it is easy to see that we obtain the usual result involving two scalars and two gravitons.

In summary, the above results show that the scalar Compton amplitudes in quadratic gravity have a better high-energy behavior in comparison with the corresponding ones calculated from general relativity~\cite{Berends:1974gk}.

\subsubsection{Compton scattering: The Merlin-fermion case}

Now let us discuss amplitudes involving gravitational Merlin particles and massless fermions. The fermions must have opposite helicity for a non-vanishing amplitude. For $3$-particle amplitudes, we obtain a product of a $3$-particle amplitude involving fermions and the gauge Merlin particle and a $3$-particle amplitude involving massless scalars and a massive vector particle. That this is correct can be explicitly verified by employing the vertices derived in Ref.~\cite{Johansson:19} for the case of massless fermions. One finds
\begin{eqnarray}
M_{3}(1^{-1/2},2^{+1/2},{\bf 3}) &=& 
i A^{\textrm{tree, HD}}_{3}[1^{-1/2},2^{+1/2},{\bf 3}]
A^{\textrm{tree,YM}}_{3}[1,2,{\bf 3}]
\nonumber\\
&=& - \sqrt{2} \frac{\langle {\bf 3} 1 \rangle \bigl[ 2 {\bf 3} \bigr] \langle {\bf 3} | 2 | {\bf 3} \bigr]}{M^2} .
\end{eqnarray}
In a similar fashion:
\begin{equation}
M_{3}(1^{+1/2},2^{-1/2},{\bf 3}) = 
-\sqrt{2} \frac{\langle {\bf 3} 2 \rangle \bigl[ 1 {\bf 3} \bigr] \langle {\bf 3} | 2 | {\bf 3} \bigr]}{M^2} .
\end{equation}
Now let us calculate the Compton scattering amplitude involving two gravitational Merlin particles and two massless fermions. We follow the same idea as above. Consider fermion $4$ as outgoing and fermion $3$ as incoming.  We consider that $h_3 = - 1/2$ and $h_4 = + 1/2$. As discussed in the gauge-theory case, we find that
\begin{equation}
A(1,2,3,4) = \sum_{k} \frac{c_k n^{(1/2)}_k}{s_k}
\end{equation}
where $n^{(1/2)}_k$ are numerators associated with the scattering with scalars that share the same algebraic properties of the $c_k$. Hence we can employ the double-copy prescription in order to calculate the associated gravity amplitudes, which implies that
\begin{equation}
M(1,2,3,4) = i \sum_{k} \frac{n^{(1/2)}_k \tilde{n}^{(0)}_k }{s_k}
\end{equation}
for the scattering involving gravitational Merlin particles and massless fermions. We find
\begin{eqnarray}
M_{4}( 3^{-1/2}, {\bf 1}, {\bf 2},4^{+1/2}) &=& 
\left( \frac{2}{M^2}\langle 3 {\bf 1} \rangle \bigl[ {\bf 1} | ({\bf 1} + 3) |{\bf 2} \rangle \bigl[ {\bf 2} 4 \bigr] \right)
\left( - \frac{ 2 }{M^2} \langle {\bf 1} | 3 | {\bf 1} \bigr]  
\langle {\bf 2} | 4 | {\bf 2} \bigr]
+ \frac{\langle {\bf 1} {\bf 2} \rangle \bigl[ {\bf 2} {\bf 1} \bigr] }{M^2} (2 ({\bf 1} \cdot 3) + M^2) \right)
\frac{i}{s}
\nonumber\\
&+& \left( \frac{2}{M^2} \langle 3 {\bf 2} \rangle \bigl[ {\bf 2} | ({\bf 2} + 3) |{\bf 1} \rangle \bigl[ {\bf 1} 4 \bigr] \right)
\left( - \frac{2}{M^2}  \langle {\bf 1} | 4 | {\bf 1} \bigr] 
\langle {\bf 2} | 3 | {\bf 2} \bigr]
+ \frac{\langle {\bf 1} {\bf 2} \rangle \bigl[ {\bf 2} {\bf 1} \bigr] }{M^2} (2 ({\bf 2} \cdot 3) + M^2) \right)
\frac{i}{u}
\nonumber\\
&+&  
i \left( \frac{1}{M^2}
 \left(  \langle {\bf 1} {\bf 2} \rangle \bigl[ {\bf 2} {\bf 1} \bigr] 
\langle 3 | ( {\bf 1} - {\bf 2} ) | 4 \bigr]
+  2 \Bigl( \langle 3 {\bf 1} \rangle \bigl[ {\bf 1} 4 \bigr] \langle {\bf 2} | (3-4) | {\bf 2} \bigr] 
- \langle 3 {\bf 2} \rangle \bigl[ {\bf 2} 4 \bigr] \langle {\bf 1} | (3-4) | {\bf 1} \bigr] \Bigr)
\right) \right)
\nonumber\\
&\times& 
\left( \frac{2}{M^2} 
\Bigl( \langle {\bf 1} {\bf 2} \rangle \bigl[ {\bf 2} {\bf 1} \bigr] ( {\bf 1} - {\bf 2} ) \cdot p_3 
-  \langle {\bf 1} | 3 | {\bf 1} \bigr] \langle {\bf 2} | 4 | {\bf 2} \bigr] 
+ \langle {\bf 2} | 3 | {\bf 2} \bigr] \langle {\bf 1} | 4 | {\bf 1} \bigr] 
\Bigr) \right)
\left( \frac{ 1 }{t} - \frac{ 2 }{t - M^2} \right) .
\end{eqnarray}
As in the scalar case, to project out non-gravitational states in the $t$-channel, one may use consistent factorization in all possible channels and then use the state projectors given in Eq.~(\ref{projectors}).

Let us analyze the high-energy behavior. Take $\epsilon({\bf 1})$ transverse and $\epsilon({\bf 2})$ longitudinal. We find that
\begin{eqnarray}
\textrm{s-channel} &\approx&- 2  \ell \cdot \epsilon(1) \Bigl( (p_3 +  p_4) \cdot \epsilon(1) \Bigr) 
\frac{i s}{M^2}
\nonumber\\
\textrm{u-channel} &\approx& - 2  \ell \cdot \epsilon(1)
\Bigl( (p_3 +  p_4) \cdot \epsilon(1) \Bigr) 
\frac{i u}{M^2}
\nonumber\\
\textrm{t-channel} &\approx& - 2 \ell \cdot \epsilon(1)  
\Bigl( (p_3 + p_4) \cdot \epsilon(1) \Bigr) \frac{i t}{M^2}
\end{eqnarray}
where $\ell =  |3 \rangle \bigl[ 4|$, as defined above. Hence at leading order
\begin{equation}
M_{4}(3^{-1/2}, 1^{T}, 2^{L}, 4^{+1/2}) \approx
- 2 \ell \cdot \epsilon(1)  
\Bigl( (p_3 + p_4) \cdot \epsilon(1) \Bigr) \frac{i}{M^2} (s+t+u).
\end{equation}
For longitudinal polarizations, we find that
\begin{eqnarray}
\textrm{s-channel} &\approx&  - \ell \cdot p_1 (s-u) \frac{i s}{M^4}
\nonumber\\
\textrm{u-channel} &\approx& -  \ell \cdot p_1(s-u) \frac{i u}{M^4}
\nonumber\\
\textrm{t-channel} &\approx& - \ell \cdot p_1 (s-u) \frac{i t}{M^4}
\end{eqnarray}
at leading order; therefore
\begin{equation}
M_{4}(3^{-1/2}, 1^{L}, 2^{L}, 4^{+1/2}) \approx
- \ell \cdot p_1 (s-u) \frac{i}{M^4} (s+t+u).
\end{equation}

\subsubsection{Compton scattering: The Merlin-photon case}

Now let us discuss amplitudes involving gravitational Merlin particles and photons. Using the vertices derived in Ref.~\cite{Johansson:19} for the case of massless fermions and photons, one can explicitly check that the $3$-particle amplitude will be given by the product of a $3$-particle amplitude involving massless fermions and the gauge Merlin particle and a $3$-particle amplitude involving massless fermions and a massive vector particle. This construction implies that both photons need to have opposite helicity to give a non-vanishing result. Furthermore, we will also consider the fermionic amplitude involving a fermion and an anti-fermion, so both can be incoming (or outgoing). One obtains
\begin{eqnarray}
M^{\textrm{tree}}_{3}({\bf 1},2^{+},3^{-}) &=& - i 
A^{\textrm{tree,HD}}_{3}[{\bf 1},2^{+1/2},3^{-1/2}] 
A^{\textrm{tree,YM}}_{3}[{\bf 1},2^{+1/2},3^{-1/2}]
= - 2 i \frac{ \langle 3 {\bf 1} \rangle \bigl[ {\bf 1} 2 \bigr] 
\langle 3 {\bf 1} \rangle \bigl[ {\bf 1} 2 \bigr] }{M^2} 
\nonumber\\
M^{\textrm{tree}}_{3}({\bf 1},2^{-},3^{+}) &=&
- 2 i \frac{ \langle 2 {\bf 1} \rangle \bigl[ {\bf 1} 3 \bigr] 
\langle 2 {\bf 1}\rangle \bigl[ {\bf 1} 3 \bigr] }{M^2}
\end{eqnarray}
Now let us calculate the Compton scattering amplitude involving two gravitational Merlin particles and two photons. We follow the same idea as above. Hence we write
\begin{equation}
M(1,2,3,4) = i \sum_{k} \frac{\tilde{n}^{(1/2)}_k n^{(1/2)}_k}{s_k}.
\end{equation}
\begin{eqnarray}
M_{4}(3^{-}, {\bf 1}, {\bf 2}, 4^{+}) &=& 
\left( \frac{2}{M^2}\langle 3 {\bf 1} \rangle \bigl[ {\bf 1} | ({\bf 1} + 3) |{\bf 2} \rangle \bigl[ {\bf 2} 4 \bigr] 
\right)^2
\frac{i}{s}
+ \left( \frac{2}{M^2} \langle 3 {\bf 2} \rangle \bigl[ {\bf 2} | ({\bf 2} + 3) |{\bf 1} \rangle \bigl[ {\bf 1} 4 \bigr] \right)^2
\frac{i}{u}
\nonumber\\
&-&  
i \left[ \frac{1}{M^2}
 \left(  \langle {\bf 1} {\bf 2} \rangle \bigl[ {\bf 2} {\bf 1} \bigr] 
\langle 3 | ( {\bf 1} - {\bf 2} ) | 4 \bigr]
+  2 \Bigl( \langle 3 {\bf 1} \rangle \bigl[ {\bf 1} 4 \bigr] \langle {\bf 2} | (3-4) | {\bf 2} \bigr] 
- \langle 3 {\bf 2} \rangle \bigl[ {\bf 2} 4 \bigr] \langle {\bf 1} | (3-4) | {\bf 1} \bigr] \Bigr)
\right) \right]^2
\nonumber\\
&\times& \left( \frac{ 1 }{t} - \frac{ 2 }{t - M^2} \right) .
\end{eqnarray}
Non-gravitational states in the $t$-channel can be projected out by using the state projectors given in Eq.~(\ref{projectors}), after demanding consistent factorization in all possible channels with three-particle amplitudes.

Let us investigate the high-energy behavior. Assume that $\epsilon({\bf 1})$ is transverse and $\epsilon({\bf 2})$ is longitudinal. We find that
\begin{eqnarray}
\textrm{s-channel} &\approx& \bigl( 2  \ell \cdot \epsilon(1) \bigr)^2 \frac{i s}{M^2}
\nonumber\\
\textrm{u-channel} &\approx& \bigl( 2  \ell \cdot \epsilon(1) \bigr)^2 \frac{i u}{M^2}
\nonumber\\
\textrm{t-channel} &\approx&  \bigl( 2  \ell \cdot \epsilon(1) \bigr)^2 \frac{i t}{M^2}
\end{eqnarray}
where $\ell =  |3 \rangle \bigl[ 4|$. Hence at leading order
\begin{equation}
M_{4}(3^{-}, 1^{T}, 2^{L}, 4^{+}) \approx
\bigl( 2  \ell \cdot \epsilon(1) \bigr)^2 \frac{i}{M^2} (s+t+u).
\end{equation}
Likewise, for longitudinal polarizations:
\begin{eqnarray}
\textrm{s-channel} &\approx& \bigl( 2  \ell \cdot p_1 \bigr)^2 \frac{i s}{M^4}
\nonumber\\
\textrm{u-channel} &\approx& \bigl( 2  \ell \cdot p_1 \bigr)^2 \frac{i u}{M^4}
\nonumber\\
\textrm{t-channel} &\approx&  \bigl( 2  \ell \cdot p_1 \bigr)^2 \frac{i t}{M^4}
\end{eqnarray}
at leading order; therefore
\begin{equation}
M_{4}( 3^{-},1^{L}, 2^{L}, 4^{+}) \approx
\bigl( 2  \ell \cdot p_1 \bigr)^2 \frac{i}{M^4} (s+t+u).
\end{equation}
As in the scalar case, we see that the presence of the Merlin modes is crucial to obtain a nice high-energy behavior for the Compton amplitudes~\cite{Berends:1974gk}.

\subsubsection{Scattering of gravitational Merlin particles}

Finally, let us study the scattering of gravitational Merlin particles. Following the double-copy prescription, we have that
\begin{equation}
M({\bf 1},{\bf 2},{\bf 3},{\bf 4}) = i \sum_{k} \frac{n^{({\bf 1})}_k \tilde{n}^{({\bf 1})}_k }{s_k}.
\end{equation}
where the bold indicates the numerator associated with the scattering involving only Merlin particles in the associated gauge theory. We find that
\begin{equation}
M_{4}( {\bf 1}, {\bf 2} \to {\bf 3}, {\bf 4}) =
i \left( \frac{(n_s^{A})^2}{s} - \frac{4(n_s^{\widetilde{A}})^2}{s - M^2} \right)
+ i \left( \frac{(n_u^{A})^2}{u} - \frac{4(n_u^{\widetilde{A}})^2}{u - M^2} \right)
+ i \left( \frac{(n_t^{A})^2}{t} - \frac{4(n_t^{\widetilde{A}})^2}{t - M^2} \right)
\end{equation}
where the numerators $n_k^{A},n_k^{\widetilde{A}}$ can be read from Eqs.~(\ref{numm1}) and~(\ref{numm2}). As in the cases studied above, we can project out non-gravitational states by using consistent factorization in all physical channels and then employ the state projectors given in Eq.~(\ref{projectors}).

Let us now discuss the high-energy limit of this amplitude. For longitudinal polarizations, we find that
\begin{eqnarray}
\textrm{s-channel} &\approx& - \frac{3i}{16 M^8} \bigl( s (u-t) + u^2 - t^2 \bigr)^2 s
\nonumber\\
\textrm{u-channel} &\approx& - \frac{3i}{16 M^8} \bigl( u (s-t) + s^2 - t^2 \bigr)^2 u
\nonumber\\
\textrm{t-channel} &\approx& - \frac{3i}{16 M^8} \bigl( t (u-s) + u^2 - s^2 \bigr)^2 t
\end{eqnarray}
at leading order. Hence we find that
\begin{equation}
M_{4}( {\bf 1}^L, {\bf 2}^L, {\bf 3}^L, {\bf 4}^L) \approx - \frac{3i}{16 M^8}
(s+t+u)^2 \bigl[ s^2 (t+u)+s \left(t^2-6 t u+u^2\right)+t u (t+u) \bigr].
\end{equation}
For three longitudinal polarizations and one transversal, one gets
\begin{eqnarray}
\textrm{s-channel} &\approx& - \frac{3i}{4 M^6} \bigl(  (s+u) p_2 \cdot \epsilon(4) - (s+t) p_1 \cdot \epsilon(4)  \bigr)^2 s
\nonumber\\
\textrm{u-channel} &\approx&   - \frac{3i}{4 M^6} \bigl( (s+u) p_2 \cdot \epsilon(4) - (s+t) p_1 \cdot \epsilon(4)  \bigr)^2 u
\nonumber\\
\textrm{t-channel} &\approx& - \frac{3i}{4 M^6} \bigl( (s+u) p_2 \cdot \epsilon(4) - (s+t) p_1 \cdot \epsilon(4)  \bigr)^2 t
\end{eqnarray}
and therefore
\begin{equation}
M_{4}( {\bf 1}^L, {\bf 2}^L, {\bf 3}^L, {\bf 4}^T ) \approx
- \frac{3i}{4 M^6} \bigl( (s+u) p_2 \cdot \epsilon(4) - (s+t) p_1 \cdot \epsilon(4)  \bigr)^2 (s+t+u) .
\end{equation}
Finally, for two longitudinal polarizations and two transversals, one gets
\begin{eqnarray}
\textrm{s-channel} &\approx&  - \frac{3i}{4 M^4} 
\bigl( - \epsilon(3) \cdot \epsilon(4) ( u  - t ) 
- 2  p_2 \cdot \epsilon(4)  p_{1} \cdot \epsilon(3)
+ 2  p_{1} \cdot \epsilon(4) p_2 \cdot \epsilon(3) \bigr)^2 s
\nonumber\\
\textrm{u-channel} &\approx& - \frac{3i}{4 M^4} 
\bigl( - \epsilon(3) \cdot \epsilon(4) (u-t) - 2 p_{2} \cdot \epsilon(4) p_1 \cdot \epsilon(3) 
+ 2 p_1 \cdot \epsilon(4)p_2 \cdot \epsilon(3) \bigr)^2 u
\nonumber\\
\textrm{t-channel} &\approx&  - \frac{3i}{4 M^4} 
\bigl( - \epsilon(3) \cdot \epsilon(4) (u-t) - 2 p_2 \cdot \epsilon(4) p_1 \cdot \epsilon(3)
+ 2 p_1 \cdot \epsilon(4) p_{2} \cdot \epsilon(3)  \bigr)^2 t .
\end{eqnarray}
Hence
\begin{equation}
M_{4}( {\bf 1}^L, {\bf 2}^L, {\bf 3}^T, {\bf 4}^T) \approx  
- \frac{3i}{4 M^4} 
\bigl( - \epsilon(3) \cdot \epsilon(4) (u-t) - 2 p_2 \cdot \epsilon(4) p_1 \cdot \epsilon(3)
+ 2 p_1 \cdot \epsilon(4) p_{2} \cdot \epsilon(3)  \bigr)^2 (s+t+u) .
\end{equation}

We find that the amplitudes do not grow with energy in all the cases analyzed above. This is the gravitational version of the statements given in Ref.~\cite{Wise:2009mi}. For this to happen it is crucial to have a Merlin exchange accompanying the associated graviton exchange in physical channels as well as taking the Merlins to be massless in the high-energy limit. If the latter does not hold, then quadratic gravity would inherit all issues regarding high-energy behavior at tree level from general relativity. A way to circumvent such difficulties is to take into account inclusive processes, as described in Refs.~\cite{Holdom:2021hlo,Holdom:2021oii}.

\subsection{A one-loop example}

Let us now calculate the one-loop amplitude for the graviton scattering process 
$h^{++} h^{++} \to h^{++} h^{++}$. The case of only physical gravitons running in the loop has already been calculated~\cite{Dunbar:95,Dunbar:97}. For completeness let us reproduce the maximal-cut calculation. As well known, the tree-level graviton amplitude vanishes for this helicity configuration; in fact, as in the case of pure YM, one can perform a decomposition in terms of supergravity amplitudes and one finds that this amplitude is just proportional to the scalar contribution~\cite{Dunbar:95}. Hence one just have to consider a complex scalar running in the loop. To be able to reconstruct the rational term, we consider that the loop momentum lives in $4 - 2\epsilon$ dimensions, which means that we can consider the loop momentum to be massive, with mass $\mu^2$, in $4$ dimensions -- exactly the same step taken for the pure YM. Therefore, one finds that 
\begin{equation}
M^{1-\textrm{loop integrand}}_{4}(\ell) \bigg|_{\textrm{quadruple cut}} = 
M^{\textrm{tree}}_{3}(p^{++}_1, - \ell_{2}, \ell_1) 
M^{\textrm{tree}}_{3}(p^{++}_2, -\ell_{3}, \ell_2) 
M^{\textrm{tree}}_{3}(p^{++}_3, -\ell_{4}, \ell_3) 
M^{\textrm{tree}}_{3}(p^{++}_4, -\ell_{1}, \ell_4)
\end{equation}
with the same cut conditions as given above. The $3$-particle amplitudes can be obtained from double-copying the corresponding $3$-particle amplitudes in YM theory:
\begin{equation}
M^{\textrm{tree}}_{3}(1,2^{++},3) = i A^{\textrm{tree,YM}}_{3}[1,2^+,3] A^{\textrm{tree,YM}}_{3}[1,2^+,3]
= i \frac{\langle \xi | p_1 | 2 \bigr]^2}{\langle \xi 2 \rangle^2} .
\end{equation}
Hence
\begin{eqnarray}
M^{1-\textrm{loop integrand}}_{4}(\ell) \bigg|_{\textrm{quadruple cut}} &=&
\bigl\{ A^{\textrm{tree}}_{3}[p^{+}_1, - \ell_{2}, \ell_1] 
A^{\textrm{tree}}_{3}[p^{+}_2, -\ell_{3}, \ell_2] 
A^{\textrm{tree}}_{3}[p^{+}_3, -\ell_{4}, \ell_3] 
A^{\textrm{tree}}_{3}[p^{+}_4, -\ell_{1}, \ell_4] \bigr\}
\nonumber\\
&\times&\bigl\{ A^{\textrm{tree}}_{3}[p^{+}_1, - \ell_{2}, \ell_1] 
A^{\textrm{tree}}_{3}[p^{+}_2, -\ell_{3}, \ell_2] 
A^{\textrm{tree}}_{3}[p^{+}_3, -\ell_{4}, \ell_3] 
A^{\textrm{tree}}_{3}[p^{+}_4, -\ell_{1}, \ell_4] \bigr\} 
\nonumber\\
&=& \mu^8 \frac{\bigl[ 12 \bigr]^2 \bigl[ 34 \bigr]^2 }{ \langle 12 \rangle^2 \langle 34 \rangle^2 }
\end{eqnarray}
where we have chosen $\xi_1 = 2$, $\xi_2 = 1$, $\xi_3 = 4$ and $\xi_4 = 3$. When we evaluate the other cuts, we discover that there are other contributions given by the permutations $4 \leftrightarrow 3$ and 
$2 \leftrightarrow 3$. Hence the amplitude is given by
\begin{equation}
M^{1-\textrm{loop}}_{4}(1^{++}, 2^{++}, 3^{++}, 4^{++}) = \frac{2 i}{(4\pi)^{2-\epsilon}} 
\left[ \frac{ s_{12} s_{23}  }{ \langle 12 \rangle \langle 23 \rangle \langle 34 \rangle \langle 41 \rangle} \right]^2 \Bigl( I_{1234}[\mu^8] + I_{1243}[\mu^8] + I_{1324}[\mu^8] \Bigr)
\end{equation}
where $2$ comes from the fact that the scalar is complex, and, to order $\epsilon^0$~\cite{Dunbar:97}
\begin{equation}
I_{1234}[\mu^8] + I_{1243}[\mu^8] + I_{1324}[\mu^8] = \frac{s_{12}^2 + s_{13}^2 + s_{14}^2}{240} .
\end{equation}
Of course, in order to check that this is indeed the correct solution, one has to evaluate a complete set of cuts.

Now let us consider the contribution coming from the gravitational Merlin particle. As in the YM case, since the amplitudes involving a single external Merlin particle vanish, we just have to consider contributions in which only Merlin lines are cut. As above, in order to implement unitarity methods to unstable particles we resort to the narrow-width approximation. The quadruple-cut reads
\begin{eqnarray}
M^{1-\textrm{loop integrand}}_{4}(\ell) \bigg|_{\textrm{quadruple cut}} &=& 
M^{\textrm{tree}}_{3}(p^{++}_1, - \ell_{2KL}, \ell^{IJ}_1) 
M^{\textrm{tree}}_{3}(p^{++}_2, -\ell_{3MN}, \ell^{KL}_2) 
\nonumber\\
&\times& M^{\textrm{tree}}_{3}(p^{++}_3, -\ell_{4PQ}, \ell^{MN}_3) 
M^{\textrm{tree}}_{3}(p^{++}_4, -\ell_{1IJ}, \ell^{PQ}_4)
\nonumber\\
&=& \bigl\{ A^{\textrm{tree, HD}}_{3}[p^{+}_1, - \ell_{2KL}, \ell^{IJ}_1] 
A^{\textrm{tree, HD}}_{3}[p^{+}_2, -\ell_{3MN}, \ell^{KL}_2] 
\nonumber\\
&\times& A^{\textrm{tree, HD}}_{3}[p^{+}_3, -\ell_{4PQ}, \ell^{MN}_3] 
A^{\textrm{tree, HD}}_{3}[p^{+}_4, -\ell_{1IJ}, \ell^{PQ}_4] \bigr\}
\nonumber\\
&\times&
\bigl\{ A^{\textrm{tree, YM}}_{3}[p^{+}_1, - \ell_{2KL}, \ell^{IJ}_1] 
A^{\textrm{tree, YM}}_{3}[p^{+}_2, -\ell_{3MN}, \ell^{KL}_2] 
\nonumber\\
&\times& A^{\textrm{tree, YM}}_{3}[p^{+}_3, -\ell_{4PQ}, \ell^{MN}_3] 
A^{\textrm{tree, YM}}_{3}[p^{+}_4, -\ell_{1IJ}, \ell^{PQ}_4] \bigr\}
\nonumber\\
&=& 256 M^{8} \left[ \frac{ s_{12} s_{23} }
{\langle 1 2 \rangle \langle 23 \rangle \langle 34 \rangle \langle 41 \rangle} \right]^2 .
\end{eqnarray}
The triple cut reads
\begin{eqnarray}
M^{1-\textrm{loop integrand}}_{4}(\ell) \bigg|_{\textrm{triple cut}} &=& 
M^{\textrm{tree}}_{3}(p^{++}_1, - \ell_{2KL}, \ell^{IJ}_1) 
M^{\textrm{tree}}_{3}(p^{++}_2, -\ell_{3MN}, \ell^{KL}_2) 
M^{\textrm{tree}}_{4}(p^{++}_4, -\ell_{1IJ}, \ell^{MN}_3, p^{++}_3)
\nonumber\\
&=& i s_{34}
\bigl\{ A^{\textrm{tree, HD}}_{3}[p^{+}_1, - \ell_{2KL}, \ell^{IJ}_1] 
A^{\textrm{tree, HD}}_{3}[p^{+}_2, -\ell_{3MN}, \ell^{KL}_2] 
A^{\textrm{tree, HD}}_{4}[p^{+}_4, -\ell_{1IJ}, \ell^{MN}_3, p^{+}_3] \bigr\}
\nonumber\\
&\times& 
\bigl\{ A^{\textrm{tree, YM}}_{3}[p^{+}_1, - \ell_{2KL}, \ell^{IJ}_1] 
A^{\textrm{tree, YM}}_{3}[p^{+}_2, -\ell_{3MN}, \ell^{KL}_2] 
A^{\textrm{tree, YM}}_{4}[p^{+}_4, \ell^{MN}_3, -\ell_{1IJ},  p^{+}_3] \bigr\} .
\nonumber\\
\end{eqnarray}
The other possible three-particle cut diagrams are obtained from this one by cyclic relabeling of the external particles. The cut conditions are given by
\begin{eqnarray}
\ell_1^2 &=& \ell^2 = M^2 
\nonumber\\
\ell_2^2 &=& (\ell + p_{1})^2 = M^2
\nonumber\\ 
\ell_3^2 &=& (\ell + p_1 + p_2)^2 = M^2 . 
\end{eqnarray}
Using previous results, one obtains
\begin{eqnarray}
M^{1-\textrm{loop integrand}}_{4}(\ell) \bigg|_{\textrm{triple cut}} &=& 
- 256 i s_{34} M^{8} \left[ \frac{ s_{12} s_{23} }{\langle 1 2 \rangle \langle 23 \rangle \langle 34 \rangle \langle 41 \rangle} \right]^2
\frac{1}{(\ell - p_4)^2 - M^2} \frac{1}{(\ell_3 + p_4)^2 - M^2}
\nonumber\\
&=& 256 i M^{8}
\left[ \frac{ s_{12} s_{23} }{\langle 1 2 \rangle \langle 23 \rangle \langle 34 \rangle \langle 41 \rangle} \right]^2
 \left[ \frac{1}{(\ell - p_4)^2 - M^2} + \frac{1}{(\ell - p_3)^2 - M^2} \right] .
\end{eqnarray}
An overall factor of $i$ comes from the $(-i)^3$ in the cut propagators. Finally, let us evaluate the two-particle cut. We have that
\begin{eqnarray}
M^{1-\textrm{loop integrand}}_{4}(\ell) \bigg|_{\textrm{double cut}} &=& 
M^{\textrm{tree}}_{4}(p^{++}_2, -\ell_{3MN}, \ell^{IJ}_1,p^{++}_1) 
M^{\textrm{tree}}_{4}(p^{++}_4, -\ell_{1IJ}, \ell^{MN}_3, p^{++}_3)
\nonumber\\
&=& - s_{12}^2
\bigl\{A^{\textrm{tree, HD}}_{3}[p^{+}_2, -\ell_{3MN}, \ell^{IJ}_1,p^{+}_1] 
A^{\textrm{tree, HD}}_{4}[p^{+}_4, -\ell_{1IJ}, \ell^{MN}_3, p^{+}_3] \bigr\}
\nonumber\\
&\times& \bigl\{ A^{\textrm{tree, YM}}_{3}[p^{+}_2, \ell^{IJ}_1,-\ell_{3MN}, p^{+}_1] 
A^{\textrm{tree, YM}}_{4}[p^{+}_4, \ell^{MN}_3, -\ell_{1IJ}, p^{+}_3] \bigr\} .
\end{eqnarray}
As above, the other possible two-particle cut diagrams are obtained from this one by cyclic relabeling of the external particles. The cut conditions are given by
\begin{eqnarray}
\ell_1^2 &=& \ell^2 = M^2 
\nonumber\\
\ell_3^2 &=& (\ell + p_{1} + p_{2})^2 = M^2 . 
\end{eqnarray}
One finds that
\begin{eqnarray}
M^{1-\textrm{loop integrand}}_{4}(\ell) \bigg|_{\textrm{double cut}} &=& - 256 s_{12}^2 M^{8}
\left[ \frac{ s_{12} s_{23} }{\langle 1 2 \rangle \langle 23 \rangle \langle 34 \rangle \langle 41 \rangle} \right]^2
\nonumber\\
&\times& \frac{1}{(\ell+p_2)^2 - M^2} \frac{1}{(\ell_3 - p_2)^2 - M^2}
\frac{1}{(\ell - p_4)^2 - M^2} \frac{1}{(\ell_3 + p_4)^2 - M^2}
\nonumber\\
&=& - 256 M^{8}
\left[ \frac{ s_{12} s_{23} }{\langle 1 2 \rangle \langle 23 \rangle \langle 34 \rangle \langle 41 \rangle} \right]^2
\nonumber\\
&\times& \left[ \frac{1}{(\ell+p_2)^2 - M^2} + \frac{1}{(\ell + p_1)^2 - M^2} \right]
\left[ \frac{1}{(\ell - p_4)^2 - M^2} + \frac{1}{(\ell - p_3)^2 - M^2} \right]
\nonumber\\
&=& - 256 M^{8}
\left[ \frac{ s_{12} s_{23} }{\langle 1 2 \rangle \langle 23 \rangle \langle 34 \rangle \langle 41 \rangle} \right]^2
\nonumber\\
&\times& \left[ \frac{1}{(\ell + p_1)^2 - M^2}  \frac{1}{(\ell - p_4)^2 - M^2} 
+ 3 \leftrightarrow 4 + 1 \leftrightarrow 2 + \{1 \leftrightarrow 2, 3 \leftrightarrow 4 \} \right] .
\end{eqnarray}
An overall minus sign comes from the $(-i)^2$ in the cut propagators.

By lifting all the cuts and evaluating the integrals we find that $1 \leftrightarrow 2$ gives the same result as 
$3 \leftrightarrow 4$. Moreover, by cyclic relabeling of the external particles, one also finds that a term with 
$3 \leftrightarrow 2$ also contributes.

In order to reconstruct the amplitude, we have to consider the rational terms. As in the YM case, this is obtained by working in $d = 4-2\epsilon$ dimensions -- in addition, if one wants to consider only gravitational Merlin modes crossing the cut, one must use a physical state projector in $D$ dimensions. By using in the double-copy construction gluon amplitudes that automatically preserves Ward identities, the Merlin physical-state projector simplifies to
\begin{equation}
P_{\mu\nu\rho\sigma} = \sum_{\textrm{pols.}} \epsilon_{\mu\nu}(-p) \epsilon_{\rho\sigma}(p) 
=  \frac{1}{2} \Bigl( \eta_{\mu\rho} \eta_{\nu\sigma} + \eta_{\mu\sigma} \eta_{\nu\rho} \Bigr)
- \frac{1}{D-1} \eta_{\mu\nu} \eta_{\rho\sigma} .
\end{equation}
The sum runs over all physical states of the gravitational Merlin. When $D \to d$, we find the same result as the four-dimensional case, except that one must shift the mass as $M^2 \to M^2 + \mu^2$ (there is, however, a difference between the four-dimensional integrand and the $d$-dimensional integrand due to the last term in the state projector~\cite{Bern:2019crd}; we ignore this difference here). The final result is given by
\begin{eqnarray}
M^{1-\textrm{loop}}_{4}(1^{++}, 2^{++}, 3^{++}, 4^{++}) &=& \frac{2 i}{(4\pi)^{2-\epsilon}} 
\left[ \frac{ s_{12} s_{23}  }{ \langle 12 \rangle \langle 23 \rangle \langle 34 \rangle \langle 41 \rangle} \right]^2 \Bigl[ I_{1234}[\mu^8] + I_{1243}[\mu^8] + I_{1324}[\mu^8] 
\nonumber\\
&+& 128 \Bigl( {\cal I}_{1234}[(M^2 + \mu^2)^4] + {\cal I}_{1243}[(M^2 + \mu^2)^4] 
+ {\cal I}_{1324}[(M^2 + \mu^2)^4]  \Bigr)
\Bigr]
\end{eqnarray}
where 
\begin{equation}
{\cal I}_{1234}[(M^2 + \mu^2)^4] = M^8 {\cal I}_{1234}[1] + 4 M^6 {\cal I}_{1234}[\mu^2]
+ 6 M^4 {\cal I}_{1234}[\mu^4] + 4 M^2 {\cal I}_{1234}[\mu^6] + {\cal I}_{1234}[\mu^8]
\end{equation}
and hence, to order $\epsilon^0$~\cite{Dunbar:97}
\begin{equation}
{\cal I}_{1234}[(M^2 + \mu^2)^4] + {\cal I}_{1243}[(M^2 + \mu^2)^4] 
+ {\cal I}_{1324}[(M^2 + \mu^2)^4] =
\frac{s_{12}^2 + s_{13}^2 + s_{14}^2}{240} + \frac{1}{2} M^4
+ M^8 ( {\cal I}_{1234}[1] + {\cal I}_{1243}[1] + {\cal I}_{1324}[1] ) .
\end{equation}
A large-mass expansion for the Merlin contribution to the amplitude reveals that this part tends to zero as 
$M \to \infty$~\cite{Dunbar:97}. Moreover, the double-copy structure of this amplitude is clearly manifest.

\section{Summary}

In this paper we have shown some progress concerning the study of higher-derivative Yang-Mills and quadratic-gravity amplitudes by explicitly including the contribution from the ghost resonance. As we have shown above, Compton amplitudes in quadratic gravity seem to have a healthy high-energy behavior. Furthermore, we have explicitly demonstrated that tree Yang-Mills amplitudes satisfy color-kinematics duality, even though the standard BCJ relations for color-ordered amplitudes are not obeyed for the scattering encompassing matter particles and Merlin modes. This type of behavior is known to exist in other interesting situations -- for instance, in Ref.~\cite{Johansson:15}, the authors have shown that there are no BCJ relations for color-kinematics satisfying all-quark massive amplitudes (we believe this was the first work where this feature was clearly emphasized). In turn, it was also observed in Ref.~\cite{Johnson:20} the absence of standard BCJ relations for massive gluons, even though the simple model put forward by the authors does not work well beyond four points.

In few words, we are exposing another nontrivial physical situation in which color-kinematics duality can hold even when standard BCJ relations do not exist. In our case, this is a consequence of the presence of quartic propagators -- in this case, the associated numerators in the amplitudes are gauge invariant and color-ordered amplitudes will not be related in the conventional BCJ sense. Nevertheless, as argued above BCJ relations are known to be valid for $n$-point amplitudes comprising at least $n-2$ external massless adjoint particles, which was not the case for some of the Compton amplitudes analyzed in this work. Hence in principle one could have expected the violation of such amplitude relations. Indeed, in general BCJ relations can be envisaged as arising from simultaneously keeping color-kinematics duality and gauge invariance of massless particles. One straightforward way to see this is to consider Einstein-Yang-Mills amplitudes with one graviton and $n-1$ gluons, as explained in some detail in Ref.~\cite{Chiodaroli:2017ngp}. In any case, the double-copy construction is still valid and we have managed to obtain associated quadratic-gravity amplitudes from the knowledge of higher-derivative QCD-like amplitudes, even though the standard KLT relations are not valid here when Merlin particles are present as external states. Finally, as remarked, for the double-copy prescription to work correctly, the normal Yang-Mills part of the map should be associated with a spontaneously broken gauge theory

We have also calculated specific one-loop amplitudes to illustrate how the generalized unitarity method can be applied when ghost resonances circulate in the loops. Here we have presented, for the first time, how such a framework can still be used for higher-derivative Yang-Mills theory and quadratic gravity. Unitarity methods and double-copy construction are powerful techniques that alleviates the rather laborious task of computing one-loop quadratic-gravity amplitudes. 

Despite the various aspects studied here, there are still facets of the present construction which are not known. For instance, at five or higher points, the inverse of the propagator matrix contains spurious poles, which usually cancel out in the double-copy prescription in virtue of the BCJ relations. It would be interesting to know whether this cancellation can also take place when one has only color-kinematics duality (assuming its validity to higher-point amplitudes). This would imply that locality and gauge invariance for massless particles in BCJ numerators would suffice to ensure the absence (or cancellation) of spurious poles. Such unphysical poles also appear in the setup put forward in Ref.~\cite{Brandhuber:2021bsf} which nevertheless can be eliminated -- this might also be the case here, which would point to an interesting connection to be unraveled between the two constructions. Otherwise, this might be a strong evidence that additional fields or interactions should be included -- in particular, the dimension-six gauge theory considered in Refs.~\cite{Johansson:18,Johansson:17} could emerge as an acceptable required representation (possibly unique). Be that as it may, we believe this introductory survey will be useful in the investigations of quadratic-gravity amplitudes -- such quantities are known to be non-analytic close to the Merlin resonance, which produces violations of causality at such an energy scale. Indeed, standard analyticity axioms of S-matrix theory are not satisfied by such amplitudes, and as such causality arises as an emergent macroscopic phenomenon in this theories~\cite{Grinstein:08,Donoghue:2019ecz,Donoghue:2021meq}. In this regard, it would be interesting to explore Shapiro's time delay in quadratic gravity, which would could allow us to understand more clearly causality violations under modern amplitude methods trends. In turn, even though there are strong evidences in favor of the stability of quadratic gravity~\cite{DM:19,Donoghue:2021eto,Salvio:2019ewf}, this is an issue that should be addressed by modern amplitude techniques. On the other hand, an interesting study would be to exploit the calculation of loop amplitudes from the perspective of leading singularities~\cite{Cachazo:2008vp}. Indeed, we know that information about the analytic structure of loop amplitude of Yang-Mills theory (and of gravity) can be captured by leading singularities and this could be potentially useful in the evaluation of loop MHV amplitudes for higher-derivative versions of such theories. Another interesting direction is to use on-shell methods to examine the high-energy behavior of gravitational amplitudes calculated within quantum quadratic gravity, a formalism where one cannot ignore the ghost resonance mass~\cite{Holdom:2021hlo,Holdom:2021oii}. These exciting prospects are key steps in the important program of a quantum-field-theory UV completion of quantum gravity and we hope to explore these calculations in the near future.

\section*{Acknowledgements} We thank John F. Donoghue, Bob Holdom, Thales Azevedo and Henrik Johansson for useful discussions, suggestions and for providing comments on the draft. This work has been partially supported by Conselho Nacional de Desenvolvimento Cient\'ifico e Tecnol\'ogico - CNPq under grant 310291/2018-6 and Funda\c{c}\~ao Carlos Chagas Filho de Amparo \`a Pesquisa do Estado do Rio de Janeiro - FAPERJ under grant E-26/202.725/2018.

\section*{Appendix A -- Quick review of massive spinors}

Let us briefly discuss the formalism for massive spinors adopted in this paper~\cite{Aoude:19,Shadmi:19,Durieux:20,Arkani-Hamed:17,Chung:19}. Since $\det p^{\alpha \dot{\alpha}} = m^2$ in the massive case, hence $p^{\alpha \dot{\alpha}}$ has rank $2$. As a consequence, it can be written as the sum of two rank-one matrices:
\begin{equation}
p^{\alpha \dot{\alpha}} = \lambda^{\alpha\, I} \tilde{\lambda}^{\dot{\alpha}}_{I} .
\end{equation}
The index $I=1,2$ indicates a doublet of the $SU(2)$ little group. We take $\det\lambda = \det\tilde{\lambda} = m$. Little group indices are raised and lowered by the $SU(2)$-invariant tensor  
$\varepsilon^{IJ}, \varepsilon_{IJ}$. One finds
\begin{eqnarray}
p^{\alpha \dot{\alpha}} &=& \lambda^{\alpha\, I} \tilde{\lambda}^{\dot{\alpha}}_{I} 
= - \lambda^{\alpha}_{I} \tilde{\lambda}^{\dot{\alpha}\,I}
= | p^{I} \rangle \bigl[p_{I} |
\nonumber\\
p_{\dot{\alpha} \alpha} &=& - \tilde{\lambda}_{\dot{\alpha}}^{I} \lambda_{\alpha\, I} 
= \tilde{\lambda}_{\dot{\alpha} \, I}  \lambda_{\alpha}^{I} 
= - | p^{I} \bigr]  \langle p_{I} |.
\end{eqnarray}
By definition, the massive spinor helicity variables obey
\begin{equation}
p^{\alpha \dot{\alpha}} \tilde{\lambda}_{\dot{\alpha}}^{I} = m \lambda^{\alpha\, I},
\,\,\,
p_{\dot{\alpha} \alpha} \lambda^{\alpha\,I}  = m \tilde{\lambda}_{\dot{\alpha}}^I .
\end{equation}
Under little group scaling, we have the following $SL(2)$ transformation:
\begin{equation}
\lambda^{I} \to W^{I}_{J} \lambda^{J}
\,\,\,
\tilde{\lambda}_{I} \to (W^{-1})_{I}^{J} \tilde{\lambda}_{J} .
\end{equation}
Massive spinor bilinears satisfy
\begin{eqnarray}
\langle \lambda^{I} \lambda_{J} \rangle &=& m \delta^{I}_{J},
\,\,\,
\langle \lambda^{I} \lambda^{J} \rangle = - m \varepsilon^{IJ},
\,\,\,
\langle \lambda_{I} \lambda_{J} \rangle = m \varepsilon_{IJ}
\nonumber\\
\bigl[ \tilde{\lambda}^{I} \tilde{\lambda}_{J} \bigr] &=& -m \delta^{I}_{J},
\,\,\,
\bigl[ \tilde{\lambda}^{I} \tilde{\lambda}^{J} \bigr] = m \varepsilon^{IJ},
\,\,\,
\bigl[ \tilde{\lambda}_{I} \tilde{\lambda}_{J} \bigr] = - m \varepsilon_{IJ} 
\nonumber\\
\lambda^{I\,\alpha} \lambda_{I\, \beta} &=&  |\lambda^{I} \rangle \langle \lambda_{I} |
= -m \delta^{\alpha}_{\beta},
\,\,\,
\tilde{\lambda}^{I}_{\dot{\alpha}} \tilde{\lambda}_{I}^{\dot{\beta}} =  
|\lambda^{I} \bigr] \bigl[ \lambda_{I} |
= m \delta_{\dot{\alpha}}^{\dot{\beta}} .
\label{relations}
\end{eqnarray}
Often we work with a bold notation to implicitly indicate symmetric combinations of the $SU(2)$ little-group indices of massive spinors. One has that
\begin{eqnarray}
p^{\alpha \dot{\alpha}} &=& | {\bf p} \rangle \bigl[ {\bf p} |
\nonumber\\
p_{\dot{\alpha} \alpha} &=& - | {\bf p} \bigr]  \langle {\bf p} |
\end{eqnarray}
and the Dirac equation reads
\begin{eqnarray}
p^{\alpha \dot{\alpha}}  | {\bf p} \bigr] &=& m | {\bf p} \rangle
\nonumber\\
p_{\dot{\alpha} \alpha} | {\bf p} \rangle &=& m | {\bf p} \bigr] 
\nonumber\\
\bigl[ {\bf p} | p_{\dot{\alpha} \alpha} &=& - m \langle {\bf p} |
\nonumber\\
\langle {\bf p}|  p^{\alpha \dot{\alpha}} &=& -m \bigl[ {\bf p} | .
\end{eqnarray}
The polarization vector associated with a massive vector boson of momentum $p$ and mass $m$ reads
\begin{equation}
\epsilon^{IJ}_{\mu} = \frac{1}{\sqrt{2}} \, \frac{ \langle {\bf p} | \sigma_{\mu} | {\bf p} \bigr]  }{m}
\end{equation}
or
\begin{equation}
\bigl[ \epsilon^{IJ} \bigr]^{\alpha \dot{\alpha}} = \sqrt{2} \, \frac{ | {\bf p} \rangle \bigl[ {\bf p} |}{m} .
\end{equation}
An implicit symmetrization on $SU(2)$ indices is to be regarded here. Such polarizations are associated with two transverse $(+, -)$ and one longitudinal ($0$) modes:
\begin{eqnarray}
\epsilon^{+}_{\mu} &=& \epsilon^{11}_{\mu} 
= \frac{1}{\sqrt{2}} \, \frac{ \langle p^1 | \sigma_{\mu} | p^1 \bigr]  }{m}
\nonumber\\
\epsilon^{-}_{\mu} &=& \epsilon^{22}_{\mu} 
= \frac{1}{\sqrt{2}} \, \frac{ \langle p^2 | \sigma_{\mu} | p^2 \bigr]  }{m}
\nonumber\\
\epsilon^{0}_{\mu} &=& \epsilon^{12}_{\mu} = \epsilon^{21}_{\mu}
= \frac{1}{2} \, \frac{ \langle p^1 | \sigma_{\mu} | p^2 \bigr] 
+ \langle p^2 | \sigma_{\mu} | p^1 \bigr]  }{m} .
\end{eqnarray}
Massive polarization vectors satisfy the ordinary normalization for the vector boson.

\section*{Appendix B -- Proof of unitarity for higher-derivative Yang-Mills theories and quadratic gravity}

Here we discuss unitarity. Following the previous work~\cite{DM:19}, we shall employ a generalization of Veltman's argument~\cite{Veltman:63} in order to discuss unitarity for LW theories. For such purposes, one typically derives the so-called cutting equation which implies unitarity. The cutting equation contains terms associated with the imaginary part of propagators, given by the cut propagators. Essentially, the cutting equation is to be identified with the generalized optical theorem associated with the process $a \to b$:
\begin{equation}
i {\cal M}(a \to b) - i {\cal M}^{*}(b \to a)  =  - \sum_{f} \int d\Pi_{f} {\cal M}^{*}(b \to f) {\cal M}(a \to f)
(2\pi)^{4} \delta^{4}(a-f)
\label{opt_theo}
\end{equation}
where $d\Pi_{f}$ is the Lorentz-invariant phase space~\cite{Schwartz:13} and the sum runs over all possible sets $f$ of intermediate states (single- and multi-particle states) and there is an overall delta function assuring energy-momentum conservation. 

Let us begin with scalar Lee-Wick models. We assume that a non-perturbative expression for the propagator in terms of a Lehmann representation is accessible:
\begin{equation}
G(p^2) = \frac{1}{\pi} \int_{0}^{\infty} ds \frac{\sigma(s)}{p^2 - s + i\epsilon}. 
\label{lehmann}
\end{equation}
for stable particles, and
\begin{equation}
\widetilde{G}(p^2) = \frac{1}{\pi} \int_{4 m^2}^{\infty} ds \frac{\widetilde{\rho}(s)}{p^2 - s + i\epsilon}. 
\label{lehmann2}
\end{equation}
for unstable particles. For stable particles, we find an isolated delta function contribution:
\begin{equation}
\sigma(s) = Z \pi \, \delta(s-m^2) + \theta(s-4 m^2) \rho(s)
\label{ss}
\end{equation}
where $m$ is the physical mass of the stable particle. On the other hand, for unstable particles there cannot be a isolated delta function contribution to the spectral function since there are no one-particle states. For the Lee-Wick field, we take
\begin{equation}
\widetilde{G}_{\textrm{LW}}(p^2) = - \frac{1}{\pi} \int_{4 m^2}^{\infty} ds 
\frac{\widetilde{\rho}(s)}{p^2 - s - i\epsilon}. 
\label{lehmannLW}
\end{equation}
For a related discussion on a modified Lehmann representation, see also Ref.~\cite{DM:19a}. A key component in our discussion is the decomposition of the propagator into positive- and negative-frequency parts. For the stable particle, one gets
\begin{equation}
i G(x-x') = \theta(x_0-x_0') G^{+}(x-x') + \theta(-x_0+x_0') G^{-}(x-x') .
\end{equation}
Using the spectral representation
\begin{equation}
G^{\pm}(x-x') =  \int \frac{d^{4} p}{(2\pi)^{3}} e^{-i p \cdot (x-x')} \theta(\pm p_0) 
\frac{\sigma(p^2)}{\pi}
\end{equation}
The Green's functions $G^{\pm}$ are known as cut propagators. Using that
\begin{equation}
i G^{\pm}(x-x';m^2) =  \pm \, 2\pi \int \frac{d^{4} p}{(2\pi)^{4}} e^{-i p \cdot (x-x')} 
\theta(\pm p_0) \delta(p^2-m^2)
 \end{equation}
one can also write
\begin{equation}
G^{\pm}(x-x') = \pm \frac{1}{\pi} \int_{0}^{\infty} ds \, \sigma(s) \, i G^{\pm}(x-x';s).
\end{equation}
In momentum space
\begin{equation}
G^{\pm}(p^2) = 2\pi \theta(\pm p_0) \int_{0}^{\infty} ds \, \delta(p^2 - s) \frac{\sigma(s)}{\pi}.  
\end{equation}
For the normal unstable particle, one finds a similar representation in terms of cut propagators:
\begin{equation}
i \widetilde{G}(x-x') = \theta(x_0-x_0') \widetilde{G}^{+}(x-x') + \theta(-x_0+x_0') \widetilde{G}^{-}(x-x')
\end{equation}
where
\begin{equation}
\widetilde{G}^{\pm}(x-x') = \pm \frac{1}{\pi} \int_{4m^2}^{\infty} ds \, \widetilde{\rho}(s) \, i G^{\pm}(x-x';s).
\end{equation}
In momentum space:
\begin{equation}
\widetilde{G}^{\pm}(p^2) = 2\pi \theta(\pm p_0) \int_{4m^2}^{\infty} ds \, \delta(p^2 - s) 
\frac{\widetilde{\rho}(s)}{\pi}.  
\end{equation}
The Lee-Wick case is more subtle. Following the above framework, we write
\begin{equation}
i \widetilde{G}_{\textrm{LW}}(x-x') = \theta(x_0-x_0') \widetilde{G}_{\textrm{LW}}^{-}(x-x') 
+ \theta(-x_0+x_0') \widetilde{G}_{\textrm{LW}}^{+}(x-x')
\end{equation}
where
\begin{equation}
\widetilde{G}_{\textrm{LW}}^{\pm}(x-x') = \mp \frac{1}{\pi} \int_{4m^2}^{\infty} ds \, 
\widetilde{\rho}(s) \, i G^{\pm}(x-x';s).
\end{equation}
In momentum space:
\begin{equation}
\widetilde{G}_{\textrm{LW}}^{\pm}(p^2) = - 2\pi \theta(\pm p_0) \int_{4m^2}^{\infty} ds \, \delta(p^2 - s) \frac{\widetilde{\rho}(s)}{\pi}.  
\end{equation}
For all cases above:
\begin{eqnarray}
-i G^{*}(x-x') &=& \theta(x_0-x_0') G^{-}(x-x') + \theta(-x_0+x_0') G^{+}(x-x')
\nonumber\\
-i \widetilde{G}^{*}(x-x') &=& \theta(x_0-x_0') \widetilde{G}^{-}(x-x') + \theta(-x_0+x_0') \widetilde{G}^{+}(x-x')
\nonumber\\
-i \widetilde{G}_{\textrm{LW}}^{*}(x-x') &=& \theta(x_0-x_0') \widetilde{G}_{\textrm{LW}}^{+}(x-x') 
+ \theta(-x_0+x_0') \widetilde{G}_{\textrm{LW}}^{-}(x-x').
\end{eqnarray}
We have already demonstrated the unitarity of scalar Lee-Wick theories by using all such propagators just discussed~\cite{DM:19}. We now wish to discuss the unitarity of higher-derivative quantum chromodynamics and also quadratic gravity. We know that, for the optical theorem to hold in general, the numerator of a propagator must be equal to the sum over physical spin states~\cite{Schwartz:13}. So this sum must also be present in the numerator of the cut propagators. Let us discuss now the case of higher-derivative Yang-Mills. By adding the following gauge-fixing term
$$
- \frac{1}{2} (\partial_{\mu} A^{\mu a}) (\Box + M^2) (\partial_{\nu} A^{\nu a})
$$
to the Lagrangian density, we find the simple propagator in momentum space:
\begin{equation}
i G^{ab}_{\mu\nu}(p) = - i \eta_{\mu\nu} \delta^{ab} \left[ \frac{1}{p^2} - \frac{1}{p^2 - M^2} \right] .
\end{equation}
A similar form was also derived in Ref.~\cite{Johansson:18}. With this we can construct the total gauge-theory propagator in the coordinate representation (in the narrow-width approximation):
\begin{eqnarray}
iG^{ab}_{\mu\nu}(x-x') &=& \eta_{\mu\nu} \delta^{ab}
\int \frac{d^{4} q}{(2\pi)^4} e^{-i q \cdot (x-x')} (-i G(q) )
+ \eta_{\mu\nu} \delta^{ab}
\int \frac{d^{4} q}{(2\pi)^4} e^{-i q \cdot (x-x')} ( -i \widetilde{G}(q) )
\nonumber\\
G(q) &=& \frac{1}{q^2 + i\epsilon},
\,\,\,\,
\widetilde{G}(q) = - \frac{1}{q^2 - M^2 - i M \Gamma} .
\end{eqnarray}
Hence the decomposition into positive- and negative-frequency parts reads
\begin{eqnarray}
i G^{ab}_{\mu\nu}(x-x') &=& \eta_{\mu\nu} \delta^{ab} 
\left[ \theta(x_0-x_0') G^{+}(x-x') + \theta(-x_0+x_0') G^{-}(x-x') \right]
\nonumber\\
&+&  \eta_{\mu\nu} \delta^{ab}
\left[ \theta(x_0-x_0') \widetilde{G}^{-}(x-x') + \theta(-x_0+x_0') \widetilde{G}^{+}(x-x') \right]
\nonumber\\
\end{eqnarray}
where $G^{\pm}(x-x')$ ($\widetilde{G}^{\pm}(x-x')$) can be written in the form discussed previously for the scalar stable (Lee-Wick) particle. In this way all the results discussed in Ref.~\cite{DM:19} now hold for the case of higher-derivative YM.

Now let us consider quadratic gravity.  All the above arguments and expressions can be applied to the case of quadratic gravity, with some modifications.  As mentioned above, in order for the optical theorem to hold the numerator of the propagator must be equal to the sum over physical spin states. Let us therefore consider the polarization sum for the graviton field. Spin-two polarization tensors $\epsilon_{\mu\nu}$ can be constructed from tensor products of the spin-one polarization vectors $\epsilon_{\mu}$. Hence the gravitational polarization sum can also be obtained from the spin-one polarization sum. In this latter case one finds~\cite{Schwartz:13}
\begin{equation}
\sum_{j_{z} = \pm 1} \epsilon_{\mu}(p,j_{z}) \epsilon^{*}_{\nu}(p,j_{z})
= - \eta_{\mu\nu} + \frac{p_{\mu} \bar{p}_{\nu} + p_{\nu} \bar{p}_{\mu} }{p \cdot \bar{p}} = - \pi_{\mu\nu}
\end{equation}
where $\bar{p}^{\mu} = (p^{0},-{\bf p})$. For the gravitational case, one can show that~\cite{Kachelriess:18}
\begin{equation}
\sum_{j_{z} = \pm 2} \epsilon_{\mu\nu}(p,j_{z}) \epsilon^{*}_{\rho\sigma}(p,j_{z}) = 
\frac{1}{2} \Bigl( \pi_{\mu\rho} \pi_{\nu\sigma} + \pi_{\mu\sigma} \pi_{\nu\rho} 
- \pi_{\mu\nu} \pi_{\rho\sigma} \Bigr) .
\end{equation}
However, for quadratic gravity we also have a massive spin-$2$ mode, the Merlin mode. For this one we have to use a massive spin sum. For a massive spin-$1$ field, the spin sum reads~\cite{Schwartz:13}
\begin{equation}
\sum_{j_{z} = 0, \pm 1} \epsilon_{\mu}(p,j_{z}) \epsilon^{*}_{\nu}(p,j_{z})
= - \eta_{\mu\nu} + \frac{p_{\mu} p_{\nu} }{M^2} = - \widetilde{\pi}_{\mu\nu} .
\end{equation}
We find the following spin sum for the massive Merlin mode:
\begin{equation}
\sum_{j_{z} = 0, \pm1, \pm 2} \epsilon_{\mu\nu}(p,j_{z}) \epsilon^{*}_{\rho\sigma}(p,j_{z}) = 
\frac{1}{2} \Bigl( \widetilde{\pi}_{\mu\rho} \widetilde{\pi}_{\nu\sigma} 
+ \widetilde{\pi}_{\mu\sigma} \widetilde{\pi}_{\nu\rho} 
- \frac{2}{3} \widetilde{\pi}_{\mu\nu} \widetilde{\pi}_{\rho\sigma} \Bigr) .
\end{equation}
Then the total gravitational propagator would have the form
\begin{equation}
D_{\mu\nu\rho\sigma}(p) = 
\frac{1}{2 p^2} \Bigl( \pi_{\mu\rho} \pi_{\nu\sigma} + \pi_{\mu\sigma} \pi_{\nu\rho} 
- \pi_{\mu\nu} \pi_{\rho\sigma} \Bigr) 
- \frac{1}{2 (p^2 - M^2)} \Bigl( \widetilde{\pi}_{\mu\rho} \widetilde{\pi}_{\nu\sigma} 
+ \widetilde{\pi}_{\mu\sigma} \widetilde{\pi}_{\nu\rho} 
- \frac{2}{3} \widetilde{\pi}_{\mu\nu} \widetilde{\pi}_{\rho\sigma} \Bigr) .
\end{equation}
Due to the Slavnov-Taylor identities, all the $p^{\mu}$ terms drop out in physical calculations. Hence we can write
\begin{equation}
\sum_{j_{z} = \pm 2} \epsilon_{\mu\nu}(p,j_{z}) \epsilon^{*}_{\rho\sigma}(p,j_{z}) \bigg|_{\textrm{physical}} = 
\frac{1}{2} \Bigl( \eta_{\mu\rho} \eta_{\nu\sigma} + \eta_{\mu\sigma} \eta_{\nu\rho} 
- \eta_{\mu\nu} \eta_{\rho\sigma} \Bigr) 
\end{equation}
for the physical graviton, and
\begin{equation}
\sum_{j_{z} = 0, \pm1, \pm 2} \epsilon_{\mu\nu}(p,j_{z}) \epsilon^{*}_{\rho\sigma}(p,j_{z}) = 
\frac{1}{2} \Bigl( \eta_{\mu\rho} \eta_{\nu\sigma} 
+ \eta_{\mu\sigma} \eta_{\nu\rho} 
- \frac{2}{3} \eta_{\mu\nu} \eta_{\rho\sigma} \Bigr) .
\end{equation}
for the massive Merlin. Hence a convenient prescription for the propagator would be
\begin{equation}
D_{\mu\nu\rho\sigma}(p) = 
\frac{1}{2 p^2} \Bigl( \eta_{\mu\rho} \eta_{\nu\sigma} + \eta_{\mu\sigma} \eta_{\nu\rho} 
- \eta_{\mu\nu} \eta_{\rho\sigma} \Bigr) 
- \frac{1}{2 (p^2 - M^2)} \Bigl( \eta_{\mu\rho} \eta_{\nu\sigma} 
+ \eta_{\mu\sigma} \eta_{\nu\rho} 
- \frac{2}{3} \eta_{\mu\nu} \eta_{\rho\sigma} \Bigr) 
\end{equation}
where the numerator of the Merlin mode is the $\eta$-dependent part of the spin-$2$ projector. A similar form was also considered in Ref.~\cite{Johansson:18}. 

We can rewrite the expressions for the polarization sums in terms of the usual projectors for symmetric second-rank tensors in momentum space~\cite{shapiro}. However, if we are only interested in the spin-two contribution, one finds that
\begin{equation}
\sum_{j_{z}} \epsilon_{\mu\nu}(p,j_{z}) \epsilon_{\rho\sigma}(p,j_{z}) 
\bigg|_{\textrm{physical, spin-2}} = 
{\cal P}^{(2)}_{\mu\nu\rho\sigma} .
\end{equation}
On the other hand, the spin-two part of the propagator can be written generically as
\begin{equation}
iD_{\mu\nu\alpha\beta}(q) = i {\cal P}^{(2)}_{\mu\nu\alpha\beta} D_2(q) .
\end{equation}
This is the only contribution to the propagator that is gauge invariant and independent of the choice of the field parametrization. One sees that the numerator of the spin-two propagator equals the spin-two part of the sum over physical polarizations. 

In the case of quadratic gravity, the propagator in the coordinate representation reads
\begin{eqnarray}
iD_{\mu\nu\alpha\beta}(x-x') &=& \frac{1}{2} \Bigl( \eta_{\mu\alpha} \eta_{\nu\beta} 
+ \eta_{\mu\beta} \eta_{\nu\alpha} - \eta_{\mu\nu} \eta_{\alpha\beta} \Bigr)
\int \frac{d^{4} q}{(2\pi)^4} e^{-i q \cdot (x-x')} i D_2(q)
\nonumber\\
&+& \frac{1}{2} \Bigl( \eta_{\mu\rho} \eta_{\nu\sigma} 
+ \eta_{\mu\sigma} \eta_{\nu\rho} 
- \frac{2}{3} \eta_{\mu\nu} \eta_{\rho\sigma} \Bigr)
\int \frac{d^{4} q}{(2\pi)^4} e^{-i q \cdot (x-x')} i \widetilde{D}_2(q)
\nonumber\\
D_2(q) &=& \frac{1}{q^2 + i \epsilon},
\,\,\,\,
\widetilde{D}_2(q) = - \frac{1}{q^2 - M^2 - i M \Gamma} .
\end{eqnarray}
Hence the decomposition into positive- and negative-frequency parts reads
\begin{eqnarray}
i D_{\mu\nu\alpha\beta}(x-x') &=& \frac{1}{2} \Bigl( \eta_{\mu\alpha} \eta_{\nu\beta} 
+ \eta_{\mu\beta} \eta_{\nu\alpha} - \eta_{\mu\nu} \eta_{\alpha\beta} \Bigr) 
\left[ \theta(x_0-x_0') D^{+}_{2}(x-x') + \theta(-x_0+x_0') D^{-}_{2}(x-x') \right]
\nonumber\\
&+&  \frac{1}{2} \Bigl( \eta_{\mu\rho} \eta_{\nu\sigma} 
+ \eta_{\mu\sigma} \eta_{\nu\rho} 
- \frac{2}{3} \eta_{\mu\nu} \eta_{\rho\sigma} \Bigr)
\left[ \theta(x_0-x_0') \widetilde{D}^{-}_{2}(x-x') + \theta(-x_0+x_0') \widetilde{D}^{+}_{2}(x-x') \right]
\end{eqnarray}
where $D^{\pm}_{2}(x-x')$ ($\widetilde{D}^{\pm}_{2}(x-x')$) can be written in the form discussed previously for the scalar stable (Lee-Wick) particle. In this way all the results discussed in Ref.~\cite{DM:19} now hold for the case of quadratic gravity. Hence Slavnov-Taylor identities (for the scattering amplitudes) together with gauge invariance (for the propagator) and the fact that unstable particles are not to be included in unitarity sums, guarantees unitarity for quadratic gravity to all orders in perturbation theory.

We can also discuss unitarity of quadratic gravity in the light of double-copy theories~\cite{KLT,Bern:08,Bern:10,Johansson:17}. That is, assuming that the double-copy relation between Yang-Mills theories and gravity is valid to all orders in perturbation theory, the unitarity of higher-derivative Yang-Mills theory implies that of quadratic gravity. 

As a final remark, we again emphasize that unstable propagators must be resummed and, if not done carefully, one will certainly face issues associated with gauge invariance and gauge-fixing parameter dependence~\cite{Papavassiliou:1995fq,Papavassiliou:1995gs,Papavassiliou:1996zn,Papavassiliou:1997fn,Papavassiliou:1997pb,Rodenburg,Lang-thesis,Denner:15}. This problem is particularly important for gauge theories and gravity. In any case, the assumption of the narrow-width approximation allows us to partially circumvent such difficulties.

\end{document}